\documentclass[11pt]{article}

\usepackage[usenames,dvipsnames,table]{xcolor}

\usepackage{jheppub} 

\usepackage{graphicx}
\usepackage{amsmath, amssymb}

\usepackage{enumerate}

\newcommand{\be}{\begin{eqnarray}}
\newcommand{\ee}{\end{eqnarray}}
\newcommand{\bea}{\begin{eqnarray}}
\newcommand{\eea}{\end{eqnarray}}

\newcommand{\nn}{\nonumber}
\newcommand{\bn}{\begin{enumerate}}
\newcommand{\en}{\end{enumerate}}

\def\Tr{\mathop{\mathrm{Tr}}\nolimits}

\def\e{\epsilon}

\def\e{\textrm{e}}

\def\cQ{{\cal Q}}

\newcommand{\udl}[1]{\mathrm{d} #1 \,}

\newcommand{\sbfunc}[1]{s_b\left( #1\right)}

\newcommand{\Gpq}[1]{\Gamma_e\left( #1\right)}

\newcommand{\tsu}{$T[SU(N)]$\,}
\newcommand{\mirrortsu}{$T[SU(N)]^\vee~$}
\newcommand{\fftsu}{$FFT[SU(N)]$\,}
\newcommand{\mirrorfftsu}{$FFT[SU(N)]^\vee~$}
\newcommand{\eusp}{$E[USp(2N)]$\,}
\newcommand{\mirroreusp}{$E[USp(2N)]^\vee~$}
\newcommand{\ffeusp}{$FFE[USp(2N)]$\,}
\newcommand{\mirrorffeusp}{$FFE[USp(2N)]^\vee~$}

\newcommand{\hb}{\mathcal{H}}
\newcommand{\cb}{\mathcal{C}}

\def\ga{\alpha}
\def\gb{\beta}
\def\gc{\gamma}

\def\gd{\delta}

\def\gs{\sigma}

\def\gr{\rho}

\def\Gp{\Phi}

\title{4d mirror-like dualities}

\preprint{}

\author[a]{Chiung Hwang}
\author[a]{Sara Pasquetti}
\author[a]{Matteo Sacchi}

\affiliation[a]{Dipartimento di Fisica, Universit\`a di Milano-Bicocca \& INFN, Sezione di Milano-Bicocca,
I-20126 Milano, Italy}

\emailAdd{chiung.hwang@unimib.it}
\emailAdd{sara.pasquetti@gmail.com} 
\emailAdd{m.sacchi13@campus.unimib.it}

\abstract{
We construct a family of $4d$ $\mathcal{N}=1$ theories that we call $E^\sigma_\rho[USp(2N)]$ which exhibit  a novel type of $4d$ IR duality  very reminiscent of the  mirror duality enjoyed 
by the $3d$ $\mathcal{N}=4$  $T^\sigma_\rho[SU(N)]$ theories.
 We obtain  the $E^\sigma_\rho[USp(2N)]$ theories from the recently introduced  $E[USp(2N)]$ theory, by following the RG flow 
initiated by  vevs labelled by partitions $\rho$ and $\sigma$ for two operators  transforming in the antisymmetric representations of the $USp(2N) \times USp(2N)$ IR symmetries of the    $E[USp(2N)]$ theory.
These vevs are the $4d$ uplift of the ones we turn on for the moment maps of $T[SU(N)]$
to trigger the flow to $T^\sigma_\rho[SU(N)]$. Indeed   the  $E[USp(2N)]$ theory, upon dimensional reduction and suitable real mass deformations, reduces to the $T[SU(N)]$ theory.
In order to study the RG flows triggered by the vevs we develop a new strategy based on the duality webs of the
 $T[SU(N)]$ and  $E[USp(2N)]$ theories.
}

\begin{document} 

\maketitle
\flushbottom

\section{Introduction}

Recently in  \cite{Pasquetti:2019hxf} it has been observed that a $4d$ $\mathcal{N}=1$ quiver theory, called \eusp,   is left invariant by the action of an infra-red (IR) duality which is  reminiscent of $3d$ $\mathcal{N}=4$ mirror symmetry \cite{Intriligator:1996ex}. This duality does not seem to be related to Seiberg dualities \cite{Seiberg:1994pq} and it appears to be of a genuinely new type.
Moreover in a suitable  $3d$ limit followed by various mass deformations, the $4d$ $\mathcal{N}=1$ \eusp theory reduces to the familiar $3d$ $\mathcal{N}=4$ $T[SU(N)]$ theory introduced in \cite{Gaiotto:2008ak}   and the $4d$ self-duality  reduces to  the mirror self-duality of $T[SU(N)]$.
This represents the first example of derivation of a $3d$ mirror duality from a $4d$ IR duality\footnote{A derivation of a $3d$ mirror duality from $6d$ has been discussed recently in \cite{Razamat:2019sea}.}. Indeed so far most of the known $3d$ IR dualities with the exception of mirror dualities have been shown to have $4d$ ancestors, which upon compactifications followed by various real mass deformations reproduce Seiberg-like dualities in $3d$ \cite{Aharony:2013dha,Aharony:2013kma,Csaki:2014cwa,Nii:2014jsa,Amariti:2015vwa,Amariti:2016kat,Benini:2017dud,Hwang:2018uyj,Benvenuti:2018bav,Amariti:2018wht,Nii:2019wxi}.

\begin{figure}[t]
	\centering
  	\includegraphics[scale=0.7]{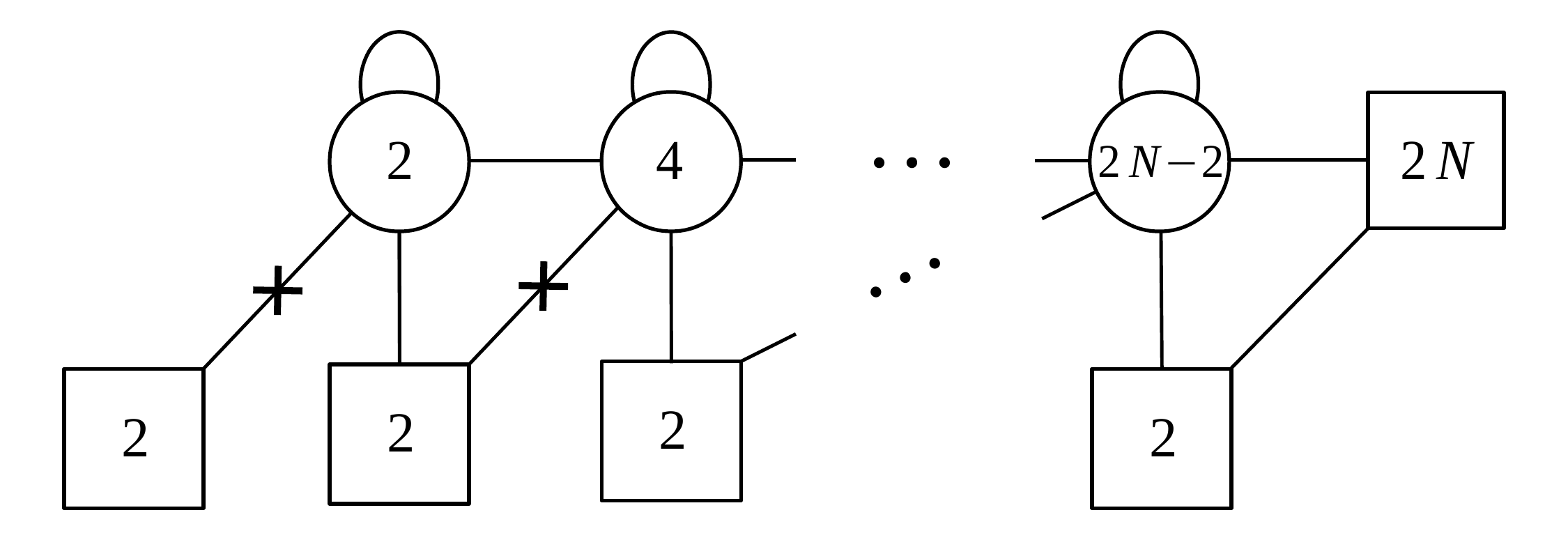} 
    \caption{Quiver diagram of the \eusp theory.}
 \label{eblockintro}
\end{figure}

In this work, starting from \eusp we will construct a family of $4d$ $\mathcal{N}=1$ theories that we call $E^\gs_\gr[USp(2N)]$, which are related by mirror-like dualities and which reduce in the $3d$ limit   to the $T^\sigma_\rho[SU(N)]$ theories introduced in \cite{Gaiotto:2008ak} that are related by mirror dualities.

The \eusp is the quiver gauge theory depicted in Figure \ref{eblockintro}, where all the nodes denote $USp(2n)$ symmetries. This theory has a $USp(2N)_x\times USp(2N)_y\times U(1)_t\times U(1)_c$ global symmetry, with the second $USp(2N)_y$ being enhanced in the IR from the $SU(2)$ symmetries of the saw\footnote{The definition of the \eusp theory we use here is slightly different from the one of \cite{Pasquetti:2019hxf}, which included an extra set of singlet
 fields flipping the meson matrix constructed with the chirals at the end of the tail and transforming in the antisymmetric representation of $USp(2N)_x$. Consequently, the self-dualities of \eusp we consider here are slightly different from those discussed in \cite{Pasquetti:2019hxf}.}.
This theory was used as a building block in \cite{Pasquetti:2019hxf} to construct more complicated four-dimensional theories that were shown to arise from the compactification of the $6d$ $\mathcal{N}=(1,0)$ rank-N E-string theory on Riemann surfaces with fluxes for the $E_8$ part of its $E_8\times SU(2)_L$ global symmetry. As such, some of these theories exhibit interesting global symmetry enhancements, according to the subgroup of the $6d$ $E_8$ global symmetry preserved by the flux.

The duality leaving the  \eusp theory  invariant acts by exchanging operators charged under $USp(2N)_x$ with those charged under $USp(2N)_y$ much like the  mirror self-duality for the $3d$ $\mathcal{N}=4$ $T[SU(N)]$ theory exchanges the Higgs branch operators in the adjoint of the flavor $SU(N)$ group with the Coulomb branch operators in the adjoint of the other $SU(N)$ group  emerging in the IR as an enhancement of the topological symmetries.
In particular two of the  \eusp operators transforming under the $USp(2N)_x$ and  $USp(2N)_y$ global symmetry and which are exchanged by the $4d$ duality reduce to the Coulomb and Higgs branch moment maps of $T[SU(N)]$ which are swapped by Mirror Symmetry. 
In this sense we consider the self-duality of \eusp, which is a $4d$ ancestor of the self-duality under Mirror Symmetry of $T[SU(N)]$, a $4d$ mirror-like self-duality.

Many other mirror dualities are known in $3d$. For example, closely related to $T[SU(N)]$ is the class of $\mathcal{N}=4$ $T^\gs_\gr[SU(N)]$ theories introduced by Gaiotto and Witten in \cite{Gaiotto:2008ak}, where $\sigma$ and $\rho$ are partitions of $N$. The $T_\rho^\sigma[SU(N)]$ theory can be realized on a brane set-up \cite{Hanany:1996ie} with $N$ D3-branes suspended between $K$ D5-branes and $L$ NS5-branes, where $K$ and $L$ are the lengths of the partitions $\gs$ and $\gr$ respectively. The integers $\sigma_i$ in $\sigma =[\sigma_1,\cdots, \sigma_K]$ are the net number of D3-branes ending on the D5-branes going from the interior to the exterior of the configuration, while the integers $\rho_i$ in  $\rho=[\rho_1,\cdots, \rho_{L}]$ are the net number of D3-branes ending on the NS5 branes again going from the interior to the exterior. 

It is then natural to wonder whether it is possible to find a $4d$ ancestor for $T^\gs_\gr[SU(N)]$ and construct a family of $4d$ theories
enjoying mirror-like dualities. As a brane realisation is not available in $4d$ we need to rely on field theory methods only.

\begin{figure}[t]
	\centering
  	\includegraphics[scale=0.7]{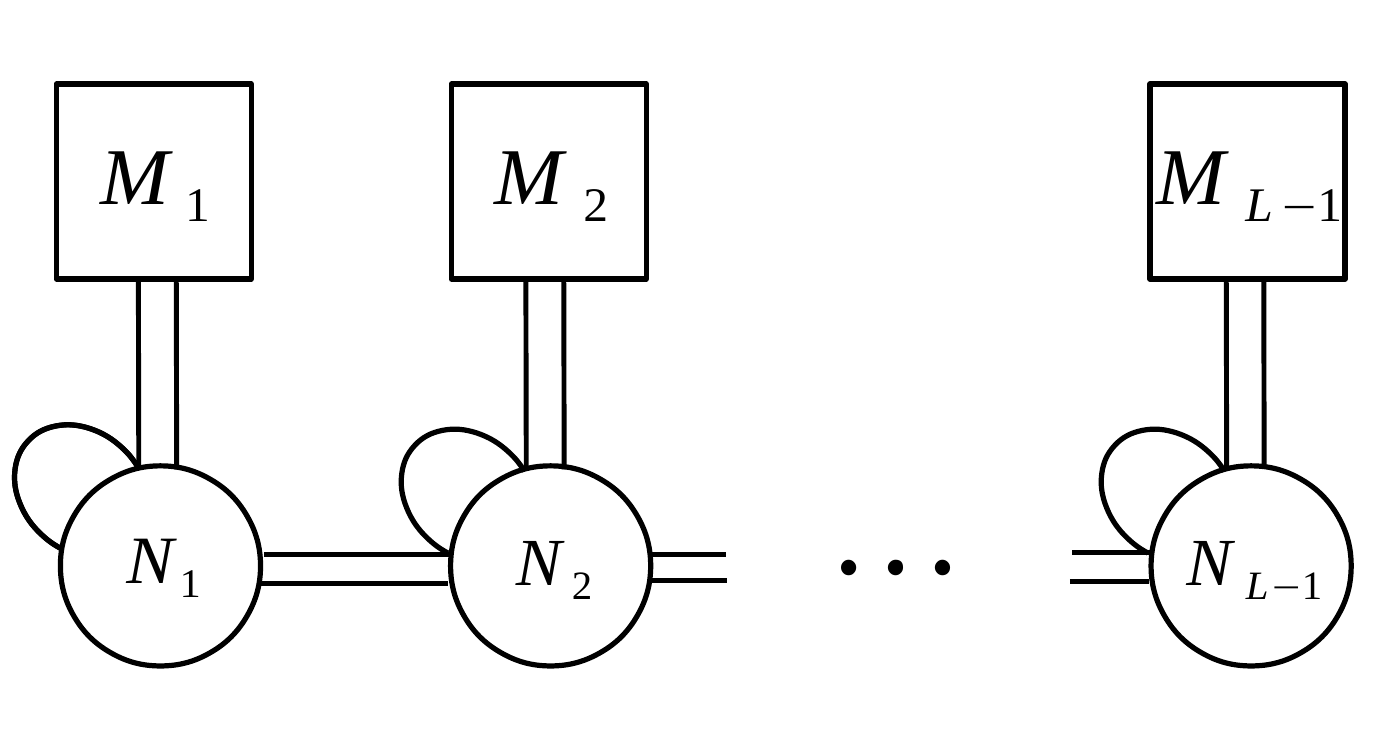} 
	\label{trsgen}
	\caption{The $T_\rho^\sigma[SU(N)]$ quiver. Ranks $N_i, M_i$ are as in eq. \eqref{rsranks}.}
\end{figure}

The structure of the $T_\rho^\sigma[SU(N)]$  quiver, with $\sigma^T<\rho$,  depicted in Figure \ref{trsgen},  is dictated by the partitions which we rewrite as
\begin{align}
\rho = \left[N^{l_N}, \dots, 1^{l_1}\right] \,, \qquad \sigma = \left[N^{k_N}, \dots, 1^{k_1}\right]
\end{align}
where some of the $l_n, \, k_m$ integers can be zero and must satisfy the conditions
\begin{gather}
\sum_{n = 1}^N n \times l_n = \sum_{m = 1}^N m \times k_m = N \,, \nn\\
L = l_1+ \dots +l_N \,, \qquad K = k_1+ \dots +k_N \,.
\end{gather}
The gauge and flavor ranks $N_i, \, M_i$ are  given by
\begin{align}
M_{L-i} &= k_i \,, \nn\\
N_{L-i} &= \sum_{j = i+1}^L \rho_j-\sum_{j = i+1}^N (j-i) k_j \,.
\label{rsranks}
\end{align}

The $T_\rho^\sigma[SU(N)]$  global symmetry group is $S(\prod_{i=1}^N U(k_i))\times S(\prod_{i=1}^N U(l_i))$.
While the factor $S(\prod_{i=1}^N U(k_i))$ acting on the Higgs branch is visibile in the UV Lagrangian, the factor
$S(\prod_{i=1}^N U(l_i))$ acting on the Coulomb branch appears only in the IR as an enhancement of the topological symmetries.
This pattern of symmetry enhancement is consistent with the prediction of  Mirror Symmetry stating that 
$T_\rho^\sigma[SU(N)]$ is  mirror dual to $T^\rho_\sigma[SU(N)]$.

The $T^\gs_\gr[SU(N)]$ theory can be reached from  the $T[SU(N)]$ theory by giving a nilpotent vev to the Higgs and the Coulomb branch moment maps labelled by $\gs$ and $\gr$ respectively. 
These vevs initiate sequential Higgs mechanisms which are quite intricate to follow\footnote{In \cite{Cremonesi:2014uva} the vev was implemented at the level of the Hilbert series by means of a residue procedure.}.
 Indeed one typically relies on the brane  realisation of the theory. 
Here we propose an alternative procedure to systematically derive $T^\gs_\gr[SU(N)]$ theories from $T[SU(N)]$ which is based on field theory methods only. We will then apply the same procedure in $4d$  to the \eusp theory to construct a new family of $4d$ theories, which we name $E^\gs_\gr[USp(2N)]$ theories, enjoying  mirror-like dualities.

\begin{figure}[t]
	\centering
  	\includegraphics[scale=0.48]{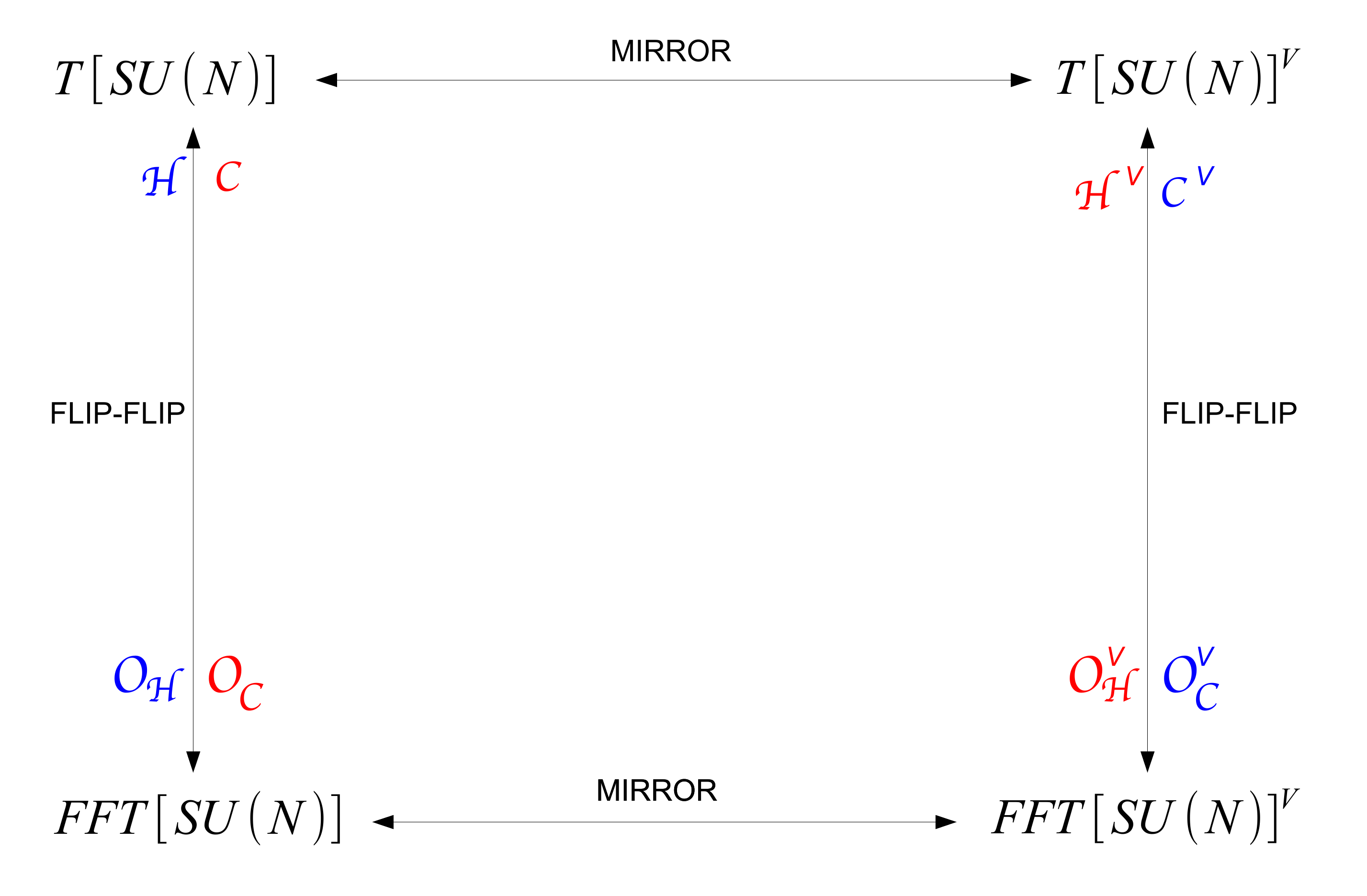} 
    \caption{Duality web for $T[SU(N)]$.}
 \label{tsunweb}
\end{figure}

Our approach relies on a web of dualities for $T[SU(N)]$ that was discussed in \cite{Aprile:2018oau}. This web, depicted in Figure \ref{tsunweb}, is constructed combining two dualities for \tsu: one is the standard self-duality under Mirror Symmetry discussed in the original paper \cite{Gaiotto:2008ak} and the other is called flip-flip duality \cite{Aprile:2018oau}. We recall that under Mirror Symmetry the Higgs and the Coulomb branch of the theory are exchanged. Hence, if we denote with $\hb$ and $\cb$ the Higgs and Coulomb branch moment maps of \tsu and with $\hb^\vee$ and $\cb^\vee$ those of the mirror dual \mirrortsu, we have the operator map
\be
\hb\quad&\leftrightarrow&\quad \cb^\vee\nn\\
\cb\quad&\leftrightarrow&\quad \hb^\vee\, .
\ee
This duality corresponds to the upper edge of the diagram  of Figure \ref{tsunweb}. 

On top of the mirror dual frame there exists another flip-flip dual frame called $FFT[SU(N)]$. The latter theory is defined starting from $T[SU(N)]$ and adding two sets of singlet fields $\mathcal{O_H}$ and $\mathcal{O_C}$ that flip both the Higgs and the Coulomb branch moment maps
\be
\mathcal{W}_{FFT[SU(N)]}=\mathcal{W}_{T[SU(N)]}+\Tr_X\left(\mathcal{O_H}\, \hb_{FF}\right)+\Tr_Y\left(\mathcal{O_C}\,\cb_{FF}\right)\, ,
\ee
where $\hb_{FF}$ and $\cb_{FF}$ denote the moment maps of the dual $FFT[SU(N)]$ and the $X,Y$ subscripts in the traces refer to the IR global $SU(N)_X\times SU(N)_Y$ symmetry groups.
The moment maps $\hb$ and $\cb$ of the original $T[SU(N)]$ theory are mapped across this duality to the two sets of flipping fields $\mathcal{O_H}$ and $\mathcal{O_C}$
\be
\hb\quad&\leftrightarrow&\quad \mathcal{O_H}\nn\\
\cb\quad&\leftrightarrow&\quad \mathcal{O_C}\, .
\ee
This duality corresponds to the left vertical edge of the diagram of Figure \ref{tsunweb}. 
As we will show the flip-flip duality can be derived by applying sequentially the Aharony duality \cite{Aharony:1997gp}.

Combining Mirror Symmetry and flip-flip duality we can find a third dual frame, which we denote by \mirrorfftsu. The superpotential of the theory is
\be
\mathcal{W}_{FFT[SU(N)]^\vee}=\mathcal{W}_{T[SU(N)]}+\Tr_Y\left(\mathcal{O}_{\mathcal{H}}^\vee\, \hb^\vee_{FF}\right)+
\Tr_X\left(\mathcal{O}_{\mathcal{C}}^\vee\,\cb^\vee_{FF}\right)\, .
\ee
The operator map between the original \tsu and \mirrorfftsu ~is
\be
\hb\quad&\leftrightarrow&\quad \mathcal{O}_{\mathcal{C}}^\vee\nn\\
\cb\quad&\leftrightarrow&\quad \mathcal{O}_{\mathcal{H}}^\vee\, .
\ee
The order in which we apply Mirror Symmetry and flip-flip duality doesn't affect the result, so that we obtain the commutative diagram of Figure \ref{tsunweb}. \\

\begin{figure}[t]
	\centering
  	\includegraphics[scale=0.48]{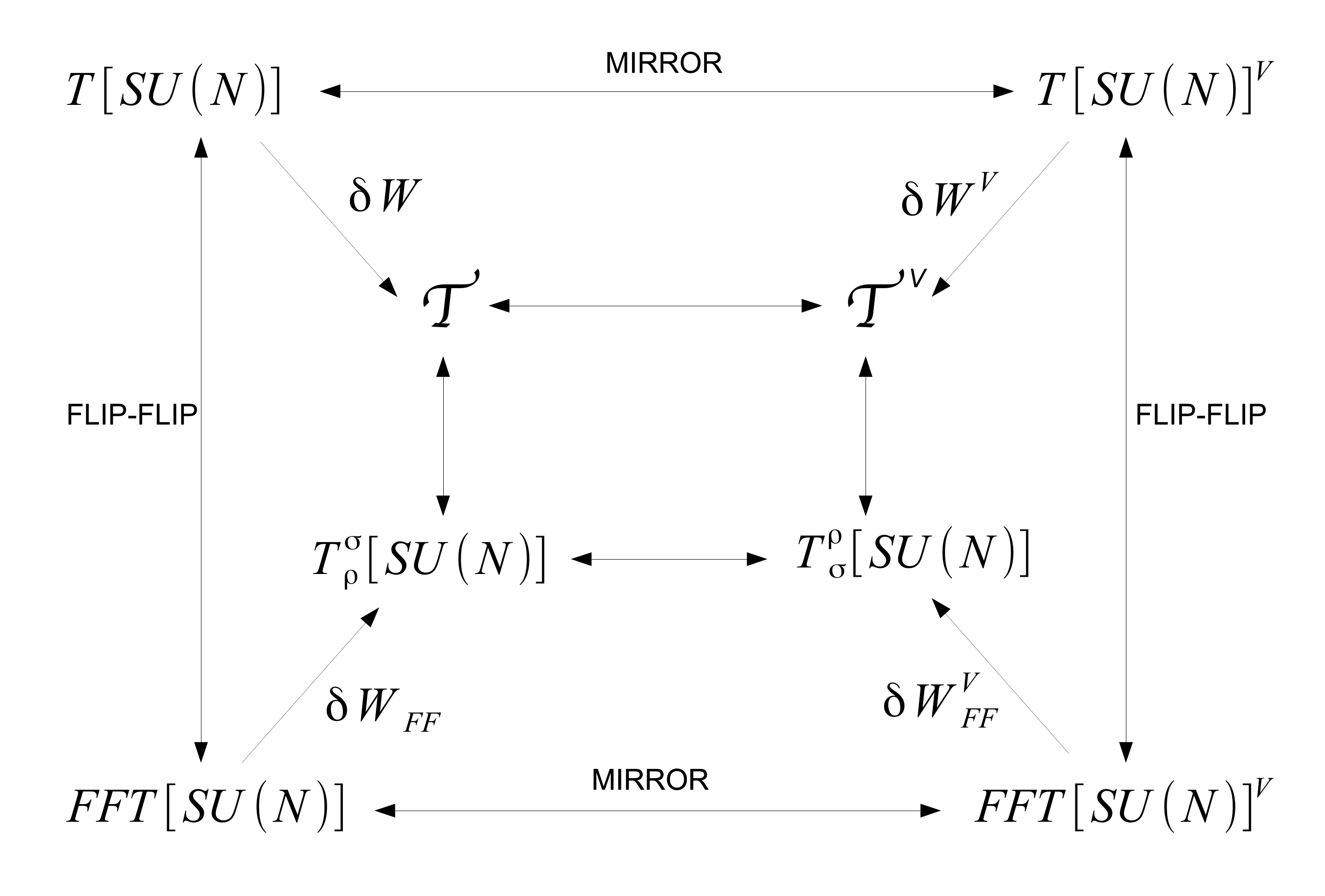} 
    \caption{Deformed duality web for $T[SU(N)]$.}
 \label{tsundefweb}
\end{figure}  

In order to study the nilpotent vev of $T[SU(N)]$, we notice that it can be implemented by adding singlets flipping some components of its moment maps and by turning them on linearly in the superpotential. The F-term equations of the singlets then fix the vev of these components of the moment maps to a non-vanishing value.
Hence the IR theory obtained turning on a vev in $T[SU(N)]$ is equivalently reached by deforming $FFT[SU(N)]$ by a linear superpotential in some of  the components of $\mathcal{O_H}$ and $\mathcal{O_C}$ and by removing those that become free after the deformation.
that is, we claim that by deforming $FFT[SU(N)]$ by:
\be
\delta\mathcal{W}_{FF}=\Tr_X\left[\left(\mathcal{J}_\gs+\mathcal{S}_\gs\right) \mathcal{O_H}\right]+\Tr_Y\left[\left(\mathcal{J}_\gr+\mathcal{T}_\gr\right) \mathcal{O_C}\right]\,,
\ee
where $\mathcal{J}_\gs$ and $\mathcal{J}_\gr$ are block diagonal Jordan matrices encoding the vev, while $\mathcal{S}_\gs$ and $\mathcal{T}_\gr$ are matrices of gauge singlets (both of these will be described in more details in the main text),
we flow to  $T^\gs_\gr[SU(N)]$ as shown in  the bottom left corner of Figure \ref{tsundefweb}.

Using the flip-flip duality, we can map this deformation into a deformation of $T[SU(N)]$ linear in the entries of the moment maps, that is, in this frame rather than turning on vevs, we turn on  mass and monopole deformations:
\be
\delta\mathcal{W}=\Tr_X\left[\left(\mathcal{J}_\gs+\mathcal{S}_\gs\right) \mathcal{H}\right]+\Tr_Y\left[\left(\mathcal{J}_\gr+\mathcal{T}_\gr\right) \mathcal{C}\right]\,.
\ee
This deformation triggers a flow to theory $\mathcal{T}$, in the upper left corner of Figure \ref{tsundefweb},
which is flip-flip dual to  $T^\gs_\gr[SU(N)]$.
We  will show that moving along the vertical edge of the web from $\mathcal{T}$ to $T^\gs_\gr[SU(N)]$ by means of the flip-flip duality  is equivalent to iteratively applying  a combination of the Aharony and one-monopole duality \cite{Benini:2017dud}.
Flowing from $T[SU(N)]$ to $\mathcal{T}$ and then to  $T^\gs_\gr[SU(N)]$ allows us to bypass the  study  of the 
sequential Higgs mechanism initiated by the vevs, which, in the case of monopole vev, is particularly complicated.

We can then apply the same procedure to the mirror dual frame. The  $T^\gr_\gs[SU(N)]$ can be obtained by deforming   \mirrorfftsu ~by a linear superpotential 
\be
\delta\mathcal{W}^\vee_{FF}=\Tr_Y\left[\left(\mathcal{J}_\rho+\mathcal{T}_\rho\right) \mathcal{O}_{\mathcal{H}}^\vee\right]+\Tr_X\left[\left(\mathcal{J}_\sigma+\mathcal{S}_\sigma\right) \mathcal{O}_{\mathcal{C}}^\vee\right]\,,
\ee
as shown in the bottom right corner,  which corresponds, in the flip-flip dual frame, to a deformation of $T[SU(N)]^\vee$ by  
\be
\delta\mathcal{W}^\vee=\Tr_Y\left[\left(\mathcal{J}_\rho+\mathcal{T}_\rho\right) \mathcal{H}^\vee\right]+\Tr_X\left[\left(\mathcal{J}_\sigma+\mathcal{S}_\sigma\right) \mathcal{C}^\vee\right]\,.
\ee
This deformation triggers a flow to theory $\mathcal{T}^\vee$, upper right  corner in Figure \ref{tsundefweb},
which is flip-flip dual to  $T^\rho_\sigma[SU(N)]$.\\

\begin{figure}[t]
	\centering
  	\includegraphics[scale=0.48]{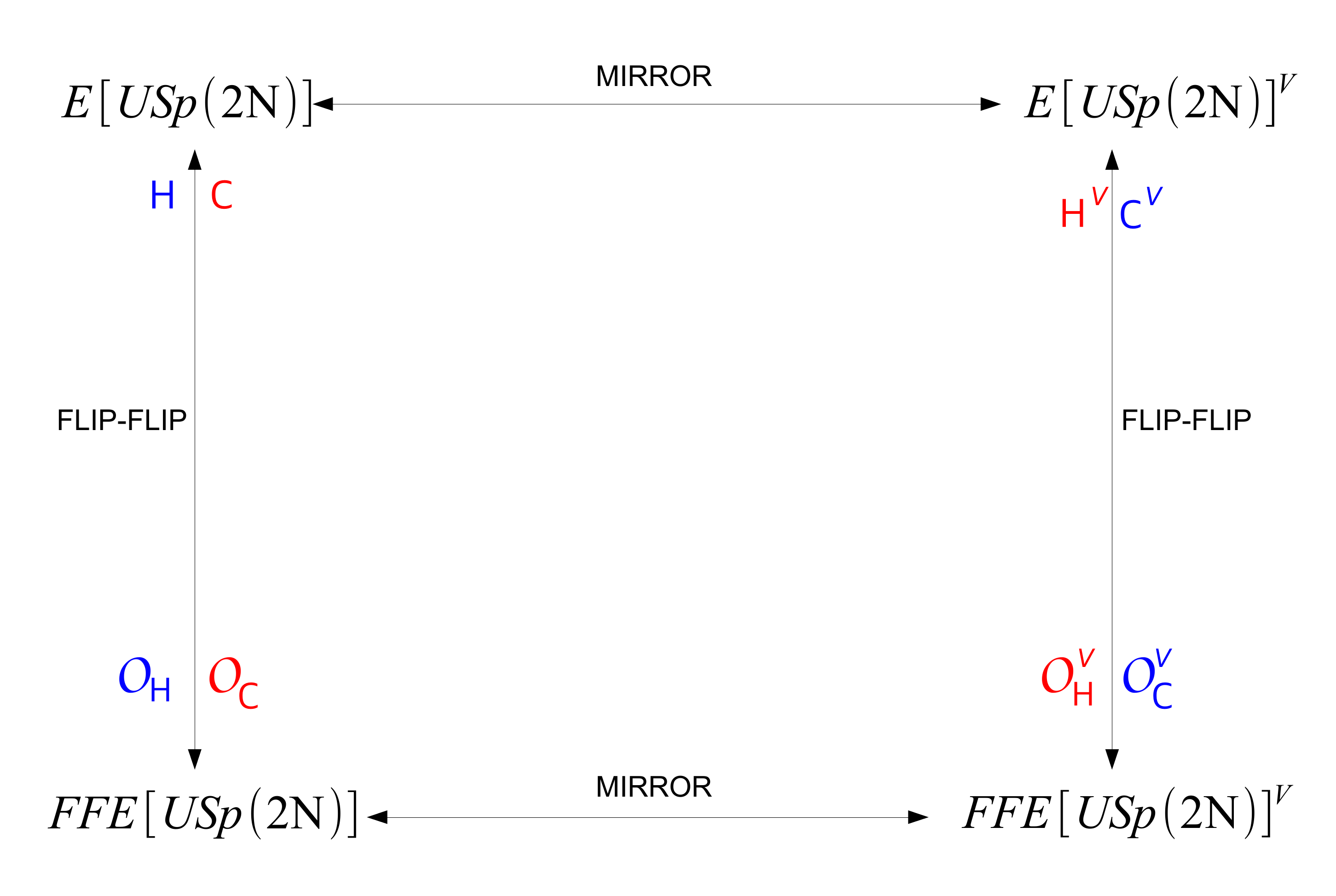} 
    \caption{Duality web for $E[USp(2N)]$.}
 \label{euspweb}
\end{figure}

Having established this alternative procedure for deriving $T^\gs_\gr[SU(N)]$ from $T[SU(N)]$ we will export it to $4d$ to construct, starting from  \eusp, a new class of theories  that we  call $E^\gs_\gr[USp(2N)]$ and that are related by  mirror-like dualities.

Indeed, also the \eusp theory enjoys a web of dualities,  similar to the $T[SU(N)]$ web, depicted in Figure \ref{euspweb}.\footnote{In \cite{Pasquetti:2019hxf} it was observed that the superconformal index of the \eusp theory  coincides with the interpolation kernel $\mathcal{K}_c(x,y)$ studied in \cite{2014arXiv1408.0305R}. The  kernel $\mathcal{K}_c(x,y)$ satisfies various highly non-trivial integral identities corresponding to the equality of the indices of the theories at the four corners of the duality web. These identities provide strong evidences for the existence of these dualities.}
This web is obtained combining the $4d$  mirror-like duality and the flip-flip duality. As we mentioned before, \eusp possesses two $USp(2N)$ global symmetries, one of which is enhanced in the IR from the $SU(2)$ symmetries of the saw.
As we will see we can construct two sets of operators transforming  in the traceless antisymmetric representation of the $USp(2N)_x$ and of the  enhanced $USp(2N)_y$ symmetry, that we denote with $\mathsf{H}$ and $\mathsf{C}$. 
In the limit in which \eusp reduces to $T[SU(N)]$  the  operators $\mathsf{H}$ and $\mathsf{C}$ reduce to the Higgs and Coulomb branch moment maps $\hb$ and $\cb$ of \tsu.
The $4d$ mirror-like duality for \eusp acts by exchanging all the operators charged under $USp(2N)_x$ with those charged under $USp(2N)_y$. It also acts non-trivially on the $U(1)_t$ symmetry, while leaving the $U(1)_c$ charges unchanged. 
In particular the operators $\mathsf{H}$ and $\mathsf{C}$ in  $E[USp(2N)]$ 
and $\mathsf{H}^\vee$ and $\mathsf{C}^\vee$ in the dual $E[USp(2N)]^\vee$ are mapped as follows:
\be
\mathsf{H}\quad&\leftrightarrow&\quad \mathsf{C}^\vee\nn\\
\mathsf{C}\quad&\leftrightarrow&\quad \mathsf{H}^\vee\, .
\ee

The flip-flip duality instead relates \eusp with \ffeusp, which is defined as \eusp  with  two extra sets of singlets $\mathsf{O_H}$ and $\mathsf{O_C}$:
\be
\mathcal{W}_{FFE[USp(2N)]}=\mathcal{W}_{E[USp(2N)]}+\Tr_x\left(\mathsf{O_H}\mathsf{H}_{FF}\right)+\Tr_y\left(\mathsf{O_C}\mathsf{C}_{FF}\right)\, ,
\ee
where the $x,y$ subscripts in the traces refer to the IR global $USp(2N)_x\times USp(2N)_y$ symmetry groups.
Across this duality, we have the operator map
\be
\mathsf{H}\quad&\leftrightarrow&\quad \mathsf{O}_{\mathsf{H}}^\vee\nn\\
\mathsf{C}\quad&\leftrightarrow&\quad \mathsf{O}_{\mathsf{C}}^\vee\, ,
\ee
meaning that flip-flip duality leaves unchanged the two $USp(2N)$ symmetries, but it acts non-trivially on the abelian global symmetries of the theory. Moreover, similarly to the $3d$ case, flip-flip duality can be derived by sequentially applying the more fundamental Intriligator--Pouliot duality \cite{Intriligator:1995ne}.

These two dualities can be combined to find a third dual frame \mirrorffeusp and to construct a duality web for \eusp, represented in Figure \ref{euspweb}, which is analogous to the one of \tsu
\be
\mathcal{W}_{FFE[USp(2N)]^\vee}=\mathcal{W}_{E[USp(2N)]}+\Tr_y\left(\mathsf{O}_\mathsf{H}^\vee\mathsf{H}_{FF}^\vee\right)+\Tr_x\left(\mathsf{O}_{\mathsf{C}}^\vee\mathsf{C}_{FF}^\vee\right)\, .
\ee
Across this last duality, we have the operator map
\be
\mathsf{H}\quad&\leftrightarrow&\quad \mathsf{O}_{\mathsf{C}}^\vee\nn\\
\mathsf{C}\quad&\leftrightarrow&\quad \mathsf{O}_\mathsf{H}^\vee\, .
\ee

\begin{figure}[t]
	\centering
  	\includegraphics[scale=0.48]{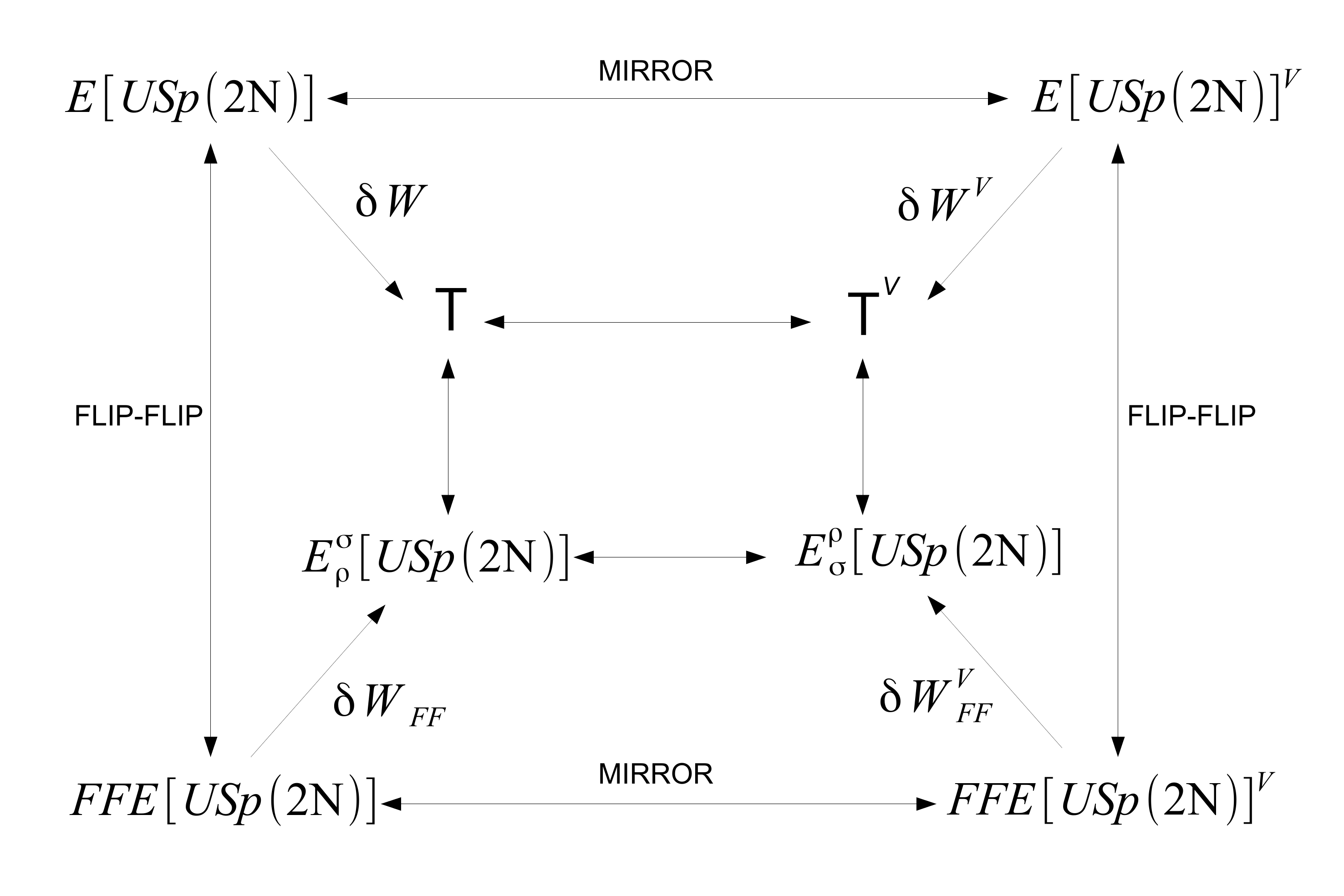} 
    \caption{Deformed duality web for $E[USp(2N)]$.}
 \label{euspweb2}
\end{figure}

\begin{figure}[t]
	\centering
  	\includegraphics[scale=0.7]{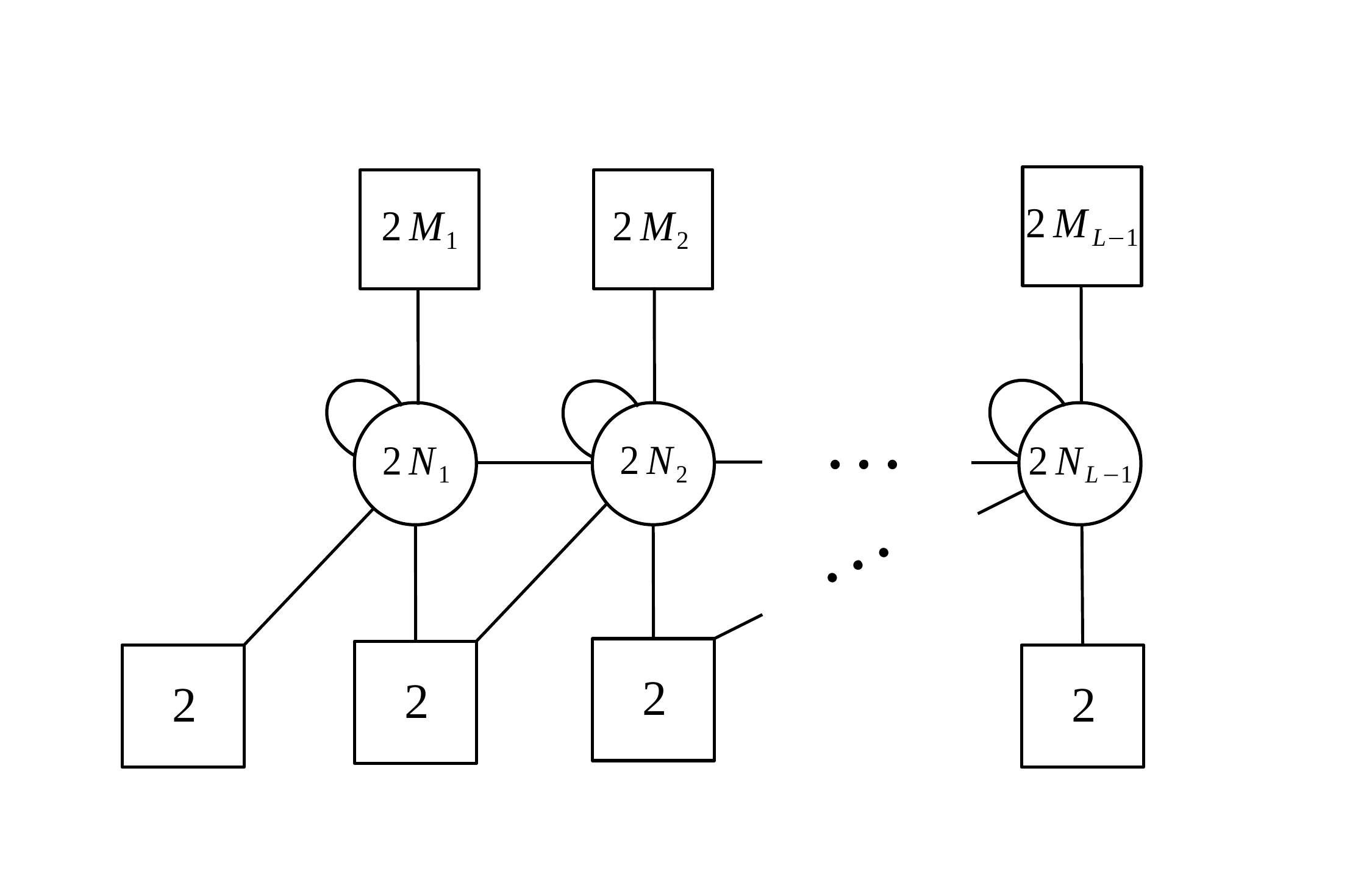} 
    \caption{Schematic structure of the $E^\gr_\gs[USp(2N)]$ theory.  Ranks $N_i, M_i$ are as in \eqref{rsranks}.}
 \label{Erhosigma}
\end{figure}

In analogy with the $3d$ case, it is natural to consider deformations of the \eusp theory triggered by vevs  of the 
operators $\mathsf{C}$ and $\mathsf{H}$.  
Studying the Higgsing initiated by such vevs is however quite tricky and in the $4d$ case  we don't have a brane realisation for \eusp. However we can implement the same procedure we described to obtain $T^\gs_\gr[SU(N)]$, starting from the \eusp web,  as sketched in Figure \ref{euspweb2}.

 We name $E^\gs_\gr[USp(2N)]$ the theories obtained turning on vevs for $\mathsf{C}$ and $\mathsf{H}$ labelled by partitions of $N$ $\rho$ and $\sigma$. They are the quiver theories with $USp(2n)$ gauge and flavor nodes depicted in Figure \ref{Erhosigma}, where the ranks $N_i$ and $M_i$ are related to the data of the partitions $\gs$ and $\gr$ as in \eqref{rsranks}.
There are also additional singlet fields which we will discuss in the main text.

Because of the vev, the two $USp(2N)$ global symmetries of \eusp are broken to subgroups, according to the particular partitions chosen. Moreover, as a consequence of the duality web we have that $E^\gs_\gr[USp(2N)]$ is dual to $E^\gr_\gs[USp(2N)]$. This duality is a $4d$ version of the mirror duality between $T^\gs_\gr[SU(N)]$ and $T^\gr_\gs[SU(N)]$. 
It implies that the $SU(2)$ symmetries of the saw of $E^\gs_\gr[USp(2N)]$ can be collected into groups that are enhanced at low energies to 
$\prod_{i=1}^NUSp(2l_i)$, so the total IR global symmetry is  $\prod_{i=1}^N USp(2k_i)\times \prod_{i=1}^N USp(2l_i)\times U(1)^2$.\\

Given the many similarities between  the $4d$ \eusp theory and its $E^\gs_\gr[USp(2N)]$ generalizations and the $3d$ \tsu and $T^\gs_\gr[SU(N)]$ theories, it is natural to wonder whether the analogy can be pushed further.
For example since  Hanany--Witten  brane set-ups \cite{Hanany:1996ie} are known for $T^\gs_\gr[SU(N)]$ one could try to 
find a  brane realization of $E^\gs_\gr[USp(2N)]$.
Moreover, the $T^\gs_\gr[SU(N)]$ moduli space is known to have a neat description in terms of  hyperKähler quotients \cite{Nakajima:1994nid}. It would be interesting to understand if also the moduli space of $E^\gs_\gr[USp(2N)]$ possesses some interesting geometric structure. To this purpose, one possibility would be to investigate limits of the superconformal index of $E^\gs_\gr[USp(2N)]$ that are analogue of the Higgs and Coulomb limits of the superconformal index of $T^\gs_\gr[SU(N)]$ studied in \cite{Razamat:2014pta}. In addition, the Coulomb limit of the superconformal index of $T^\gs_\gr[SU(N)]$ takes the form of Hall-Littlewood polynomials \cite{Cremonesi:2014kwa}, so a possible $4d$ version of this limit for the superconformal index of $E^\gs_\gr[USp(2N)]$ may lead to an interesting generalization of these polynomials.

Another possible direction is to use $E^\gs_\gr[USp(2N)]$ as a building block to construct more complicated $4d$ $\mathcal{N}=1$ theories by gauging its non-abelian global symmetries, which may have interesting IR properties. 
In this spirit, some models involving the \eusp theory as a component have been investigated in \cite{Pasquetti:2019hxf,gmsz}.

Finally, it would be interesting to find more examples of $4d$ $\mathcal{N}=1$ IR dualities of the mirror type we discuss here. For example, it would be interesting to find a $4d$ uplift of the star-shaped quivers and of their mirror duals \cite{Benini:2010uu}.\\

The rest of the paper is organized as follows. In Section \ref{sectsunweb} we review the definition of $T[SU(N)]$ and $T^\gs_\gr[SU(N)]$ theories and we discuss the procedure for deriving the deformed web from the duality web of $T[SU(N)]$, which allows us to systematically construct $T^\gs_\gr[SU(N)]$ mirror pairs starting from the self-duality of $T[SU(N)]$. In Section \ref{sec3} we review the definition of \eusp theory and we introduce its duality web. Finally,  we discuss the deformed duality web for \eusp and we introduce the $E^\gs_\gr[USp(2N)]$ theory with its mirror dual. The main text is supplemented with appendices containing details on the partition function computations.

\section{\boldmath$3d$ Mirror Symmetry and \boldmath$T^\gs_\gr[SU(N)]$ theories}
\label{sectsunweb}

\subsection{$T[SU(N)]$  duality web}

\begin{figure}[t]
	\centering
  	\includegraphics[scale=0.6]{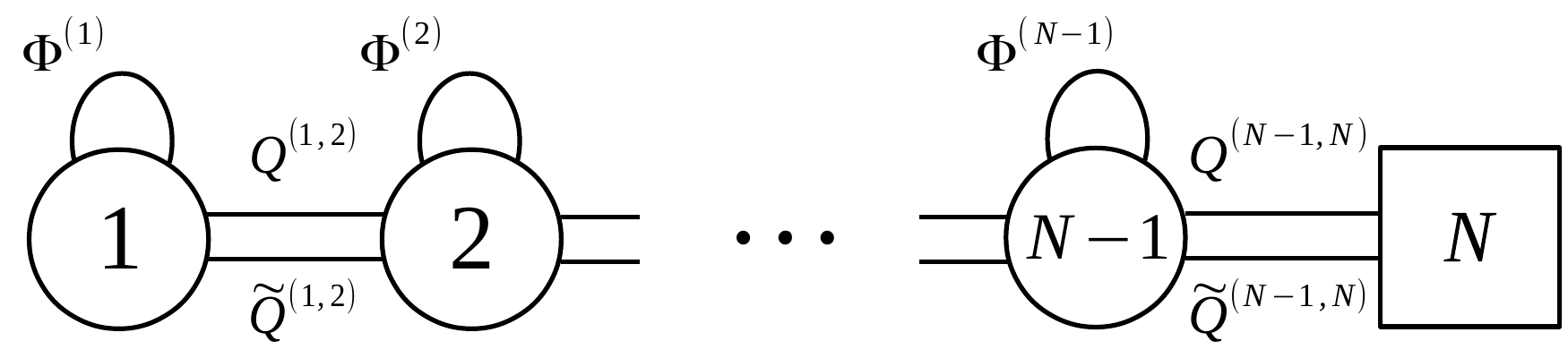} 
    \caption{Quiver diagram for \tsu in $\mathcal{N}=2$ notation. Round nodes denote gauge symmetries and square nodes denote global symmetries. Single lines denote chiral fields in representations of the nodes they are connecting. In particular,  lines between adjacent nodes denote chiral fields in the bifundamental representations of the two nodes symmetries, while arcs denote chiral fields in the adjoint representation of the corresponding node symmetry.}
 \label{tsunquiver}
\end{figure}

The $T[SU(N)]$ theory admits a Lagrangian description in terms of the quiver in Figure \ref{tsunquiver}. The gauge group of the theory is $\prod_{i=1}^{N-1}U(i)$ and each factor is represented by a round node in the quiver. We will use $\mathcal{N}=2$ notation, where each gauge node carries a vector multiplet and a chiral multiplet $\Phi^{(i)}$ in the adjoint representation of the corresponding gauge symmetry. The matter content of the theory consists also of bifundamental chiral fields $Q^{(i,i+1)}_{ab}$ and $\tilde{Q}^{(i,i+1)}_{\tilde{a}\tilde{b}}$ represented in the quiver by lines connecting 
adjacent nodes, which come from $\mathcal{N}=4$ hypermultiplets\footnote{In our conventions, the bifundamentals $Q^{(i,i+1)}_{ab}$ 
transform in the representation $\Box\otimes \overline{\Box}$ of $U(i)\times U(i+1)$ and the bifundamental 
$\tilde{Q}^{(i,i+1)}_{ab}$ transform in the representation $\overline{\Box}\otimes\Box$.}. For $i=N-1$ these are actually fundamental fields of the $U(N-1)$ gauge node and they transform under an $SU(N)_X$ global symmetry, which is represented in Figure \ref{tsunquiver} by a square node. 
In $\mathcal{N}=2$ notation the superpotential of the theory is 

\be
\mathcal{W}_{T[SU(N)]}= \sum_{i=1}^{N-1}  \Tr_i \left[ \Phi^{(i)} \left( \Tr_{i+1} \mathbb{Q}^{(i,i+1)}-\Tr_{i-1} \mathbb{Q}^{(i-1,i)} \right) \right]\, ,
\label{superPini}
\ee
where we are following the same conventions of \cite{Aprile:2018oau}, that is we defined the matrix of bifundamentals $\mathbb{Q}^{(i,i+1)}=Q^{(i,i+1)}_{ab} \widetilde{Q}^{(i,i+1)}_{\tilde{a}\tilde{b}}$ connecting the $U(i)$ to the $U(i+1)$ gauge node. 
On the first node $\mathbb{Q}^{(0,1)}=0$. 
Traces $\Tr_i$ are taken in the adjoint of $i$-th gauge node, except for $i=N$ which corrsponds to the trace $\Tr_X$ over the global symmetry $SU(N)_X$. 
The manifest global symmetry of $T[SU(N)]$ is $SU(N)_X\times U(1)^{N-1}$. The $U(1)$ factors corresponding to the topological symmetry of each gauge node are actually enhanced to the second $SU(N)_Y$ symmetry in the IR.  
For each Cartan in the two $SU(N)$ global symmetries 
we can turn on real masses. The most suitable parametrization of these masses consists of turning on $2N$ parameters $X_n$ and $Y_n$ with $n=1,\cdots,N$ and imposing the tracelessness conditions $\sum_{n=1}^NX_n=\sum_{n=1}^NY_n=0$.

We will turn on a real mass for the $U(1)_{m_A}=U(1)_{C-H}$ axial symmetry 
where $C$ and $H$ are the generators of the Cartans $U(1)_C\subset SU(2)_C$ and $ U(1)_H\subset SU(2)_H$
of the $\mathcal{N}=4$ R-symmetry  $SU(2)_C\times SU(2)_H$, so our theories will have $\mathcal{N}=2^*$ supersymmetry  \cite{Tong:2000ky}.  We will then take the UV R-symmetry as  $R_0 = C + H$. In the IR the R-symmetry 
can mix with other abelian symmetries, but since the topological symmetry is non-abelian, 
$R_0$ will only mix with $U(1)_{m_A}$. Denoting with $r$ the mixing coefficient and with $q_A$ the charge under $U(1)_{m_A}$, we have
\be
R=R_0+q_A\,r\,.
\ee
Our choice for the parametrization of $U(1)_{m_A}$ and $U(1)_{R}$ is summarised in Table \ref{chargestsun}. The exact value of $r$ corresponding to the IR superconformal R-symmetry can be fixed by F-extremization \cite{Jafferis:2010un}. As we did for the non-abelian symmetries, we can turn on a real mass $\mathrm{Re}(m_A)$ for the axial symmetry. It is also useful to define the following holomorphic combination:
\be
m_A=\mathrm{Re}(m_A)+i\frac{Q}{2}r\, .
\ee

Summing up, the complete IR global symmetry of the $\mathcal{N}=2^*$ version of \tsu is
\be
SU(N)_X\times SU(N)_Y\times U(1)_{m_A}\, .
\ee
The chiral fields of the theory transform under these symmetries according to Table \ref{chargestsun}.
\begin{table}[t]
\centering
\scalebox{1}{
\begin{tabular}{c|ccc|c}
{} & $SU(N)_X$ & $SU(N)_{Y}$ & $U(1)_{m_A}$ & $U(1)_{R}$  \\ \hline
$Q^{(i-1,i)}$ & $\bullet$ & $\bullet$ & $1$ & $r$ \\
$\tilde{Q}^{(i-1,i)}$ & $\bullet$ & $\bullet$ & $1$ & $r$ \\
$Q^{(N-1,N)}$ & $\bf N$ & $\bullet$ & $1$ & $r$ \\
$\tilde{Q}^{(N-1,N)}$ & $\bf \bar{N}$ & $\bullet$ & $1$ & $r$ \\ 
$\Gp^{(i)}$ & $\bullet$ & $\bullet$ & $-2$ & $2-2r$ \\\hline
$\hb$ & $\bf N^2-1$ & $\bullet$ & $2$ & $2r$ \\
$\cb$  & $\bullet$ & $\bf N^2-1$ & $-2$ & $2-2r$
\end{tabular}}
    \caption{Charges and representations of the chiral fields and of the chiral ring generators of \tsu under the global symmetries. In the table $i=1,\cdots,N-1$ and $Q^{(0,1)}=\tilde{Q}^{(0,1)}=0$.}
 \label{chargestsun}
\end{table}

The generators of the chiral ring are the Higgs branch (HB) and the Coulomb branch (CB) moment maps $\hb$ and $\cb$. The HB moment map is
\be
\hb= \mathcal{Q} -\frac{1}{N}\Tr_X\mathcal{Q}\ 
\ee
with $\mathcal{Q}$ the $N\times N$ meson matrix
\be\label{meson_def_ori}
\cQ_{ij}= \Tr_{N-1} \mathbb{Q}^{(N-1,N)}\,.
\ee
The CB branch moment map is instead generated by  $\Tr_i\,\Phi^{(i)}$ and monopole operators with magnetic flux vectors $(m_1,\ldots m_{N-1})$, where $m_i$ denotes the unit of flux for the topological $U(1)$ of the $i$-th node. In particular
monopole operators defined with fluxes of the form $(0^i, (\pm1)^j, 0^k)$, 
where $0$ and $1$ are repeated with integer multiplicities $i$, $j$, and $k$ 
such that $i+j+k=N-1$, have the same R-charge of the adjoint chiral fields and the same charge under $U(1)_{m_A}$. 
We then collect these  $N(N-1)$ monopoles and the traces of the $N-1$ adjoint chirals
into a single $N\times N$ traceless matrix.
 For $N=4$ this matrix reads
\be\label{1monopolematrix}
\cb\equiv 
\left(\begin{array}{ccccccc}
0									&\ & \mathfrak{M}^{(1,0,0)} 							&\ & \mathfrak{M}^{(1,1,0)} 		&\ & \mathfrak{M}^{(1,1,1)}	\\[.15cm]	
\mathfrak{M}^{(\textrm{-}1,0,0)}      				&\ & 0									&\ & \mathfrak{M}^{(0,1,0)} 		&\ & \mathfrak{M}^{(0,1,1)}  \\[.15cm]	
\mathfrak{M}^{(\textrm{-}1,\textrm{-}1,0)}			&\ &  	\mathfrak{M}^{(0,\textrm{-}1,0)}				&\ & 	0				&\ & \mathfrak{M}^{(0,0,1)} 	 \\[.15cm]		
\mathfrak{M}^{(\textrm{-}1,\textrm{-}1,\textrm{-}1)}		&\ &  	\mathfrak{M}^{(0,\textrm{-}1,\textrm{-}1)}		&\ &\mathfrak{M}^{(0,0,\textrm{-}1)}				&\ & 0 
\end{array}\right)+ \sum_{i=1}^{3} \Tr_i\Phi^{(i)} \mathcal{D}_i\, ,
\ee
where $\mathcal{D}_i$ are traceless diagonal generators of $SU(N)_Y$. The operator $\cb$ constructed in this way transforms in the adjoint representation of $SU(N)_Y$ and thus corresponds to the moment map for this enhanced symmetry.

In Table \ref{chargestsun} we also report the charges and representations under the global symmetries of the chiral ring generators $\hb$ and $\cb$ according to our parametrization of $U(1)_{m_A}$ and $U(1)_{R_0}$.
Notice that these charges are consistent with the operator map dictated by Mirror Symmetry
which in this case corresponds to a self-duality of the theory, under which the operators of the HB and the CB are exchanged. Hence, the nilpotency of $\hb$, which follows by the F-term equations of \eqref{superPini}, together with the operator map of Mirror Symmetry implies that also the matrix $\cb$ is nilpotent.

The main tool we will use to study \tsu, its duality frames and their deformations related to $T^\gs_\gr[SU(N)]$ is the supersymmetric partition function on $S^3_b$ \cite{Jafferis:2010un,Hama:2010av,Hama:2011ea}. For $T[SU(N)]$, this will be a function of the parameters in the Cartan of the global symmetry group, which we denoted as $X_n$, $Y_n$ and $m_A$. Indeed, the partition function depends only on the holomorphic combination of the real mass for the $U(1)_{m_A}$ abelian symmetry and the mixing coefficient with the trial R-symmetry $U(1)_{R_0}$ \cite{Jafferis:2010un}.
With these conventions, the partition function of $T[SU(N)]$ can be written recursively as
\begin{equation}
\makebox[\linewidth][c]{\scalebox{0.95}{$
\begin{split}
\mathcal{Z}_{T[SU(N)]}(\vec X;\vec Y;m_A)&=\int\udl{\vec{z}^{(N-1)}_{N-1}}\e^{2\pi i(Y_{N-1}-Y_N)\sum_{i=1}^{N-1}z_i^{(N-1)}}\prod_{i,j=1}^{N-1}\sbfunc{-i\frac{Q}{2}+\left(z^{(N-1)}_i-z^{(N-1)}_j\right)+2m_A}\\
 	&\times\prod_{i=1}^{N-1}\prod_{n=1}^N\sbfunc{i\frac{Q}{2}\pm(z^{(N-1)}_i-X_n)-m_A}\mathcal{Z}_{T[SU(N-1)]}\left(\vec z^{(N-1)};Y_1, \cdots , Y_{N-1};m_A\right)\, ,
\label{tsunPF}
\end{split}$}}
\end{equation}
where we defined the measure of integration for the $m$-th $U(n)$ gauge groups on $S^3_b$ including both the contribution of the $\mathcal{N}=2$ vector multiplet and the Weyl symmetry factor
\be
\udl{\vec{z}^{(m)}_{n}}=\frac{1}{n!}\frac{\prod_{i=1}^{n}\udl{z^{(m)}_i}}{\prod_{i<j}^{n}\sbfunc{i\frac{Q}{2}\pm\left(z^{(m)}_i-z^{(m)}_j\right)}}\,.
\ee

In  \cite{Aprile:2018oau} it has been observed that \tsu possesses several duality frames that can be summarized in the commutative diagram of Figure \ref{tsunweb}. 
One frame is the one obtained applying Mirror Symmetry, which we denote by \mirrortsu. As we mentioned before, \tsu is self-dual under this duality, which acts non-trivially on the chiral ring generators of the theory. In particular, it exchanges the operators charged under $SU(N)_X$ with those charged under $SU(N)_Y$. If we consider the $\mathcal{N}=2^*$ deformation of \tsu, Mirror Symmetry also acts flipping the sign of the $U(1)_{m_A}$ charges as well as the mixing coefficient of the R-symmetry with this abelian symmetry $r\rightarrow1-r$. In terms of the mass parameter $m_A$, we have
\be
m_A\rightarrow i\frac{Q}{2}-m_A\, .
\label{mAchange}
\ee
In other words, using Table \ref{chargestsun} we have the following operator map:
\be
\hb\quad&\leftrightarrow&\quad \cb^\vee\nn\\
\cb\quad&\leftrightarrow&\quad \hb^\vee\, .
\ee
At the level of the $S^3_b$ partition function, Mirror Symmetry for \tsu translates into the following non-trivial integral identity
\be
\mathcal{Z}_{T[SU(N)]}(\vec X;\vec Y;m_A)=\mathcal{Z}_{T[SU(N)]}\left(\vec Y;\vec X;i\frac{Q}{2}-m_A\right)=\mathcal{Z}_{T[SU(N)]^\vee}(\vec X; \vec Y;m_A)\, .\nn\\
\label{idmirror}
\ee
This identity can be proven using the fact that $\mathcal{Z}_{T[SU(N)]}$ is an eigenfunction of the trigonometric Ruijsenaars-Schneider model \cite{Bullimore:2014awa}.

On top of the mirror dual frame, \tsu has another interesting dual which was named flip-flip dual \fftsu in \cite{Aprile:2018oau}. This theory is $T[SU(N)]$  with two extra sets of singlet fields $\mathcal{O_H}$ and $\mathcal{O_C}$ flipping the HB and CB moment maps
\be
\mathcal{W}_{FFT[SU(N)]}=\mathcal{W}_{T[SU(N)]}+\Tr_X\left(\mathcal{O_H}\, \hb_{FF}\right)+\Tr_Y\left(\mathcal{O_C}\,\cb_{FF}\right)\, ,
\ee
where $\hb_{FF}$ and $\cb_{FF}$ denote the HB and CB moment maps of $FFT[SU(N)]$. 
Flip-flip duality acts trivially on the non-abelian global symmetries of \tsu, while it acts on $U(1)_{m_A}$ and $U(1)_R$ exactly as Mirror Symmetry \eqref{mAchange}. The operators are accordingly mapped as
\be
\hb\quad&\leftrightarrow&\quad \mathcal{O_H}\nn\\
\cb\quad&\leftrightarrow&\quad \mathcal{O_C}\, .
\label{opmapflipflip}
\ee
This duality implies another non-trivial integral identity satisfied by $\mathcal{Z}_{T[SU(N)]}$
\be
\mathcal{Z}_{T[SU(N)]}(\vec X;\vec Y;m_A)&=&\prod_{n,m=1}^N\frac{\sbfunc{i\frac{Q}{2}+(X_n-X_m)-2m_A}}{\sbfunc{i\frac{Q}{2}+(Y_n-Y_m)-2m_A}}\mathcal{Z}_{T[SU(N)]}\left(\vec X;\vec Y;i\frac{Q}{2}-m_A\right)\nn\\
&=&\mathcal{Z}_{FFT[SU(N)]}(\vec X;\vec Y;m_A)\, ,
\label{idflipflip}
\ee
which can also be proven using the trigonometric Ruijsenaars-Schneider model eigenvalue equation \cite{Aprile:2018oau, Zenkevich:2017ylb}.\\

The flip-flip duality can be also derived  by iteratively applying  the Aharony duality (see Appendix \ref{funddualities} for a review) along the tail:
\begin{itemize}
\item At the first iteration we start from the  $U(1)$ node, whose adjoint chiral is just a singlet.
Aharony duality has the effect of making the adjoint chiral field of the adjacent $U(2)$ node  massive, hence we can apply again  the Aharony  duality on it. We continue applying iteratively the Ahaorny duality until we reach the last $U(N-1)$  node.
Notice that since every $U(n)$ node sees $2n$ flavors, the ranks do not change when we apply the duality.
Moreover some of the singlet fields expected from the Aharony duality are massive (because of the R-charge assignement) and no new links  between nodes are created.

\item At the second iteration we start again from the $U(1)$ node and proceed along the tail, but this time we stop at the second last node
$U(N-2)$.

\item At the third iteration we start again from the $U(1)$  node and proceed along the tail stopping at the  
$U(N-3)$ node.

\item We iterate this procedure for a total of $N-1$ times, meaning that we apply Aharony duality $N(N-1)/2$ times. 

\item The singlet fields flipping the mesons and the monopoles appearing in the Aharony duality reconstruct the singlet matrices $\mathcal{O_H}$ and $\mathcal{O_C}$.
\end{itemize}
We checked this  procedure in the $N=3$ case, by applying the integral identity for Aharony duality \eqref{aha} to the $S^3$ partition function in Appendix \ref{ff3}.\\

By combining Mirror Symmetry and flip-flip duality we can reach a third duality frame \mirrorfftsu, which again corresponds to \tsu with two sets of singlet fields $\mathcal{O}_\mathcal{H}^\vee$ and $\mathcal{O}_\mathcal{C}^\vee$ flipping the HB and CB moment maps $\hb^\vee_{FF}$ and $\cb^\vee_{FF}$
\be
\mathcal{W}_{FFT[SU(N)]^\vee}=\mathcal{W}_{T[SU(N)]}+\Tr_Y\left(\mathcal{O}_{\mathcal{H}}^\vee\, \hb^\vee_{FF}\right)+\Tr_X\left(\mathcal{O}_{\mathcal{C}}^\vee\,\cb^\vee_{FF}\right)\, ,
\ee
but in this case the duality acts exchanging $SU(N)_X$ and $SU(N)_Y$, while leaving unchanged $U(1)_{m_A}$ and $U(1)_R$\footnote{In \cite{Aprile:2018oau} this kind of duality was called spectral duality.}
The operator map between the original \tsu and \mirrorfftsu is
\be
\hb\quad&\leftrightarrow&\quad \mathcal{O}_{\mathcal{C}}^\vee\nn\\
\cb\quad&\leftrightarrow&\quad \mathcal{O}_{\mathcal{H}}^\vee\, .
\ee

\subsection{From $T[SU(N)]$ to $T_\rho^\sigma[SU(N)]$ using the web}
\label{trhosigmaweb}

$T_\rho^\sigma[SU(N)]$ can be obtained as a deformation of $T[SU(N)]$ corresponding to giving nilpotent vevs  labelled by partitions $\gs$ and $\gr$ of $N$ to the moment maps:
\be
\langle\mathcal{C}\rangle=\mathcal{J}_\gr \,, \qquad \langle\mathcal{H}\rangle=\mathcal{J}_\gs\,,
\ee
where $\mathcal{J}_\gr$ and $\mathcal{J}_\gs$ are $N\times N$ block diagonal matrices with each block being a Jordan matrix that can be uniquely determined after specifying the partitions $\gs$ and $\gr$
\be
\mathcal{J}_\gr=\bigoplus_{i=1}^L\mathbb{J}_{\gr_i}=\left( \begin{array}{c|c|c|c}\  \mathbb{J}_{\gr_1} \ & \ 0_{\gr_1\times \gr_2} \ &  \cdots \ & \ 0_{\gr_{1}\times \gr_L}\\ \hline
\ 0_{\gr_2\times \gr_1} \ & \  \mathbb{J}_{\gr_2} \ &   \cdots \ & \ 0_{\gr_{2}\times \gr_L}\\ \hline
\  \ &\   \ &   \ddots \ &  \\ \hline
\ 0_{\gr_{L}\times \gr_1} & \  0_{\gr_{L}\times \gr_2} \ &   \cdots \ & \ \mathbb{J}_{\gr_L}\\
\end{array}\right),\qquad \mathbb{J}_{\gr_i}= \underbrace{ \left(\begin{array}{ccccc} 0 &1& \ldots & \ldots & 0 \\ 0 & 0 & 1 & \ldots  & 0  \\ \vdots & \vdots & &  & \vdots  \\ 0 & 0 & \ldots & 0 & 0 \end{array}\right) }_{\gr_i} \, .
\label{jordanmatrices}
\ee
These vevs trigger a sequential higgsing. The higgsing procedure is in general very difficult to study, in particular when the vev is for the monopole operators contained in $\cb$. 

As we explained in the introduction we will follow an alternative procedure  based on the duality web of \tsu we reviewed in the previous section.
First of all we observe that the vev  can be implemented by adding two sets of $N^2-1$ flipping fields $\mathcal{O_H}$ and $\mathcal{O_C}$ that couple to the meson and monopole matrices, which is the same as considering  $FF[TSU(N)]$, 
 and turning on linearly in the superpotential some of their entries, depending on the partitions $\gs$ and $\gr$. Some of the components of $\mathcal{O_H}$ and $\mathcal{O_C}$ remain massless and correspond to a decoupled free sector of the low energy theory. Hence, we remove them by adding some additional singlets $\mathcal{S}_\gs$ and $\mathcal{T}_\gr$ that flip them \cite{Gadde:2013fma,Agarwal:2014rua,Benvenuti:2017kud}. In order to do so, $\mathcal{S}_\gs$ and $\mathcal{T}_\gr$ have to be $N\times N$ traceless matrices whose transpose commute with the Jordan matrices $\mathcal{J}_\gs$ and $\mathcal{J}_\gr$ respectively.

 For a generic nilpotent vev, the deformation taking $FF[TSU(N)]$  to $T_\rho^\sigma[SU(N)]$  is
 \be
\gd\mathcal{W}_{FF}=Tr_X\left[(\mathcal{J}_\gs+\mathcal{S}_\gs) \,\mathcal{O_H}\right]+\Tr_Y\left[(\mathcal{J}_\gr+\mathcal{T}_\gr) \,\mathcal{O_C}\right]\, .
\label{TrhosigmaVEV}
\ee
Using the operator map \eqref{opmapflipflip} we can then translate the deformation of \fftsu into a deformation of \tsu which is linear in some of the components of $\hb$ and $\cb$
\be
\delta\mathcal{W}=\Tr_X\left[(\mathcal{J}_\gs+\mathcal{S}_\gs) \,\hb\right]+\Tr_Y\left[(\mathcal{J}_\gr+\mathcal{T}_\gr) \,\cb\right]\, .
\label{TrhosigmaMASSMONO}
\ee
This is a mass and linear monopole deformation of \tsu that leads to an IR theory that we denoted with $\mathcal{T}$ in Figure \ref{tsundefweb}. This deformation is easier to study than the vev of \tsu, but the price we have to pay is that we end up not directly with $T^\gs_\gr[SU(N)]$ but its flip-flip dual $\mathcal{T}$.

We propose that to implement the flip-flip duality moving from  $\mathcal{T}$ to $T^\gs_\gr[SU(N)]$
we can generalise the strategy to move from $T[SU(N)]$  to $FFT[SU(N)]$,  where we applied iteratively the Aharony duality. Here since some of the nodes will have a linear monopole superpotential we will  use a combination of  Aharony duality and the one-monopole duality \cite{Benini:2017dud} (see also Appendix \ref{funddualities}), depending on whether a monopole is turned on in the superpotential at the node we are considering.\\

For simplicity we will restrict to the case where one of the two partitions is trivial. We first consider the case where $\sigma=[1^N]$, which corresponds to turning on a nilpotent vev labelled by a partition $\gr$ for the CB moment map $\cb$ leading to $T_\gr[SU(N)]$.
In the flip-flip dual frame, this deformation corresponds to the following deformation of \tsu:
\be 
\delta\mathcal{W}=\Tr_X\left[\left(\mathcal{J}_{[1^N]}+\mathcal{S}_{[1^N]}\right) \hb\right]+\Tr_Y\left[\left(J_\gr+\mathcal{T}_\gr\right) \cb\right]\, .
\ee
Here  $\mathcal{J}_{[1^N]}$ is the null matrix, while $\mathcal{S}_{[1^N]}$ and $\mathcal{T}_\gr$ are matrices of gauge singlets whose transposes commute with $\mathcal{J}_{[1^N]}$ and $\mathcal{J}_\gr$ respectively, so in particular $\mathcal{S}_{[1^N]}$ is an arbitrary $N\times N$ traceless matrix which is completely flipping the HB moment map $\hb$.

This deformation leads to theory $\mathcal{T}$ whose global symmetry will be the product of $SU(N)_X$ and of the subgroup of $SU(N)_Y$ preserved by the vev, which can be at most broken  to $S(U(1)^L)$ when all the entries $\gr_i$ of the partition are different. Instead, when some of the entries coincide the corresponding $U(1)$ factors combine and are enhanced in the infrared. More precisely, for a generic partition of the form $\gr=[N^{l_N},\cdots,1^{l_1}]$ the IR CB global symmetry will be broken to\footnote{Notice that when we write the partition as $\gr=[N^{l_N},\cdots,1^{l_1}]$, some of the $l_i$ will in general be zero. The corresponding factor in the CB global symmetry is just an empty group.}
\be SU(N)_Y \to S\left(\prod_{i=1}^N U(l_i)\right)\, 
\ee which is precisely the CB symmetry of $T_\gr[SU(N)]$. Correspondingly at the level of partition functions we will  introduce the following fugacities 
\be
Y_i, \quad {\rm with} \quad i=1,\cdots,N \quad \to \quad  Y^{(1)}_{i_1}, Y^{(2)}_{i_2},\cdots \quad{\rm with } \quad i_s=1,\cdots,l_s
\ee
and similarly, when also $\sigma$ is non-trivial, we introduce
\be
X_j, \quad {\rm with} \quad j=1,\cdots,N \quad \to \quad  X^{(1)}_{j_1}, X^{(2)}_{j_2},\cdots \quad{\rm with } \quad j_r=1,\cdots,k_r\,.
\ee
We can then reach $T_\gr[SU(N)]$ implementing the flip-flip duality by applying sequentially Aharony and one-monopole duality. Below we illustrate  this procedure  in the case of a next-to-maximal vev corresponding to partition $\rho=[N-1,1]$ and for the partition $\gr=[2,1^2]$.\\

On the mirror dual side, we will have a nilpotent vev labelled by a partition $\gr$ for the HB moment map $\hb^\vee$  leading to $T^\gr[SU(N)]$.   In the flip-flip dual frame this vev corresponds to the following deformation of $T[SU(N)]^\vee$:
\be 
\delta\mathcal{W}^\vee=\Tr_Y\left[\left(\mathcal{J}_\gr+\mathcal{T}_\gr\right) \hb^\vee\right]+\Tr_X\left[\left(\mathcal{J}_{[1^N]}+\mathcal{S}_{[1^N]}\right) \cb^\vee\right]\, .
\ee
Since this is a purely massive deformation we can find a Lagrangian description for the theory $\mathcal{T}^\vee$ which we flow to by integrating out the massive fields. 
 $\mathcal{T}^\vee$ is the same quiver as $T[SU(N)]^\vee$ but with less flavors attached to the last $U(N-1)$ node.
The number of remaining massless flavors coincides with the length $L$ of the partition $\gr$ and each of them interacts with a different power of the adjoint chiral $\Gp^{(N-1)}$ of the last gauge node. Because of this superpotential coupling the HB $SU(N)_Y$ global symmetry of $T[SU(N)]^\vee$ will be generically broken down to $S(U(1)^L)$, but if some of the $\gr_i$ are equal we can form blocks of chirals  transforming under a larger symmetry group since they interact with the same power of $\Gp^{(N-1)}$. Hence, for a  partition of the form $\gr=[N^{l_N},\cdots,1^{l_1}]$ the resulting interaction is
\bea
 \Tr_{N-1} \left[ \Phi^{(N-1)} \left( \Tr_{Y}\tilde q^{(N-1,N)}  q^{(N-1,N)} \right) \right]
\quad \to\ && \sum_{i=1}^L\Tr_{N-1}\left[\tilde{q}_i\left(\Gp^{(N-1)}\right)^{\gr_i}q_i\right]\nn\\
&&=\sum_{m=1}^N\Tr_{N-1}\left[\left(\Gp^{(N-1)}\right)^{m}\Tr_{Y^{(m)}}\left(\tilde{q}_{m}q_{m}\right)\right]\, ,\nn\\
\label{gbw}
\eea
where we renamed as $q_m, \tilde q_m$ the massless chirals at the $U(N-1)$ gauge node in the fundamental and anti-fundamental representation of each $U(l_m)$ factor, with $m=1,\cdots,N$. In particular, for the values of $m$ for which $l_m=0$ we don't have any chiral field. We also introduced the notation $\Tr_{Y^{(i)}}$ for the trace over the $i$-th factor in this global symmetry group.

 The full superpotential will be \be
\mathcal{W}_{\mathcal{T}^\vee}&=&\mathcal{W}_{T[SU(N-1)]}-\Tr_{N-1}\left(\Gp^{(N-1)}\Tr_{N-2}
\tilde q^{(N-2,N-1)}  q^{(N-2,N-1)} \right)\nn\\
&+&\sum_{m=1}^N\Tr_{N-1}\left[\left(\Gp^{(N-1)}\right)^{m}\Tr_{Y^{(m)}}\left(\tilde{q}_{m}q_{m}\right)\right]+\left.\Tr_Y\left(\mathcal{T}_\gr\, \hb^\vee\right)\right|_{eom}+\Tr_X\left(\mathcal{S}_{[1^N]}\,\cb^\vee\right)\nn\\
\ee
and the global symmetry will be $S(\prod_{i=1}^NU(l_i))$. 
The subscript $eom$ refers to the fact that after imposing the F-terms equations only some of the components of $\hb^\vee$ will survive.

From $\mathcal{T}^\vee$  we can  reach $T^\gr[SU(N)]$ by implementing the flip-flip duality, which in this case is equivalent to applying Aharony duality only since we have no monopole superpotential.
Below we illustrate this procedure for the partitions  $\rho=[N-1,1]$ and $\gr=[2,1^2]$.

\subsubsection{$\rho=[N-1,1]$ and $\sigma=[1^N]$}

\subsubsection*{Flow to $T_{[N-1,1]}[SU(N)]$}
We define theory $\mathcal{T}$ as the theory obtained from $T[SU(N)]$ via the deformation:
\be 
\delta\mathcal{W}=\Tr_X\left[\left(\mathcal{J}_{[1^N]}+\mathcal{S}_{[1^N]}\right) \hb\right]+\Tr_Y\left[\left(\mathcal{J}_{[N-1,1]}+\mathcal{T}_{[N-1,1]}\right) \cb\right]\, .
\label{deltaWN-11}
\ee
The matrix $J_{[1^N]}$ is simply the null matrix and, consequently, $\mathcal{S}_{[1^N]}$ is a generic $N\times N$ traceless matrix. 
Instead by requiring that the transpose of $\mathcal{T}_{[N-1,1]}$ commutes with $J_{[N-1,1]}$ we find its non-vanishing entries: 
\be
\mathcal{J}_{[N-1,1]}+\mathcal{T}_{[N-1,1]}=\left( \begin{array}{cccc|c}\  \mathcal{T}_1 \ & \ 1 \ &  \cdots \ & \ 0 \  & \ 0\\ 
\ \mathcal{T}_2 \ & \  \mathcal{T}_1 \ & 1 \ & & \ \vdots  \\ 
\ \vdots \ & \ \ddots \ & \ \ddots \ & \ 1 \ & \ 0  \\ 
\mathcal{T}_{N-1} \ & & \ \mathcal{T}_2 \ & \ \mathcal{T}_1 \ & \ {\mathcal{T}}_+ \\ \hline
\mathcal{T}_- \ & \ 0 \ & \ \cdots \ & \ 0 \ & \ -(N-1)\mathcal{T}_1
\end{array}\right)\,.
\ee
More explicitly, the superpotential deformation is
\be
\delta\mathcal{W}=\Tr_X\left(\mathcal{S}_{[1^N]}\hb\right)+\Tr_Y\left(\mathcal{T}_{[N-1,1]} \cb\right)+\mathfrak{M}^{(1,0,\cdots,0)}+\mathfrak{M}^{(0,1,0\cdots,0)}+\cdots++\mathfrak{M}^{(0,\cdots,1,0)}\, .\nn\\
\label{linsf}
\ee
The linear monopole deformation  at the first $N-2$ nodes  breaks the topological and the axial symmetries to a combination, implying the constraint on the fugacities
\be
Y_i-Y_{i-1}=2 m_A \quad {\rm for} \quad i=2,\cdots N-1\,,
\ee
which can be solved by
\be
Y_i=Y_1+2(i-1)m_A,\qquad i=1,\cdots N-1\,.
\label{solconstrN-11}
\ee
From this we can easily determine the charges of the singlets $\mathcal{T}_i$ and $\mathcal{T}^\pm$.
Before imposing the constraint on the fugacities the charges of the entry $(i,j)$ of the moment map martix $\mathcal{C}$ under the Cartan  
$\prod_{i=1}^{N-1} U(1)_{Y_i}\subset SU(N)_Y$ and under $U(1)_{m_A}$ can be read off  from the coefficeints of $Y_i$ and $m_A$ in the combination 
\be
Y_j-Y_i-2m_A\, .
\ee
Imposing the constraint \eqref{solconstrN-11} on this combination we can extract the charges under the residual symmetry $SU(N)_X
\times U(1)_{Y} \times U(1)_{m_A}$, where $U(1)_Y$ is a combination of $U(1)_{Y_1}$, $U(1)_{Y_N}$ and $U(1)_{m_A}$

\begin{table}[h]
\centering
\scalebox{1}{
\begin{tabular}{c|ccc|c}
{} & $U(1)_{Y}$  & $SU(N)_X$ & $U(1)_{m_A}$ & $U(1)_{R}$  \\ \hline
$\mathcal{T}_i$ & 0 & $\bullet$  & $2i$ & $2r i$ \\
$\mathcal{T}_-$ & $-1$ & $\bullet$  & $N$ & $N r$ \\
$\mathcal{T}_+$ & 1 & $\bullet$  & $N$ & $Nr$ \\
$\mathcal{S}_{[1^N]}$ & 0 & $\bf N^2-1 $  & $-2$ & $2-2r$ \\
\end{tabular}}
\end{table}

From theory $\mathcal{T}$ we want to move along the vertical edge of the web and reach $T_{[N-1,1]}[SU(N)]$. 
This is achieved by applying iteratively either the Aharony or the one-monopole duality, depending on whether the node we are considering has a linear monopole superpotential or not. In this case, we apply $N-2$ times the one-monopole duality starting from the first node until we reach the $U(N-2)$ node.
Since this duality is always applied to a $U(n)$ gauge node with $n+1$ flavors, which corresponds to the case dual to a WZ model, its effect is to sequentially confine the nodes of the quiver. This phenomenon is  known as sequential confinement \cite{Benvenuti:2017kud,Giacomelli:2017vgk,Aprile:2018oau}. 

In particular the effect of the linear monopole deformation in  \eqref{linsf},
 but without the first two terms involving the singlets $\mathcal{T}_i$, $\mathcal{T}_\pm$ and $\mathcal{S}_{[1^N]}$,
 was analysed in great detail in \cite{Aprile:2018oau}.
 There it was shown that after confining the first $N-2$ nodes one reaches a $U(N-1)$ theory with $N$ flavors and superpotential:
\be
\mathcal{W}=-\sum_{k=1}^{N-1} \frac{(-1)^k}{k} \gamma_k \Tr[\mathcal{Q}^k]\,,
\ee
where the singlets $\gamma_k$ flip the traces of powers of the meson $\mathcal{Q}$ and have $R[\gamma_k]=2(1-k r)$.
The chiral ring of this theory in addition to the $\gamma_i$ contains the fundamental $U(N-1)$ monopoles with $R[\mathfrak{M}^\pm]=2-Nr$ and the traceless meson matrix $\mathcal{Q}-\tfrac{\Tr\mathcal{Q}}{N}$ of R-charge $2r$.

To complete our flip-flip prescription we need to apply the Aharony duality to the remaining $U(N-1)$ node. We arrive at a $U(1)$ theory with $N$ flavors and three sets of singlets: $\sigma^\pm$ with R-charge $2-Nr$ flipping the  fundamental $U(1)$ monopoles, $F_{ij}$ with R-charge $2r$  flipping the meson matrix (with trace) and singlets $\theta_k$
with $k=1,\cdots N-1$,  with R-charge $2-2rk$  flipping the traces of powers of the matrix $F_{ij}$.

When we consider the full deformation in  \eqref{linsf}, including singlets $\mathcal{T}_i, \mathcal{T}^\pm$ and  $\mathcal{S}_{[1^N]}$,  the singlets  $\sigma^\pm$, $\theta_k$ and the traceless part of $F_{ij}$ becomes massive. The trace part of $F_{ij}$, which we call $\Phi=\Tr(F_{ij})$, instead reconstructs the  $\mathcal{N}=4$ superpotential
\be
\mathcal{W}_{T_{[N-1,1]}[SU(N)]}=\Gp\sum_{i=1}^N\tilde{P}^iP_i\, ,
\ee
so we arrive at theory $T_{[N-1,1]}[SU(N)]$ which is $\mathcal{N}=4$ SQED with $N$ flavors.

\subsubsection*{Flow to $T^{[N-1,1]}[SU(N)]$}
 
 Theory $\mathcal{T}^\vee$, the mirror dual of $\mathcal{T}$, is obtained by the following   deformation of $T[SU(N)]^\vee$ 
\be
&&\delta\mathcal{W}^\vee=\Tr_Y\left[\left(\mathcal{J}_{[N-1,1]}+\mathcal{T}_{[N-1,1]}\right) \hb^\vee\right]+\Tr_X\left[\left(\mathcal{J}_{[1^N]}+\mathcal{S}_{[1^N]}\right) \cb^\vee\right]\, .
\ee
We can  integrate out the massive fields to get a quiver theory 
with increasing ranks of the gauge groups as in $T[SU(N)]$, but with only two flavors at the end of the tail which interact differently with the adjoint chiral of the $U(N-1)$ gauge node, plus some residual flipping fields originally coming from $\mathcal{S}_{[1^N]}$ and $\mathcal{T}_{[N-1,1]}$
\be
\mathcal{W}_{\mathcal{T}^\vee}&=&\mathcal{W}_{T[SU(N-1)]^\vee}-\Tr_{N-1}\left(\Gp^{(N-1)}\Tr_{N-2}
\tilde q^{(N-2,N-1)}  q^{(N-2,N-1)} \right)\nn\\
&+&\Tr_{N-1}\left[\tilde{q}_1\Gp^{(N-1)}q_1+\tilde{q}_2\left(\Gp^{(N-1)}\right)^{N-1}q_2+\mathcal{T}_-\,\tilde{q}_1q_2+\mathcal{T}_+\tilde{q}_2q_1+\sum_{i=1}^{N-1}\mathcal{T}_i\,\tilde{q}_2\left(\Gp^{(N-1)}\right)^{i-1}q_2\right] \nn \\
&+&\Tr_X\left[\mathcal{S}_{[1^N]}\,\cb_{\mathcal{T}^\vee}\right]\, ,
\ee
where $\cb_{\mathcal{T}^\vee}$ is the CB moment map of theory $\mathcal{T}^\vee$, which is constructed as in \tsu.

\begin{figure}[t]
	\centering
	\makebox[\linewidth][c]{
  	\includegraphics[scale=0.32]{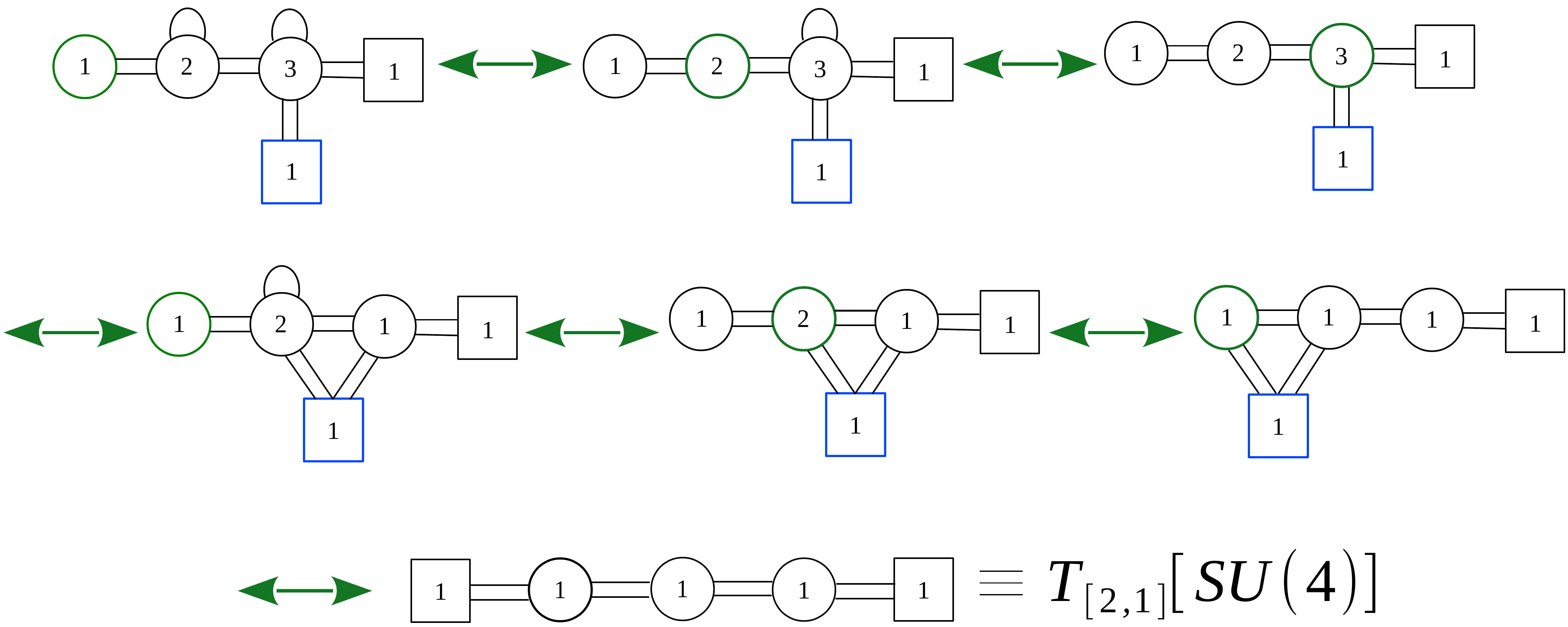} }
  	\caption{Quiver representation of the iterative application of Aharony duality in the case $N=4$. We highlighted in green the gauge node to which we apply the duality at each step. We only sketch the main steps and neglect gauge singlets; taking into account the $\mathcal{S}_{[1^N]}$ and $\mathcal{T}_{[N-1,1]}$ singlets from the beginning, all the remaining ones are only those corresponding to adjoint chirals for $U(1)$ gauge nodes.}
  	\label{TN-11mass}
\end{figure}

To reach $T^{[N-1,1]}[SU(N)]$ we now have to implement the flip-flip duality which amounts to apply Aharony duality sequentially.
This derivation is carried out explicitly at the level of the sphere partition function in the $N=3$ case in Appendix \ref{appflipflipaha}, while here we only discuss its main steps which  are sketched in Figure \ref{TN-11mass}.

\begin{itemize}
\item At the first iteration we start from the $U(1)$ gauge node and proceed applying the Aharony duality along the tail.
Since the first $N-2$ nodes are $U(n)$ nodes with $2n$ flavors,  the gauge group doesn't change when we apply the duality
and because of the charge assignments no new links are created.
The last $U(N-1)$ node however sees $N$ flavors, so when we apply Aharony duality it becomes a $U(1)$ gauge node. 
A new link is created connecting one of the two flavor nodes (the blue one in the picture) to the second last gauge node.

\item At the second iteration we start again from the leftmost $U(1)$ gauge node and go along the whole tail, but this time we stop at the second last node. Because of the result of the previous iteration, this is now a $U(N-2)$ gauge node with $N-1$ flavors, so when we apply Aharony duality it becomes a $U(1)$ node.
Now the blue flavor node gets attached to the $U(N-2)$ gauge node, while the link with the rightmost $U(1)$ gauge node is removed.

\item We iterate this procedure $N-1$ times, meaning that we apply Aharony duality $N(N-1)/2$ times and we arrive to the abelian $U(1)^{N-1}$ linear quiver with exactly  $\mathcal{N}=4$ superpotential. 

\item There are no extra singlets, since they became massive because of $\mathcal{S}_{[1^N]}$ and $\mathcal{T}_{[N-1,1]}$.

\end{itemize}

 The final results is a linear quiver with $N-1$ $U(1)$ gauge nodes, connected by  bifundamental flavors $p^{(i-1,i)}$, $\tilde{p}^{(i-1,i)}$. The first and last nodes are also connected  to  fundamental flavors $p^{(0,1)}$, $\tilde{p}^{(0,1)}$ and $p^{(N-1,N)}$, $\tilde{p}^{(N-1,N)}$. The superpotential consists of the standard $\mathcal{N}=4$ interaction with the adjoint chiral fields
\be
\mathcal{W}_{T^{[N-1,1]}[SU(N)]}=\sum_{i=1}^{N-1}\Gp^{(i)}\left(\tilde{p}^{(i,i+1)}p^{(i,i+1)}+\tilde{p}^{(i-1,i)}p^{(i-1,i)}\right)\,.
\ee
This theory is indeed dual to the $\mathcal{N}=4$ SQED with $N$ flavors according to abelian Mirror Symmetry and it corresponds to $T^{[N-1,1]}[SU(N)]$.

\subsubsection{$\rho=[2,1^2]$ and $\sigma=[1^4]$}
\label{sec2.4.1}

\subsubsection*{Flow to $T_{[2,1^2]}[SU(4)]$}

We start analyzing the vev for the CB moment map as a monopole deformation in the flip-flip dual theory plus flipping fields
\be
\delta\mathcal{W}=\Tr_X\left[\left(\mathcal{J}_{[1^4]}+\mathcal{S}_{[1^4]}\right) \hb\right]+\Tr_Y\left[\left(\mathcal{J}_{[2,1^2]}+\mathcal{T}_{[2,1^2]}\right) \cb\right]\, .
\label{superpotT211CB}
\ee
In this case the Jordan matrix encoding the nilpotent deformation is
\be
\mathcal{J}_{[2,1^2]}=\left( \begin{array}{cc|c|c}
0 & 1 & 0 & 0 \\
0 & 0 & 0 & 0 \\ \hline
0 & 0 & 0 & 0 \\ \hline
0 & 0 & 0 & 0
\end{array}\right)
\label{jordan211}
\ee
and consequently the matrix of singlets that we need to add is
\be
\mathcal{T}_{[2,1^2]}=\left( \begin{array}{cc|c|c}
\ga_2 & 0 & 0 & 0 \\
\ga_1 & \ga_2 & \tilde{\gc}_1 & \tilde{\gc}_2 \\ \hline
\gc_1 & 0 & \gb_{33} & \gb_{34} \\ \hline
\gc_2 & 0 & \gb_{43} & -2\ga_2-\gb_{33}
\end{array}\right)\, ,
\label{Tsinglets211}
\ee
while $\mathcal{S}_{[1^4]}$ is an abritrary $4\times 4$ traceless matrix. Hence, the deformation $\gd\mathcal{W}$ corresponds to turning on linearly the positive fundamental monopole of the first $U(1)$ gauge node of $T[SU(4)]$
\be
\mathcal{W}_{\mathcal{T}}=\mathcal{W}_{T[SU(4)]}+\mathfrak{M}^{(1,0,0)}+\Tr_X\left(\mathcal{S}_{[1^4]}\, \hb\right)+\Tr_Y\left(\mathcal{T}_{[2,1^2]}\,\cb\right)\, .
\ee
This monopole deformation breaks the $SU(4)_Y$ global symmetry down to $U(1)_{Y^{(1)}}\times SU(2)_{Y^{(2)}}$.

 In terms of the real masses $Y_n$, the superpotential term we added implies the constraint
\be
Y_2=Y_1+2m_A\, .
\label{realmassconstr}
\ee
Moreover, it will be useful to also redefine the $Y_1$ real mass by
\be
Y_1\rightarrow Y_1-m_A\, .
\label{Y1shift}
\ee
The residual symmetry is then parametrized by
\be
&&Y^{(1)}=Y_1\nn\\
&&Y^{(2)}_1=Y_3+Y_1\nn\\
&&Y^{(2)}_2=Y_4+Y_1\,,
\label{redefinitionY211}
\ee
The charges and representations of the chiral fields of the theory are the same as those of $T[SU(4)]$ since the deformation only affected the monopole operators. The gauge singlets in $\mathcal{T}_{[2,1^2]}$ transform under the global symmetries as follows\footnote{With $\gb$ we collectively denote the singlets $\gb_{33}$, $\gb_{34}$, $\gb_{43}$ that form a triplet of $SU(2)$. Similarly $\gc$, $\tilde{\gc}$ are made of the singlets $\gc_1$, $\gc_2$, $\tilde{\gc}_1$, $\tilde{\gc}_2$ and transform as two doublets under $SU(2)$.}
\begin{table}[h]
\centering
\scalebox{1}{
\begin{tabular}{c|cccc|c}
{} & $SU(4)_X$ & $U(1)_{Y_1}$ & $SU(2)_{Y_3,Y_4}$  & $U(1)_{m_A}$ & $U(1)_{R_0}$  \\ \hline
$\ga_1$ & $\bullet$ & 0 & $\bullet$  & 4 & 0 \\
$\ga_2$ & $\bullet$ & 0 & $\bullet$  & 2 & 0 \\
$\gb$ & $\bullet$ & 0 & ${\bf 3}$  & 2 & 0 \\
$\gc$, $\tilde{\gc}$ & $\bullet$ & $\pm1$ & ${\bf 2}$  & 3 & 0 \\
$\mathcal{S}_{[1^4]}$ & ${\bf 15}$ & 0 & $\bullet$  & $-2$ & 2 \\
\end{tabular}}
\end{table}

\newpage
\noindent where $U(1)_{Y_1}$ and $SU(2)_{Y_3,Y_4}$ denote the symmetries after imposing the superpotential constraint \eqref{realmassconstr}--\eqref{Y1shift}, but before the redefinition \eqref{redefinitionY211}. 
This will be performed at the very end of the derivation of the flip-flip dual of theory $\mathcal{T}$, coinciding with $T_{[2,1^2]}[SU(4)]$.

We can study the deformation at the level of the $S^3_b$ partition function of theory $\mathcal{T}$, which can be obtained imposing \eqref{realmassconstr} and \eqref{Y1shift} on $\mathcal{Z}_{T[SU(4)]}$
\be
&&\mathcal{Z}_{\mathcal{T}}=\mathcal{B}\int\udl{\vec{z}_3^{(3)}}\e^{2\pi i(Y_3-Y_4)\sum_{i=1}^3z_i^{(3)}}\prod_{i,j=1}^3\sbfunc{-i\frac{Q}{2}+(z_i^{(3)}-z_j^{(3)})+2m_A}\nn\\
&&\qquad\times\prod_{i=1}^3\prod_{n=1}^4\sbfunc{i\frac{Q}{2}\pm(z_i^{(3)}-X_n)-m_A}\int\udl{\vec{z}_2^{(2)}}\e^{2\pi i(Y_1-Y_3+m_A)\sum_{a=1}^2z_a^{(2)}}\nn\\
&&\qquad\times\prod_{a,b=1}^2\sbfunc{-i\frac{Q}{2}+(z_a^{(2)}-z_b^{(2)})+2m_A}\prod_{a=1}^2\prod_{i=1}^3\sbfunc{i\frac{Q}{2}\pm(z_a^{(2)}-z_i^{(3)})-m_A}\nn\\
&&\qquad\times\int\udl{z_1^{(1)}}\e^{-4\pi im_Az^{(1)}}\sbfunc{-i\frac{Q}{2}+2m_A}\prod_{a=1}^2\sbfunc{i\frac{Q}{2}\pm(z^{(1)}-z_a^{(2)})-m_A}\, ,
\label{211p}
\ee
where $\mathcal{B}$ is the contribution of the singlets
\be
\mathcal{B}&=&\prod_{n,m=1}^4\sbfunc{-i\frac{Q}{2}+(X_n-X_m)+2m_A}\sbfunc{i\frac{Q}{2}-2m_A}\sbfunc{i\frac{Q}{2}-4m_A}\nn\\
&\times&\prod_{\ga,\gb=3}^4\sbfunc{i\frac{Q}{2}+(Y_\ga-Y_\gb)-2m_A}\prod_{\ga=3}^4\sbfunc{i\frac{Q}{2}\pm(Y_1-Y_\ga)-3m_A}\, .
\label{singlets211}
\ee

\begin{figure}[t]
	\centering
	\makebox[\linewidth][c]{
  	\includegraphics[scale=0.36]{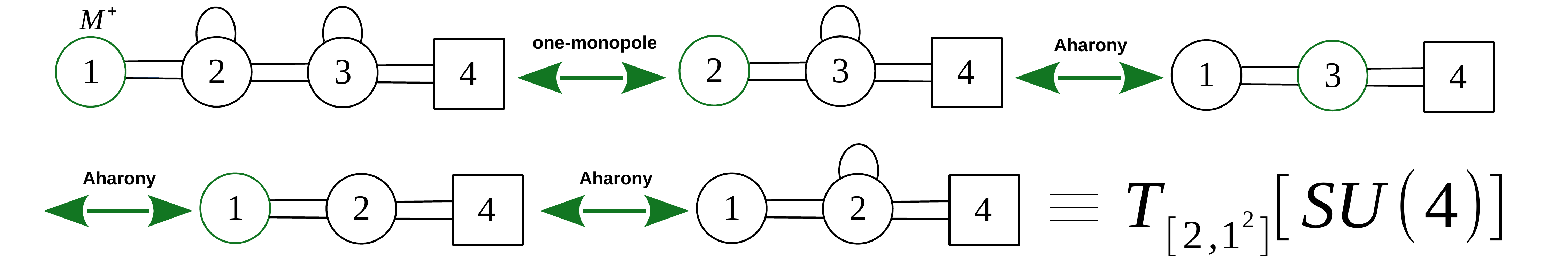} }
    \caption{Quiver representation of the sequential application of Aharony and one-monopole duality that leads to $T_{[2,1^2]}[SU(4)]$ starting from its flip-flip dual $\mathcal{T}$.}
 \label{T211mono}
\end{figure}

As mentioned in our previous general discussion, from  $\mathcal{T}$  we can reach the flip-flip dual theory 
$T_{[2,1^2]}[SU(4)]$ by sequentially applying Aharony and one-monopole duality. We show this explicitly for this particular case at the level of the sphere partition function in Appendix \ref{appA.2}, while here we only outline the main steps of the derivation sketched in Figure \ref{T211mono}.

We begin by applying the one-monopole duality to the $U(1)$ gauge node in \eqref{211p}. This node confines yielding a quiver theory  with no monopoles turned on:
  \be
\mathcal{Z}_{\mathcal{T}}&=&\mathcal{B}\,\sbfunc{-i\frac{Q}{2}+2m_A}\sbfunc{-i\frac{Q}{2}+4m_A}\int\udl{\vec{z}_3^{(3)}}\e^{2\pi i(Y_3-Y_4)\sum_{i=1}^3z_i^{(3)}}\nn\\
&\times&\prod_{i,j=1}^3\sbfunc{-i\frac{Q}{2}+(z_i^{(3)}-z_j^{(3)})+2m_A}\prod_{i=1}^3\prod_{n=1}^4\sbfunc{i\frac{Q}{2}\pm(z_i^{(3)}-X_n)-m_A}\nn\\
&\times&\int\udl{\vec{z}_2^{(2)}}\e^{2\pi i(Y_1-Y_3)\sum_{a=1}^2z_a^{(2)}}\prod_{a=1}^2\prod_{i=1}^3\sbfunc{i\frac{Q}{2}\pm(z_a^{(2)}-z_i^{(3)})-m_A}\, .
\ee
From this frame we  proceed by iteratively applying Aharony duality until we reach the flip-flip dual frame\footnote{Note that as a consequence of the sequential application of the Aharony and the one-monopole duality, the fugacities for the topological symmetries are permuted and appear in the opposite order compared to the definition of the original $T[SU(4)]$ partition function. For this reason, we call the index \eqref{pfTSU4211} as  $\mathcal{Z}_{T_{[2,1^2]}[SU(4)]}(\vec X ;\vec Y^{(2)},Y^{(1)};m_A)$ instead of $\mathcal{Z}_{T_{[2,1^2]}[SU(4)]}(\vec X ;Y^{(1)},\vec Y^{(2)};m_A)$. Indeed we can't use the  $SU(4)_Y$ Weyl symmetry to reorder the two set of fugacities $Y^{(1)}$ and $Y^{(2)}$ since this is not a symmetry of $T_{[2,1^2]}[SU(4)]$.}:
\be
&&\mathcal{Z}_{\mathcal{T}}=\int\udl{\vec{z}_{2}^{(3)}}\e^{2\pi i(2Y^{(1)}-Y^{(2)}_1)\sum_{i=1}^2z_i^{(3)}}\prod_{i,j=1}^2\sbfunc{i\frac{Q}{2}+(z_i^{(3)}-z_j^{(3)})-m_A}\nn\\
&&\quad\times\prod_{i=1}^2\prod_{n=1}^4\sbfunc{\pm(z_i^{(3)}+X_n)+m_A}\int\udl{z_{1}^{(2)}}\e^{2\pi i(Y^{(2)}_1-Y^{(2)}_2)z^{(2)}}\sbfunc{i\frac{Q}{2}-2m_A}\nn\\
&&\quad\times\prod_{i=1}^2\sbfunc{\pm(z^{(2)}-z_i^{(3)})+m_A}=\mathcal{Z}_{T_{[2,1^2]}[SU(4)]}(\vec X ;\vec Y^{(2)},Y^{(1)};m_A)\, .
\label{pfTSU4211}
\ee
In this last expression we also introduced the proper $U(1)_{Y^{(1)}}\times SU(2)_{Y^{(2)}}$ fugacities defined in \eqref{redefinitionY211}. This is precisely the partition function of $T_{[2,1^2]}[SU(4)]$.

\subsubsection*{Flow to $T^{[2,1^2]}[SU(4)]$}

We now move to analyzing the deformation in the mirror dual theory. This corresponds to a vev for the HB moment map which we can study as a mass deformation of $T[SU(4)]^\vee$ plus flipping fields
\be 
\delta\mathcal{W}^\vee=\Tr_Y\left[\left(\mathcal{J}_{[2,1^2]}+\mathcal{T}_{[2,1^2]}\right) \hb^\vee\right]+\Tr_X\left[\left(\mathcal{J}_{[1^4]}+\mathcal{S}_{[1^4]}\right) \cb^\vee\right]\, ,
\ee
where $\mathcal{T}_{[2,1^2]}$ is the matrix \eqref{Tsinglets211}. 
The mass deformation breaks the $SU(4)_Y$ global symmetry associated to the HB of $T[SU(4)]^\vee$ down to $U(1)_{Y^{(1)}}\times SU(2)_{Y^{(2)}}$. We parametrize these symmetries with the fugacities $Y^{(1)}$, $Y^{(2)}_\ga$ defined as in \eqref{realmassconstr}--\eqref{Y1shift}--\eqref{redefinitionY211}.
After integrating out the massive fields, we end up with a quiver similar to $T[SU(4)]^\vee$, but with only three flavors at the end of the tail coupling to different powers of the adjoint chiral field of the last node and extra flipping fields:
\be
\mathcal{W}_{\mathcal{T}^\vee}&=&\mathcal{W}_{T[SU(3)]}-\Tr_{3}\left(\Gp^{(3)}\Tr_{2}
\tilde q^{(2,3)}  q^{(2,3)} \right)+\Tr_3\left(\Gp^{(3)}\tilde{q}_1q_1\right)+\Tr_3\left[\left(\Gp^{(3)}\right)^2\Tr_{Y^{(2)}}\left(\tilde{q}_2q_2\right)\right]+\nn\\
&+&\left.\Tr_Y\left(\mathcal{T}_{[2,1^2]} \hb^\vee\right)\right|_{eom}+\Tr_X\left(\mathcal{S}_{[1^4]}\cb^\vee\right)\, .
\ee
where $\Tr_{Y^{(2)}}$ is the trace with respect to the $SU(2)_{Y^{(2)}}$ symmetry which is manifest in this frame of the web 
and
\be
\left.\Tr_Y\left(\mathcal{T}_{[2,1^2]} \hb^\vee\right)\right|_{eom}&=&\ga_1\Tr_3\left(\tilde{q}_1q_1\right)+\ga_2\Tr_3\Tr_{Y^{(2)}}\left(\tilde{q}_2q_2\right)\nn\\
&+&\Tr_{Y^{(2)}}\left(\gb\, \mathcal{H}^{(2)}\right)+\Tr_{Y^{(2)}}\left[\gc\Tr_3\left(\tilde{q}_2q_1\right)\right]+\Tr_{Y^{(2)}}\left[\tilde{\gc}\Tr_3\left(\tilde{q}_1q_2\right)\right]\, ,\nn\\
\ee
where we defined the $SU(2)_{Y^{(2)}}$ moment map 
\be
\mathcal{H}^{(2)}=\Tr_3\left(\tilde{q}_2q_2\right)-\frac{1}{2}\Tr_{Y^{(2)}}\Tr_3\left(\tilde{q}_2q_2\right)\,.
\ee

The three-sphere partition function of this theory can be obtained from the one of $T[SU(4)]^\vee$ imposing the constraint on the fugacities \eqref{realmassconstr} and \eqref{Y1shift}, simplifying the contribution of the massive fields thanks to the relation $\sbfunc{x}\sbfunc{-x}=1$ and adding the contribution of the singlets $\mathcal{T}_{[2,1^2]}$ and $\mathcal{S}_{[1^N]}$
\be
&&\mathcal{Z}_{\mathcal{T}^\vee}=\mathcal{B}\int\udl{\vec{z}_3^{(3)}}\e^{2\pi i(X_3-X_4)\sum_{i=1}^3z_i^{(3)}}\prod_{i,j=1}^3\sbfunc{i\frac{Q}{2}+(z_i^{(3)}-z_j^{(3)})-2m_A}\nn\\
&&\qquad\times\prod_{i=1}^3\sbfunc{\pm(z_i^{(3)}-Y_1)+2m_A}\prod_{\ga=3}^4\sbfunc{\pm(z_i^{(3)}-Y_\ga)+m_A}\int\udl{\vec{z}_2^{(2)}}\e^{2\pi i(X_2-X_3)\sum_{a=1}^2z_a^{(2)}}\nn\\
&&\qquad\times\prod_{a,b=1}^2\sbfunc{i\frac{Q}{2}+(z_a^{(2)}-z_b^{(2)})-2m_A}\prod_{a=1}^2\prod_{i=1}^3\sbfunc{\pm(z_a^{(2)}-z_i^{(3)})+m_A}\nn\\
&&\qquad\times\int\udl{z_1^{(1)}}\e^{2\pi i(X_1-X_2)z^{(1)}}\sbfunc{i\frac{Q}{2}-2m_A}\prod_{a=1}^2\sbfunc{\pm(z^{(1)}-z_a^{(2)})+m_A}\, .
\ee
where $\mathcal{B}$ is the contribution of the singlets defined in \eqref{singlets211}.

\begin{figure}[t]
	\centering
	\makebox[\linewidth][c]{
  	\includegraphics[scale=0.3]{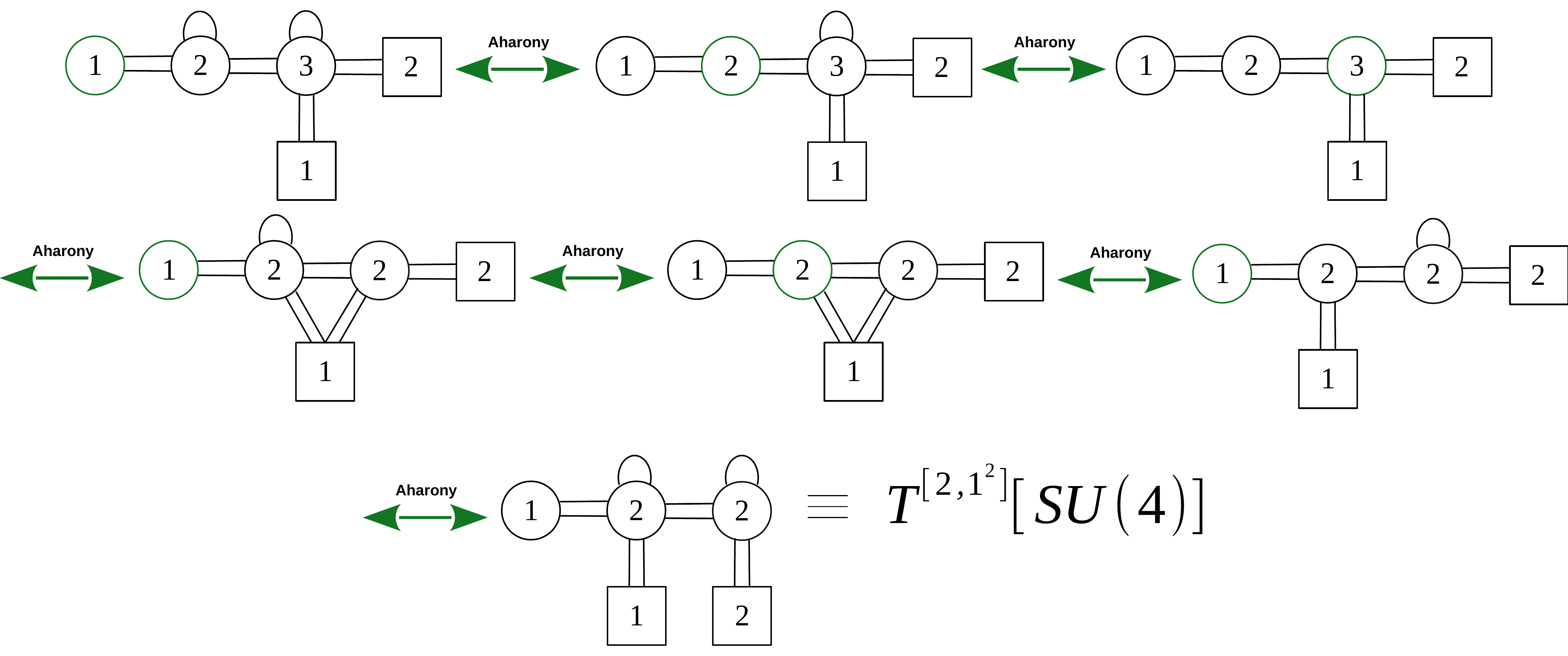} }
  	\caption{Quiver representation of the sequential application of Aharony duality that leads to $T^{[2,1^2]}[SU(4)]$ starting from its flip-flip dual $\mathcal{T}^\vee$. }
  	\label{T211mass}
\end{figure}

Again we want to find the flip-flip dual frame of this theory since we know that it will coincide with $T^{[2,1^2]}[SU(4)]$ and we claim that it can be obtained by sequentially applying Aharony duality only, as in this case there is no monopole superpotential. This derivation is carried out explicitly for this particular case at the level of the sphere partition function in Appendix \ref{appA.2}, while here we just report the final result, where we introduced the new fugacities \eqref{redefinitionY211}\footnote{{ Again, the labelling of the topological parameters  $X_n$ is in the opposite order compared to the original $T[SU(4)]^\vee$ partition function. This time, however, the permutations of $X_n$ belong to the Weyl symmetry of the $SU(4)_X$ global symmetry. Thus, the partition function is invariant under such permutations, so we just call it $\mathcal{Z}_{T^{[2,1^2]}[SU(4)]}(Y^{(1)},\vec Y^{(2)};\vec X;i\frac{Q}{2}-m_A)$ without specifying a particular order of $X_n$.}}
\be
\mathcal{Z}_{\mathcal{T}^\vee}&=&\e^{4\pi i(X_1+X_2)Y^{(1)}}\int\udl{\vec{z}_{2}^{(3)}}\e^{2\pi i(X_1-X_2)\sum_{i=1}^2z_i^{(3)}}\prod_{i,j=1}^2\sbfunc{-i\frac{Q}{2}+(z_i^{(3)}-z_j^{(3)})+2m_A}\nn\\
&\times&\prod_{i=1}^2\prod_{\ga=1}^2\sbfunc{i\frac{Q}{2}\pm(z_i^{(3)}+Y^{(2)}_\ga)-m_A}\int\udl{\vec{z}_{2}^{(2)}}\e^{2\pi i(X_2-X_3)\sum_{a=1}^2z^{(2)}_a}\nn\\
&\times&\prod_{a,b=1}^2\sbfunc{-i\frac{Q}{2}+(z^{(2)}_a-z^{(2)}_b)+2m_A}\prod_{a=1}^2\sbfunc{i\frac{Q}{2}\pm(z^{(2)}_a+Y^{(1)})-m_A}\nn\\
&\times&\prod_{i=1}^2\sbfunc{i\frac{Q}{2}\pm(z^{(2)}_a-z_i^{(3)})-m_A}\sbfunc{-i\frac{Q}{2}+2m_A}\int\udl{z^{(1)}_{1}}\e^{2\pi i(X_3-X_4)z^{(1)}}\nn\\
&\times&\prod_{a=1}^2\sbfunc{i\frac{Q}{2}\pm(z^{(1)}-z^{(2)}_a)-m_A}=\mathcal{Z}_{T^{[2,1^2]}[SU(4)]}(Y^{(1)},\vec Y^{(2)};\vec X;i\frac{Q}{2}-m_A)\, .
\label{pfTSU4211mirror}
\ee

This is precisely the partition function of $T^{[2,1^2]}[SU(4)]$, which is the quiver theory depicted at the end of Figure \ref{T211mass} where all the fields interact with the $\mathcal{N}=4$ superpotential. The presence of the contact terms in the prefactor is essential in order for the partition function of $T_{[2,1^2]}[SU(4)]$ in \eqref{pfTSU4211} to match with the one of $T^{[2,1^2]}[SU(4)]$ in \eqref{pfTSU4211mirror}. Indeed, from the equality of the partition functions \eqref{idmirror} of $T[SU(4)]$ and $T[SU(4)]^\vee$ and the results of the manipulations we just explained it follows the equality of the partition functions associated to the Mirror Symmetry relating $T_{[2,1^2]}[SU(4)]$ and $T^{[2,1^2]}[SU(4)]$
\be
\mathcal{Z}_{T_{[2,1^2]}[SU(4)]}(\vec X;\vec Y^{(2)},Y^{(1)};m_A)=\mathcal{Z}_{T^{[2,1^2]}[SU(4)]}(Y^{(1)},\vec Y^{(2)};\vec X;i\frac{Q}{2}-m_A)\, ,
\ee
where the parameter $m_A$ is mapped to $i\frac{Q}{2}-m_A$ across the duality, as required by Mirror Symmetry \eqref{mAchange}.
\\

\section{\boldmath$4d$ mirror-like dualities and \boldmath$E^\gs_\gr[USp(2N)]$ theories}
\label{sec3}

\subsection{\eusp duality web }
\label{euspgeneral}

\begin{figure}[t]
	\centering
  	\includegraphics[scale=0.65]{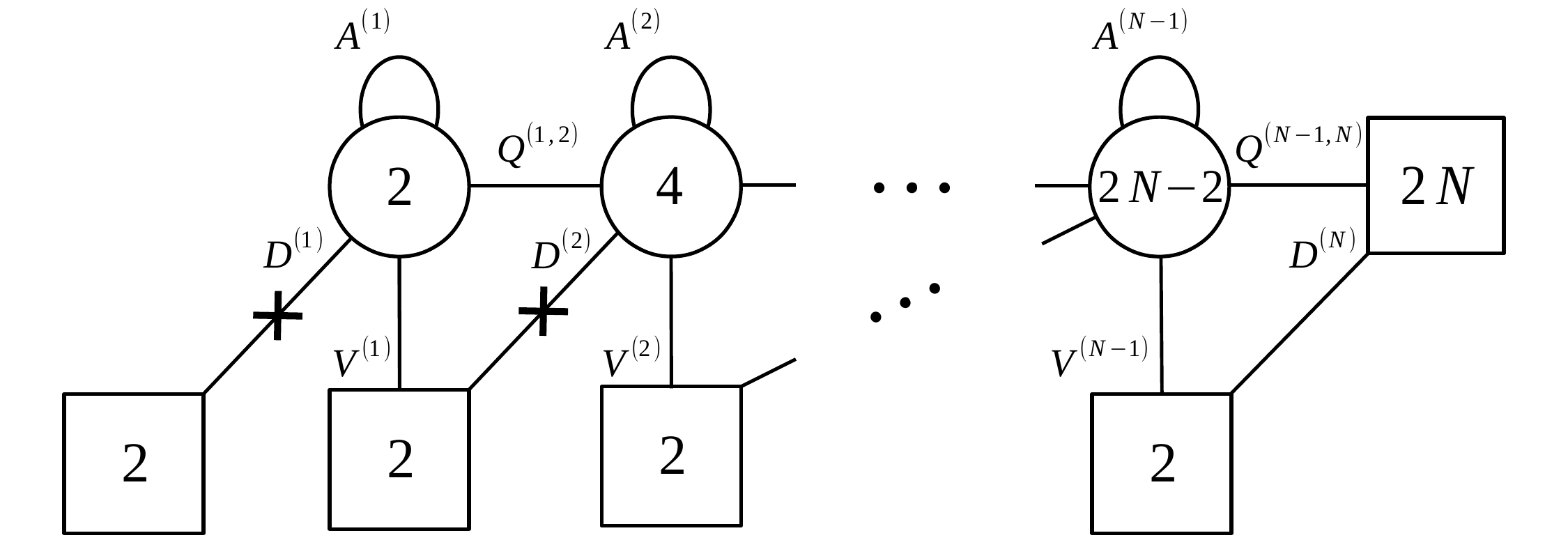} 
   \caption{Quiver diagram for \eusp. Round nodes denote gauge symmetries and square nodes denote global symmetries. Single lines denote chiral fields in representations of the nodes they are connecting. In particular, lines between adjacent nodes denote a chiral field in the bifundamental representation of the two nodes symmetries, while arcs denote chiral fields in the antisymmetric representation of the corresponding node symmetry. Crosses represent the singlets $\gb_n$ that flip the diagonal mesons.}
  	\label{euspfields}
\end{figure}

In this section we review the \eusp theory and its duality web, which were first discussed in \cite{Pasquetti:2019hxf}.
\eusp is a $4d$ $\mathcal{N}=1$ theory which admits a Lagrangian description in terms of the quiver represented in Figure \ref{euspfields}. 
The gauge group is $\prod_{n=1}^{N-1}USp(2n)$ and the matter content consists of the following chiral fields in the singlet, fundamental, bifundamental and antisymmetric representation\footnote{In contrast with \cite{Pasquetti:2019hxf}, we define \eusp without the set of singlets in the traceless antisymmetric representation of the $USp(2N)_x$ flavor symmetry, 
flipping the meson matrix,  and without the singlet $\gb_N$ flipping the last diagonal meson.}:
\begin{enumerate}[$\bullet$]
\item a chiral field $Q^{(n,n+1)}$ in the bifundamental representation of $USp(2n)\times USp(2(n+1))$, with $n=1,\cdots,N-1$;
\item two chiral fields $D^{(n)}_\ga$ in the fundamental representation of $USp(2n)$, which form a doublet of the $n$-th $SU(2)$ flavor symmetry of the saw, with $n=1,\cdots,N$;
\item two chiral fields $V^{(n)}_\ga$ in the fundamental representation of $USp(2n)$, which form a doublet of the $(n+1)$-th $SU(2)$ flavor symmetry of the saw, with $n=1,\cdots,N-1$;
\item a chiral field $A^{(n)}$ in the antisymmetric representation of $USp(2n)$, with $n=1,\cdots,N-1$;
\item a gauge singlet $\gb_n$ that is coupled to the gauge invariant meson built from $D^{(n)}$ through a superpotential which will be discussed momentarily.
\end{enumerate}
In order to write the superpotential in a compact form, we define
\be
\mathbb{Q}^{(n,n+1)}_{abij}=Q^{(n,n+1)}_{ai}Q^{(n,n+1)}_{bj}
\ee
The superpotential consists of three main types of interactions: a cubic coupling between the bifundamentals and the antisymmetrics, another cubic coupling between the chirals in each triangle of the quiver and finally the flip terms with the singlets $\beta_n$ coupled to the   diagonal mesons
\be
\mathcal{W}_{E[USp(2N)]}&=&\sum_{n=1}^{N-1}\Tr_{n}\left[A^{(n)}\left(\Tr_{n+1}\mathbb{Q}^{(n,n+1)}-\Tr_{n-1}\mathbb{Q}^{(n-1,n)}\right)\right]\nn\\
&+&\sum_{n=1}^{N-1}\Tr_{y_{n+1}}\Tr_{n}\Tr_{n+1}\left(V^{(n)}Q^{(n,n+1)}D^{(n+1)}\right)+\sum_{n=1}^{N-1} \gb_n\Tr_{y_n}\Tr_{n}\left(D^{(n)}D^{(n)}\right)\, .\nn\\
\label{superpoteusp}
\ee
The traces are labelled as follows: $\Tr_n$ denotes the trace over the color indices of the $n$-th gauge node, while $\Tr_{y_n}$ denotes the trace over the $n$-th $SU(2)$ flavor symmetry. Notice that for $n=N$ we have the trace over the $USp(2N)_x$ flavor symmetry, which we will also denote by $\Tr_N=\Tr_x$. All the traces are defined including the $J$ antisymmetric tensor of $USp(2n)$
\be
J=\mathbb{I}_n\otimes i\,\gs_2\, .
\ee
For example, given a $2n\times 2n$ matrix $A$, we define
\be
\Tr\left(A\right)= J_{ij} A^{ij}\,.
\ee

\begin{figure}[t]
	\centering
  	\includegraphics[scale=0.65]{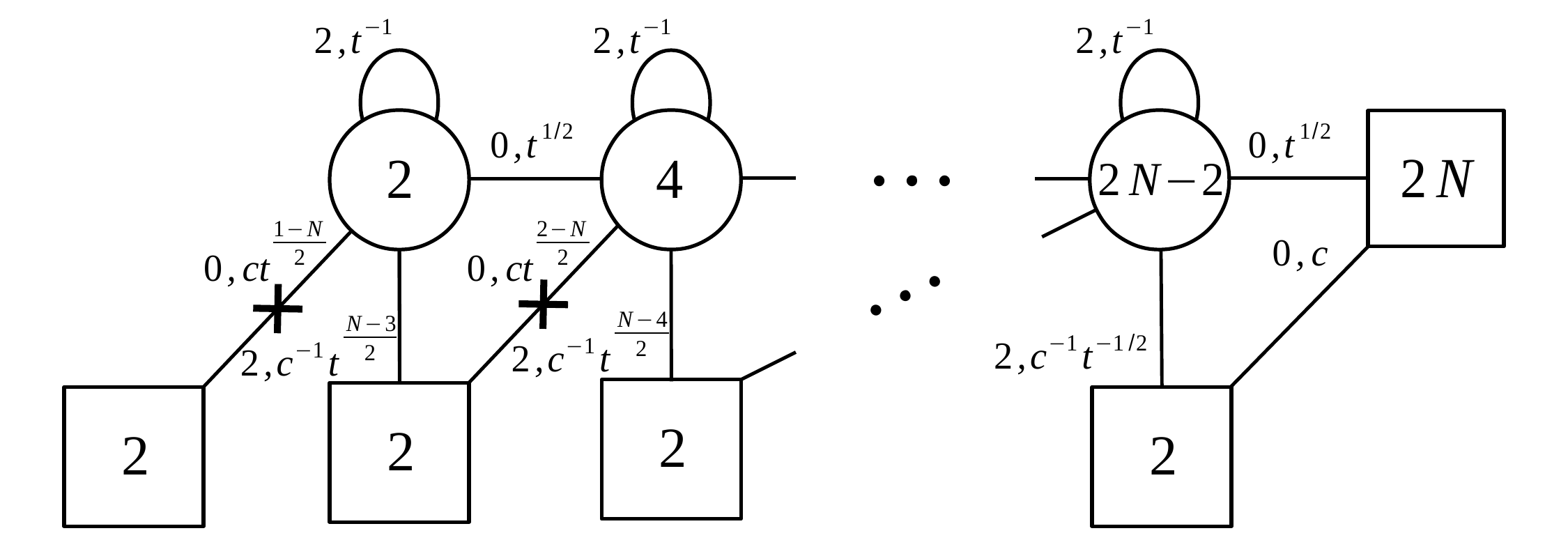} 
   \caption{Trial R-charges and charges under the abelian symmetries. The power of $c$ is the charge under $U(1)_c$, while the power of $t$ is the charge under $U(1)_t$.}
  	\label{euspfugacities}
\end{figure}

In this Lagrangian description the following non-anomalous global symmetry is manifest:
\be
USp(2N)_x\times\prod_{n=1}^NSU(2)_{y_n}\times U(1)_t\times U(1)_c\, .
\ee
This symmetry gets actually enhanced in the IR to
\be
USp(2N)_x\times USp(2N)_y\times U(1)_t\times U(1)_c\, .
\ee
In \cite{Pasquetti:2019hxf} this enhancement was argued studying the gauge invariant operators, which re-arrange into representations of the enhanced $USp(2N)_y$ symmetry, and using infra-red dualities. Indeed, as we will review shortly, there exists a dual frame of \eusp where $USp(2N)_y$ is manifest, while $USp(2N)_x$ is enhanced.

We assign trial R-charge, which we denote as $R_0$, zero to the fields $Q^{(n,n+1)}$ and $D^{(n)}$, and $R_0$ charge two to the fields $\gb_n$, $A^{(n)}$ and $V^{(n)}$.   This is not the superconformal R-symmetry, but it is anomaly free and consistent with the superpotential \eqref{superpoteusp}. Moreover, we define the $U(1)_c$ and $U(1)_t$ symmetries by assigning charges $0$ and $1/2$ to $Q^{(N-1,N)}$ and $1$ and $0$ to $D^{(N)}$. The charges of all the other chiral fields are then fixed by the superpotential and by the requirement that $U(1)_R$ is not anomalous at each gauge node, where $U(1)_R$ is defined taking into account the possible mixing of the abelian symmetries with the UV R-symmetry $U(1)_{R_0}$
\be
R=R_0+\mathfrak{c}q_c+\mathfrak{t}q_t\, ,
\label{Rdef}
\ee
where $q_c$ and $q_t$ are the charges under the two $U(1)$ symmetries and $\mathfrak{c}$ and $\mathfrak{t}$ are mixing coefficients. Among this two parameter family of R-charges, we can determine the exact superconformal one by $a$-maximization \cite{Intriligator:2003jj}. The charges of all the chiral fields under the two $U(1)$ symmetries as well as their trial R-charges in our conventions are summarized in Figure \ref{euspfugacities}. 

\begin{figure}[t]
	\centering
		\includegraphics[scale=0.6]{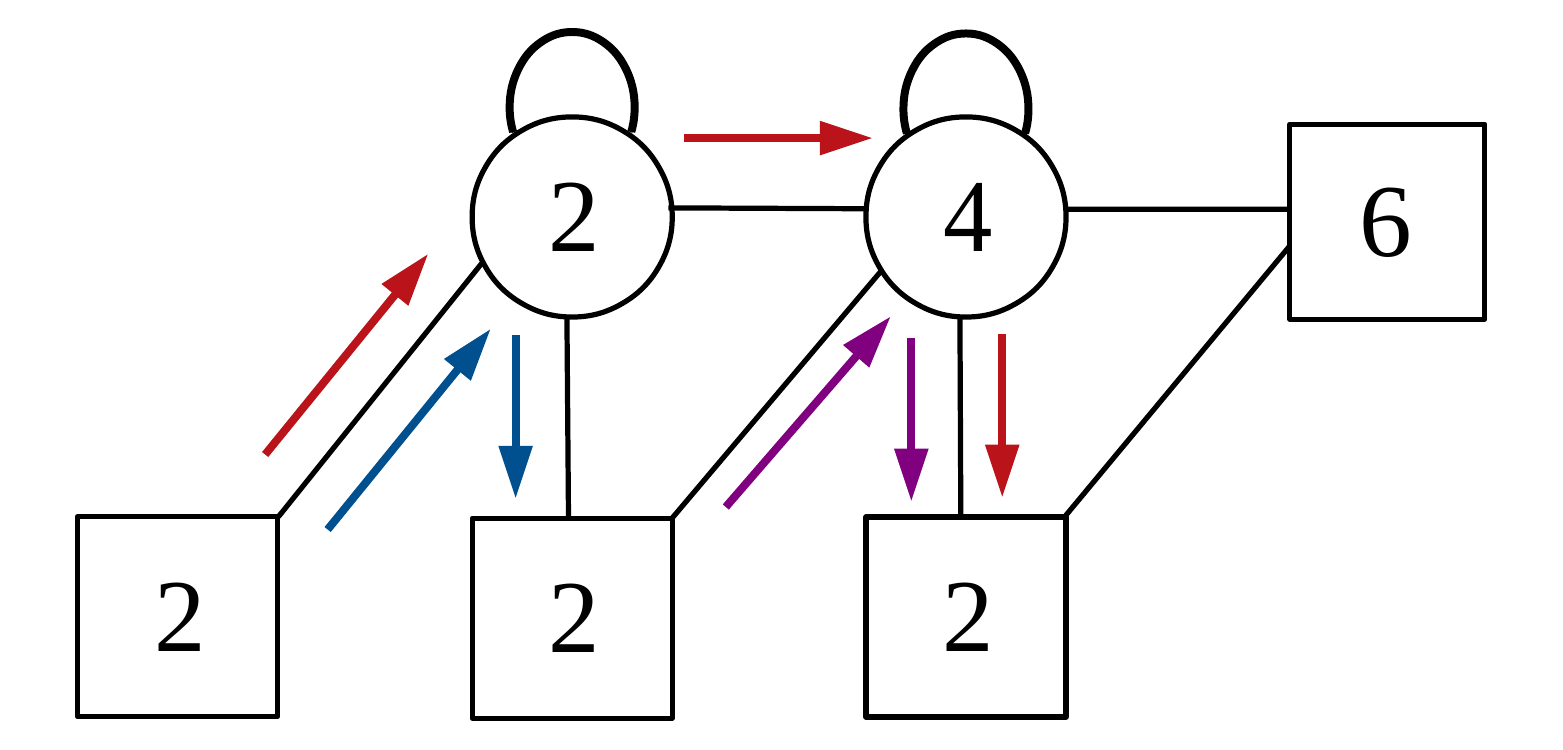}
	\caption{$SU(2)\times SU(2)$ bifundamental operators contributing to $\mathsf{C}$ in the $N=3$ case. }
	\label{Aop}
\end{figure}

The gauge invariant operators of \eusp that will be important for us are of three main types. First, we have two operators, which we denote by $\mathsf{H}$ and $\mathsf{C}$, in the traceless antisymmetric representation of $USp(2N)_x$ and $USp(2N)_y$ respectively. The first one is just the meson matrix
\be
\mathsf{H}=\Tr_{N-1}\left[Q^{(N-1,N)}Q^{(N-1,N)}-\frac{1}{N}\Tr_X\left(Q^{(N-1,N)}Q^{(N-1,N)}\right)\right]\, .
\ee
This operator has also $U(1)_c$ and $U(1)_t$ charge $0$ and $1$ respectively and trial R-charge $0$. The operator $\mathsf{C}$ is instead constructed collecting different gauge invariant operators, $N-1$ of them are singlets under the non-abelian global symmetries while the others are in the bifundamental representations of all the possible pairs of $SU(2)$ manifest symmetries of the saw. These have indeed the same charges under the abelian symmetries and the same trial R-charge and together they reconstruct the traceless antisymmetric representation of the enhanced $USp(2N)_y$ according to the branching rule under the subgroup $SU(2)^N\subset USp(2N)$
\be
{\bf N(2N-1)-1}\rightarrow (N-1)\times({\bf 1},\cdots,{\bf 1})\oplus \left[({\bf 2},{\bf 2},{\bf 1},\cdots,{\bf 1})\oplus\text{(all possible permutations)}\right]\, .\nn\\
\ee
The $N-1$ singlets are the traces of the antisymmetric chirals at each gauge node
\be
\Tr_{n}A^{(n)},\qquad n=1,\cdots,N-1\, ,
\ee
while the bifundamentals are constructed starting from one diagonal flavor, going along the tail with an arbitrary number of bifundamentals $Q^{(n,n+1)}$ and ending on a vertical chiral, with all the needed contractions of color indices (see Figure \ref{Aop}). All these operators have $U(1)_c$ and $U(1)_t$ charge $0$ and $-1$ respectively and trial R-charge $2$.

\begin{figure}[t]
	\centering
			\includegraphics[scale=0.6]{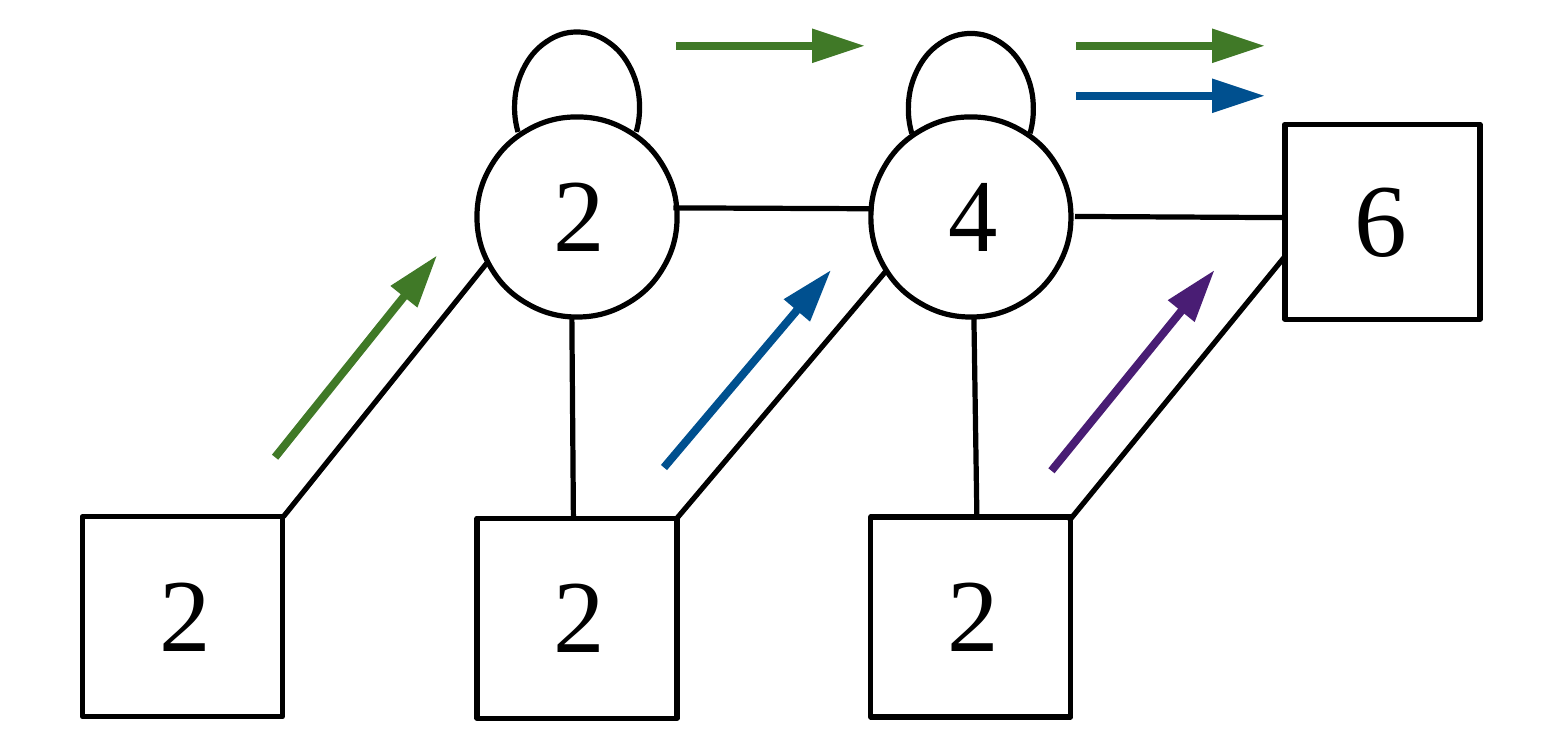}
		\caption{Operators contributing to $\Pi$ in the $N=3$ case.}
	\label{PIop}\end{figure}

There is also an operator $\Pi$ in the bifundamental representation of $USp(2N)_x\times USp(2N)_y$. This is constructed collecting $N$ operators in the fundamental representation of $USp(2N)_x$ and of each of the $SU(2)$ symmetries according to the branching rule under $SU(2)^N\subset USp(2N)$
\be
{\bf 2N}\rightarrow({\bf 2},{\bf 1},\cdots,{\bf 1})\oplus({\bf 1},{\bf 2},{\bf 1},\cdots,{\bf 1})\oplus\cdots\oplus({\bf 1},\cdots,{\bf 1},{\bf 2})\, .
\ee
These $N$ operators are obtained starting with one diagonal flavor and going along the tail with all the remaining bifundamentals ending on $Q^{(N-1,N)}$ (see Figure \ref{PIop}). All these operators have $U(1)_c$ and $U(1)_t$ charge $1$ and $0$ respectively and trial R-charge $0$.

Finally, we have some gauge invariant operators that are also singlets under the non-abelian global symmetries and are only charged under $U(1)_c$ and $U(1)_t$. Those that will be important for us are the chiral singlets $\beta_n$ and the mesons constructed with the vertical chirals and dressed with powers of the antisymmetrics. We can collectively denote these operators with
\be
B_{ij}=\begin{cases}
\gb_{N-j+1} & i=1,\quad j=2,\cdots, N \\
\Tr_{N-j}\left[\left(A^{(N-j)}\right)^{i-2}V^{(N-j)}V^{(N-j)}\right] & i=2,\cdots, N,\quad i+j\le N+1
\end{cases}\, .
\label{Bij}
\ee
These operators have $U(1)_c$ charge $-2$, $U(1)_t$ charge $j-i$ and trial R-charge $2i$.
The charges and representations of all these operators under the global symmetry are given in Table \ref{eusptable}.

In \cite{Pasquetti:2019hxf} it was shown that $E[USp(2N)]$ has a limit to the $T[SU(N)]$ theory \cite{Gaiotto:2008ak}. More precisely, the limit consists of first reducing \eusp to $3d$ and then taking a series of real mass deformations. The $3d$ limit results in an $\mathcal{N}=2$ theory with exactly the same structure, but with the fundamental monopole of $USp(2n)$ turned on at each gauge node, so the $3d$ theory has the same global symmetry of \eusp. Then we take a real mass deformation combined with a Coulomb branch deformation that breaks all the $USp(2n)$ gauge and global symmetries to $U(n)$. The resulting theory is the $M[SU(N)]$ theory studied in \cite{SP2}\footnote{The three-dimensional $M[SU(N)]$ theory was introduced in \cite{SP2} from a completely different perspective by exploiting a connection between $S^2\times S^1$ partition functions for $3d$ $\mathcal{N}=2$ theories and $2d$ free field correlators first proposed in \cite{SP1}.}.
The second real mass deformation, which reduces $M[SU(N)]$ to $T[SU(N)]$,
has the effect of integrating out all the fields charged under the original $U(1)_c$ symmetry of \eusp and restoring the topological symmetry at each node, thus removing the monopole superpotential.

Among the other operators, $\Pi$ and $B_{ij}$ become massive while the traceless antisymmetric operators $\mathsf{H}$, $\mathsf{C}$ of \eusp reduce to the adjoint operators $\mathcal{H}$, $\mathcal{C}$ of \tsu. Indeed, we can embed $U(1)\times SU(N)\subset USp(2N)$ and the traceless antisymmetric of $USp(2N)$ accordingly decomposes as
\be
{\bf N(2N-1)-1}\rightarrow({\bf N^2-1})^0\oplus\left({\bf \frac{N(N-1)}{2}}\right)^2\oplus\left({\bf\overline{ \frac{N(N-1)}{2}}}\right)^{-2}\,.
\ee
The real mass deformation makes the fields charged under the $U(1)$ part massive and leaves only the adjoint of $SU(N)$ components of $\mathsf{H}$ and $\mathsf{C}$ massless, which we identify with $\mathcal{H}$ and $\mathcal{C}$.\\

\begin{table}[t]
\centering
\scalebox{1}{
\begin{tabular}{c|cccc|c}
{} & $USp(2N)_x$ & $USp(2N)_y$ & $U(1)_t$ & $U(1)_c$ & $U(1)_{R_0}$ \\ \hline
$\mathsf{H}$ & ${\bf N(2N-1)-1}$ & $\bullet$ & $1$ & 0 & 0 \\
$\mathsf{C}$ & $\bullet$ & ${\bf N(2N-1)-1}$ & $-1$ & 0 & 2 \\
$\Pi$ & $\bf N$ & $\bf N$ & 0 & $+1$ & 0 \\
$B_{ij}$ & $\bullet$ & $\bullet$ & $j-i$ & $-2$ & $2i$
\end{tabular}}
\caption{Trasnformation rules of the \eusp operators.}
\label{eusptable}
\end{table}

One of our main tools for studying \eusp, its dualities and deformations will be the supersymmetric index \cite{Romelsberger:2005eg,Kinney:2005ej,Dolan:2008qi} (see also \cite{Rastelli:2016tbz} for a review). This will depend on fugacities for the $USp(2N)_x\times USp(2N)_y\times U(1)_c\times U(1)_t$ global symmetries that we accordingly denote by $x_n$, $y_n$, $c$ and $t$. It can be expressed with the following recursive definition:
\begin{equation}
\makebox[\linewidth][c]{\scalebox{1}{$
\begin{split}
&\mathcal{I}_{E[USp(2N)]}(\vec x;\vec y;t,c)  \\
&=\Gpq{pq\,c^{-2}t}\prod_{n=1}^N\Gpq{c\,y_N^{\pm1}x_n^{\pm1}}\oint\udl{\vec{z}_{N-1}^{(N-1)}} \Gpq{p q\, t^{-1}}^{N-1} \prod_{i < j}^{N-1} \Gpq{p q \,t^{-1} z^{(N-1)}_i{}^{\pm1} z^{(N-1)}_j{}^{\pm1}} \\
&\times \prod_{i=1}^{N-1}\frac{\prod_{n=1}^N\Gpq{t^{1/2}z^{(N-1)}_i{}^{\pm1}x_n^{\pm1}}}{\Gpq{t^{1/2}c\,y_N^{\pm1}z^{(N-1)}_i{}^{\pm1}}} \mathcal{I}_{E[USp(2(N-1))]}\left(z^{(N-1)}_1,\cdots,z^{(N-1)}_{N-1};y_1,\cdots,y_{N-1};t,t^{-1/2}c\right)\, ,
\label{indexEN}
\end{split}$}}
\end{equation}
with the base of the iteration defined as
\be
\mathcal{I}_{E[USp(2)]}(x;y;c) =\Gpq{c\,y^{\pm1}x^{\pm1}}\,.
\ee
We also defined the integration measure of the $m$-th $USp(2n)$ gauge node as
\be
\udl{\vec{z}^{(m)}_n}=\frac{\left[(p;p)(q;q)\right]^n}{2^n n!}\prod_{i=1}^n\frac{\udl{z^{(m)}_i}}{2\pi i\,z^{(n)}_i} \frac{1}{\prod_{i<j}^{n}\Gpq{z^{(m)}_i{}^{\pm1}z^{(m)}_j{}^{\pm1}}\prod_{i=1}^{n}\Gpq{z^{(m)}_i{}^{\pm2}}}\,.
\ee
This index is defined using the assignment of R-charges as depicted in Figure \ref{euspfugacities}. If one wishes to use another non-anomalous assignment of R-charges then the parameters should be redefined as,
\be
c\to  c\, (pq)^{\mathfrak{c}/2}, \qquad t\to t\, (pq)^{\mathfrak{t}/2}\,,
\label{fugdef}
\ee
where $\mathfrak{c}$ and $\mathfrak{t}$ are the mixing coefficients appearing in \eqref{Rdef}. 
As pointed out in \cite{Pasquetti:2019hxf}, the expression \eqref{indexEN} coincides with the interpolation kernel $\mathcal{K}_c(x,y)$ studied in \cite{2014arXiv1408.0305R}, where many integral identities for this function were proven which support the dualities of \eusp that we are going to review.

Indeed, \eusp enjoys a web of dualities that is completely analogous to the one of \tsu and that we sketched in Figure \ref{euspweb}. First of all, we have a dual frame we denote by \mirroreusp where the $USp(2N)_x$ and $USp(2N)_y$ symmetries are exchanged and the $U(1)_t$ fugacity is mapped to
\be
t\rightarrow\frac{pq}{t}\,,
\label{redeft}
\ee
which means that all the charges under $U(1)_t$ are flipped and that the mixing coefficient is redefined as $\mathfrak{t}\to 2-\mathfrak{t}$.
In other words, \eusp is self-dual with a non-trivial map of the gauge invariant operators
\be
\mathsf{H}\quad&\leftrightarrow&\quad \mathsf{C}^\vee\nn\\
\mathsf{C}\quad&\leftrightarrow&\quad \mathsf{H}^\vee\nn\\
\Pi\quad&\leftrightarrow&\quad \Pi^\vee\nn\\
B_{ij}\quad&\leftrightarrow&\quad B_{ji}^\vee\, .
\label{opmap4dmirror}
\ee
We will refer to this duality as a $4d$ version of Mirror Symmetry, since it reduces to the self-duality of \tsu under Mirror Symmetry upon taking the dimensional reduction limit we mentioned above.
At the level of the index we have the following identity:
\be
\mathcal{I}_{E[USp(2N)]}(\vec x;\vec y;t,c)=\mathcal{I}_{E[USp(2N)]}(\vec y;\vec x;p q/t,c)\, ,
\label{selfduality}
\ee
which has been proven in Theorem 3.1 of \cite{2014arXiv1408.0305R} and which reduces to the identity \eqref{idmirror} for the mirror self-duality of \tsu in a suitable limit. This duality strongly supports the enhancement to $USp(2N)_y$, since this symmetry is explicitly manifest in the \mirroreusp dual frame.

On top of the mirror dual frame we have a second frame we denote by \ffeusp, which is defined as \eusp plus two sets of singlets $\mathsf{O_H}$ and $\mathsf{O_C}$ flipping the two operators $\mathsf{H}_{FF}$ and $\mathsf{C}_{FF}$
\be
\mathcal{W}_{FFE[USp(2N)]}=\mathcal{W}_{E[USp(2N)]}+\Tr_x\left(\mathsf{O_H}\mathsf{H}_{FF}\right)+\Tr_y\left(\mathsf{O_C}\mathsf{C}_{FF}\right)\, .
\ee
In this case the $USp(2N)_x$ and $USp(2N)_y$ symmetries are left unchanged, while only the $U(1)_t$ fugacity transforms as in \eqref{redeft}. The operator map is indeed
\be
\mathsf{H}\quad&\leftrightarrow&\quad \mathsf{O}_{\mathsf{H}}\nn\\
\mathsf{C}\quad&\leftrightarrow&\quad \mathsf{O}_{\mathsf{C}}\nn\\
\Pi\quad&\leftrightarrow&\quad\Pi_{FF}\nn\\
B_{ij}\quad&\leftrightarrow&\quad B_{FF,\,ji}\, .
\ee
We will refer to this duality as a $4d$ version of flip-flip duality, since it reduces to the flip-flip duality of \tsu upon taking the same dimensional reduction limit discussed in \cite{Pasquetti:2019hxf}.

 In analogy with the three-dimensional case, this flip-flip dual frame can be reached by iteratively applying Intriligator--Pouliot duality \cite{Intriligator:1995ne} to \eusp with the same strategy described for the flip-flip duality of \tsu in Section \ref{sectsunweb}\footnote{It should be noted that the Aharony duality used in the derivation of the flip-flip dual of \tsu can be obtained from a dimensional reduction limit of Intriligator--Pouliot duality, as shown in \cite{Benini:2017dud}. This limit is the same that relates \eusp and \tsu.}:

\begin{itemize}
\item At the first iteration we start from the  $USp(2)$ node, whose antisymmetric  chiral is just a singlet.
Intriligator--Pouliot duality has the effect of making the antisymmetric chiral field of the adjacent $USp(4)$ node  massive, so that we can then apply again  the Intriligator--Pouliot   duality on it. We continue applying iteratively the Intriligator--Pouliot duality until we reach the last $USp(2(N-1))$  node.
Notice that since every $USp(2n)$ node sees $4n+4$ chirals the ranks do not change when we apply the duality.
Moreover some of the singlet fields expected from the Intriligator--Pouliot  duality are massive (because of the R-charge assignement) and no new links  between gauge nodes are created.

\item At the second iteration we start again from the $USp(2)$ node and proceed along the tail, but this time we stop at the second last node
$USp(2(N-2))$.

\item We iterate this procedure for a total of $N-1$ times, meaning that we apply Intriligator--Pouliot duality $N(N-1)/2$ times. 

\item The singlet fields  appearing in the Intriligator--Pouliot  duality reconstruct the singlet matrices $\mathsf{O_H}$ and $\mathsf{O_C}$.
\end{itemize}

At the level of the supersymmetric index, the flip-flip duality is encoded in the following integral identity:
\be
\mathcal{I}_{E[USp(2N)]}(\vec x;\vec y;t,c)&=&\Gpq{t}^{N} \Gpq{p q t^{-1}}^{N}\prod_{n<m}^N\Gpq{t x_n^{\pm1}x_m^{\pm1}}\Gpq{p q t^{-1} y_n^{\pm1}y_m^{\pm1}}\nn\\
&\times&\mathcal{I}_{E[USp(2N)]}(\vec x;\vec y;p q/t,c)\,,
\label{flipflipselfduality}
\ee
which is proven in Proposition 3.5 of \cite{2014arXiv1408.0305R} and can be alternatively derived by  applying iteratively as explained above the integral identity \eqref{IP} for Intriligator--Pouliot duality. We  show this in  Appendix \ref{4dff}.

Finally, we can combine the two previous dualities to find a third dual frame and complete the duality web of Figure \ref{euspweb}. We denote this frame by \mirrorffeusp and its superpotential is
\be
\mathcal{W}_{FFE[USp(2N)]^\vee}=\mathcal{W}_{E[USp(2N)]}+\Tr_y\left(\mathsf{O}_\mathsf{H}^\vee\mathsf{H}_{FF}^\vee\right)+\Tr_x\left(\mathsf{O}_{\mathsf{C}}^\vee\mathsf{C}_{FF}^\vee\right)\, .
\ee
Across this duality the $USp(2N)_x$ and $USp(2N)_y$ symmetries are exchanged, while $U(1)_t$ is left unchanged. Accordingly we have the operator map
\be
\mathsf{H}\quad&\leftrightarrow&\quad \mathsf{O}_{\mathsf{C}}^\vee\nn\\
\mathsf{C}\quad&\leftrightarrow&\quad \mathsf{O}_\mathsf{H}^\vee\nn\\
\Pi\quad&\leftrightarrow&\quad\Pi_{FF}^\vee\nn\\
B_{ij}\quad&\leftrightarrow&\quad B^\vee_{FF,ij}\, .
\ee
\\

\subsection{From $E[USp(2 N)]$ to $E_\rho^\sigma[USp(2 N)]$ using the web}

Now we would like to find a more general class of $4d$ $\mathcal{N}=1$ theories enjoying  mirror-like dualities. An obvious strategy to follow is to turn on vevs labelled by partitions $\rho = [\rho_1,\dots,\rho_N] = \left[N^{l_N},\dots,1^{l_1}\right]$ and $\sigma = [\sigma_1,\dots,\sigma_N] = \left[N^{k_N},\dots,1^{k_1}\right]$ for the operators $\mathsf{H}$ and $\mathsf{C}$. 
As we discussed above, the operators $\mathsf{H}$ and $\mathsf{C}$ reduce in the $3d$ limit followed by a suitable real mass deformation to the $3d$ moment maps $\mathcal{H}$ and $\mathcal{C}$. It is then easy to guess which $4d$ deformations of $E[USp(2 N)]$ reduce in the $3d$ limit to the nilpotent deformations depending on the partitions $\gr$ and $\gs$ of $SU(N)$  we turned on for $T[SU(N)]$. These are the deformations we are looking for and they correspond to the following vevs:
\be
\langle\mathsf{C}\rangle=\mathsf{J}_\gr \,, \qquad \langle\mathsf{H}\rangle=\mathsf{J}_\gs
\ee
where $\mathsf{J}_\gs$ and $\mathsf{J}_\gr$ are the antisymmetric matrices
\be
\mathsf{J}_\gr=\frac{1}{2}\left(J_\gr-J_\gr^T\right)\,,
\ee
where
\be
J_\gr = i\gs_2 \otimes (\mathbb J_{\gr_1} \oplus \dots \oplus \mathbb J_{\gr_L})
\ee
and $\mathbb J_{\gr_i}$ are the Jordan matrices we defined in \eqref{jordanmatrices}\footnote{Notice that the vevs we are considering are not labelled by partitions of $USp(2N)$, but by partitions of the $SU(N)$ part of $U(1)\times SU(N)\subset USp(2N)$. This choice is due to the fact that we want to mimic the deformation we perform in $3d$ and find models that reduce to $T^\gs_\gr[SU(N)]$.
}.
We call $E_\rho^\sigma[USp(2 N)]$ the theories we reach at the end of the flow triggered by such vevs, after suitably removing some extra massless fields, as we will discuss.

Again we can think that the vevs for  $\mathsf{H}$ and $\mathsf{C}$  are implemented by F-terms when we turn on  linear
deformations in $\mathsf{O}_{\mathsf{H}}$ and in $\mathsf{O}_{\mathsf{C}}$ in the flip-flip frame. We can then use the same strategy described in the $3d$ case, but this time using the $4d$ duality web of Figure \ref{euspweb2} and map these deformations across flip-flip duality, so that they become
mass deformations of $E[USp(2 N)]$. Finally we move back to the flip-flip dual frame, using sequentially the  Intriligator--Pouliot duality to reach $E_\rho^\sigma[USp(2 N)]$.\\

More precisely we consider the following deformation of $E[USp(2N)]$:
\begin{align}
\label{eq:def}
\delta \mathcal W&=  \mathrm{Tr}_x \left[(\mathsf{J}_\sigma+\mathsf S_\sigma) \cdot \mathsf H\right] + \mathrm{Tr}_y \left[(\mathsf{J}_\rho+\mathsf T_\rho) \cdot \mathsf C\right] +\sum_{\{(i,j) \neq (1,1)|1 \leq i \leq \sigma_j, \, 1 \leq j \leq \rho_i\}} \mathsf{O}_{B}^{ij} \, B_{ij}\,.
\end{align}
We have introduced extra gauge singlet chiral multiplets flipping some operators of the original $E[USp(2 N)]$ theory that would represent a massless free sector of the theory after the deformation. Note that the role of $\mathsf S_\sigma$ and $\mathsf T_\rho$ is the same as that of $\mathcal S_\sigma$ and $\mathcal T_\rho$ in $3d$, which flip part of the antisymmetric mesonic operators remaining massless in the presence of the mass terms, but in $4d$ they are determined requiring that they are traceless antisymmetric matrices commuting with the matrices $\mathsf{J}_\gs$ and $\mathsf{J}_\gr$ respectively. In addition, there are other gauge singlet fields $\mathsf{O}_B^{ij}$ which flip the operators $B_{ij}$ we defined in \eqref{Bij}\footnote{These extra $\mathsf{O}_{B}^{ij}$ singlets were absent in the $3d$ case. Indeed, they are charged under $U(1)_c$, which means that they are massive and integrated out in the limit leading to \tsu.}.

\begin{figure}[tbp]
\centering
\includegraphics[scale=0.5]{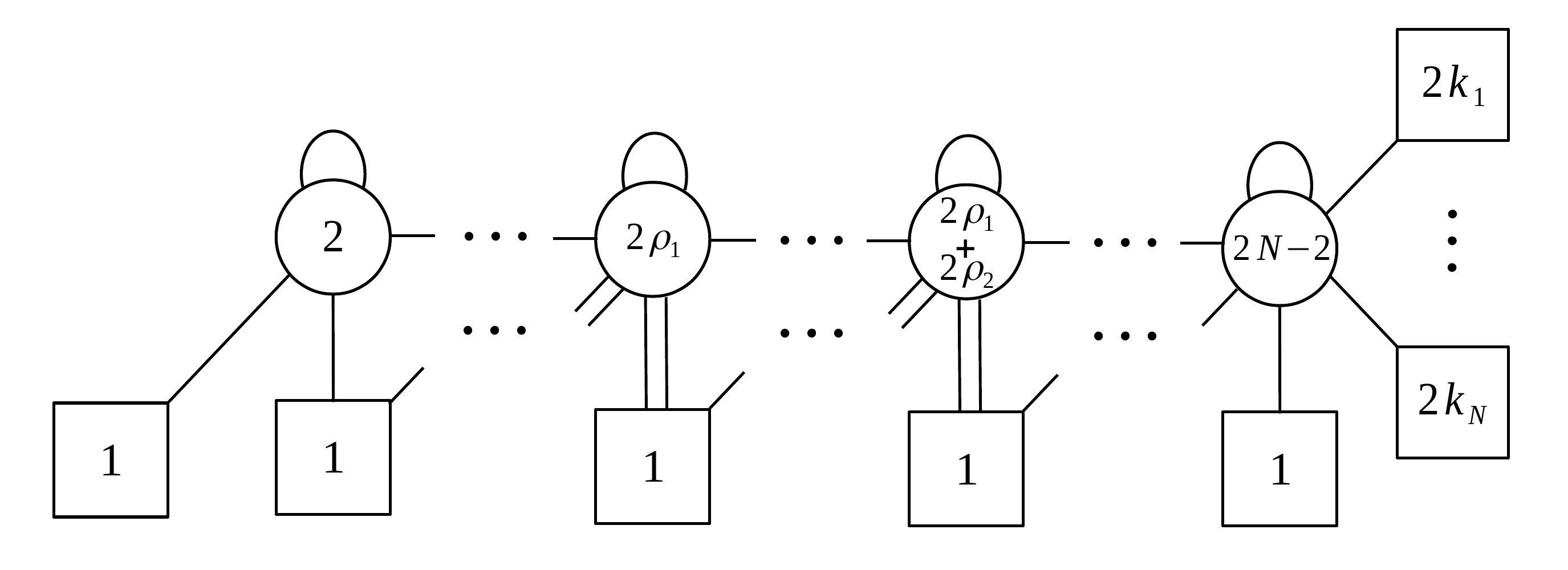}
\caption{\label{fig:T_A} The quiver diagram representation of the deformed theory $\mathsf T$. We have double lines in the saw only for the gauge nodes at positions $\rho_1, \, \rho_1+\rho_2, \, \dots, \, \sum_{i = 1}^{N-1} \rho_i$. The mirror-like dual theory, which is denoted by $\mathsf T^\vee$, has the same  diagram with $\rho$ and $\sigma$ exchanged.}
\end{figure}

The superpotential \eqref{eq:def} triggers a flow to a new theory $\mathsf T$.
Due to this superpotential term, the $USp(2 N)_x$ global symmetry of the original $E[USp(2 N)]$ theory is now broken to
\begin{align}
USp(2 N)_x \quad \longrightarrow \quad \prod_{m = 1}^N USp(2 k_m)_{x^{(m)}} \,.
\end{align}
Likewise, the $USp(2 N)_y$ global symmetry is also broken to
\begin{align}
USp(2 N)_y \quad \longrightarrow \quad \prod_{n = 1}^N USp(2 l_n)_{y^{(n)}} \,.
\end{align}
This IR symmetry  will become manifest in the mirror dual Lagrangian.
Correspondingly at the level of supersymmetric indices we will  introduce the following fugacities 
\be
&&x_i, \quad {\rm with} \quad i=1,\cdots,N \quad \to \quad  x^{(1)}_{i_1}, x^{(2)}_{i_2},\cdots \quad{\rm with } \quad i_m=1,\cdots,k_m\nn\\
&&y_i, \quad {\rm with} \quad i=1,\cdots,N \quad \to \quad  y^{(1)}_{i_1}, y^{(2)}_{i_2},\cdots \quad{\rm with } \quad i_n=1,\cdots,l_n\,.
\ee
We denote by $\Tr_{x^{(m)}}$ and $\Tr_{y^{(n)}}$ respectively the traces over $USp(2k_m)_{x^{(m)}}$ and $USp(2l_n)_{y^{(n)}}$ indices.

Moreover, the mass terms in \eqref{eq:def} make some of the chiral multiplets of $E[USp(2 N)]$ massive and being integrated out. First, let us look at the chirals in the saw. Due to the mass terms, only the followings among the original set of $D^{(n)}$ and $V^{(n)}$ remain massless:
\begin{align}
\begin{aligned}
\label{eq:massless dv}
& D^{(n)}_1 \,, \qquad n = \rho_1, \, \rho_1+\rho_2, \, \dots, \, N \,, \\
& D^{(n)}_2 \,, \qquad n = 1, \, \dots, \, N \,, \\
& V^{(n)}_1 \,, \qquad n = 1, \, \dots, \, N \,, \\
& V^{(n)}_2 \,, \qquad n = \rho_1, \, \rho_1+\rho_2, \, \dots, \, \sum_{i = 1}^{L-1} \rho_i \,.
\end{aligned}
\end{align}
Second, in $E[USp(2 N)]$ there are $2 N$ fundamental chirals $Q^{(N-1,N)}$ attached to the last gauge node. Again due to the mass terms in \eqref{eq:def}, only $2 K$ of them remain massless. We rename as $Q_m, \tilde Q_m$ the massless chirals at the $USp(2(N-1))$ gauge node in the fundamental  representation of each $USp(2k_m)$ factor, with $m=1,\cdots,N$. In particular, for the values of $m$ for which $k_m=0$ we don't have any chiral field.
Their interaction with the antysymmetric $A^{(N-1)}$ is:
\begin{align}
\mathrm{Tr}_{N-1} \left[A^{(N-1)} \mathsf{H}\right] \quad \longrightarrow \quad \sum_{m = 1}^N \mathrm{Tr}_{N-1} \left[\left(A^{(N-1)}\right)^m \Tr_{x^{(m)}}Q_mQ_m\right] \,. \label{eq:W_f} 
\end{align}
The quiver diagram of $\mathsf T$  is drawn in Figure \ref{fig:T_A}.

\begin{figure}[tbp]
\centering
\includegraphics[scale=0.6]{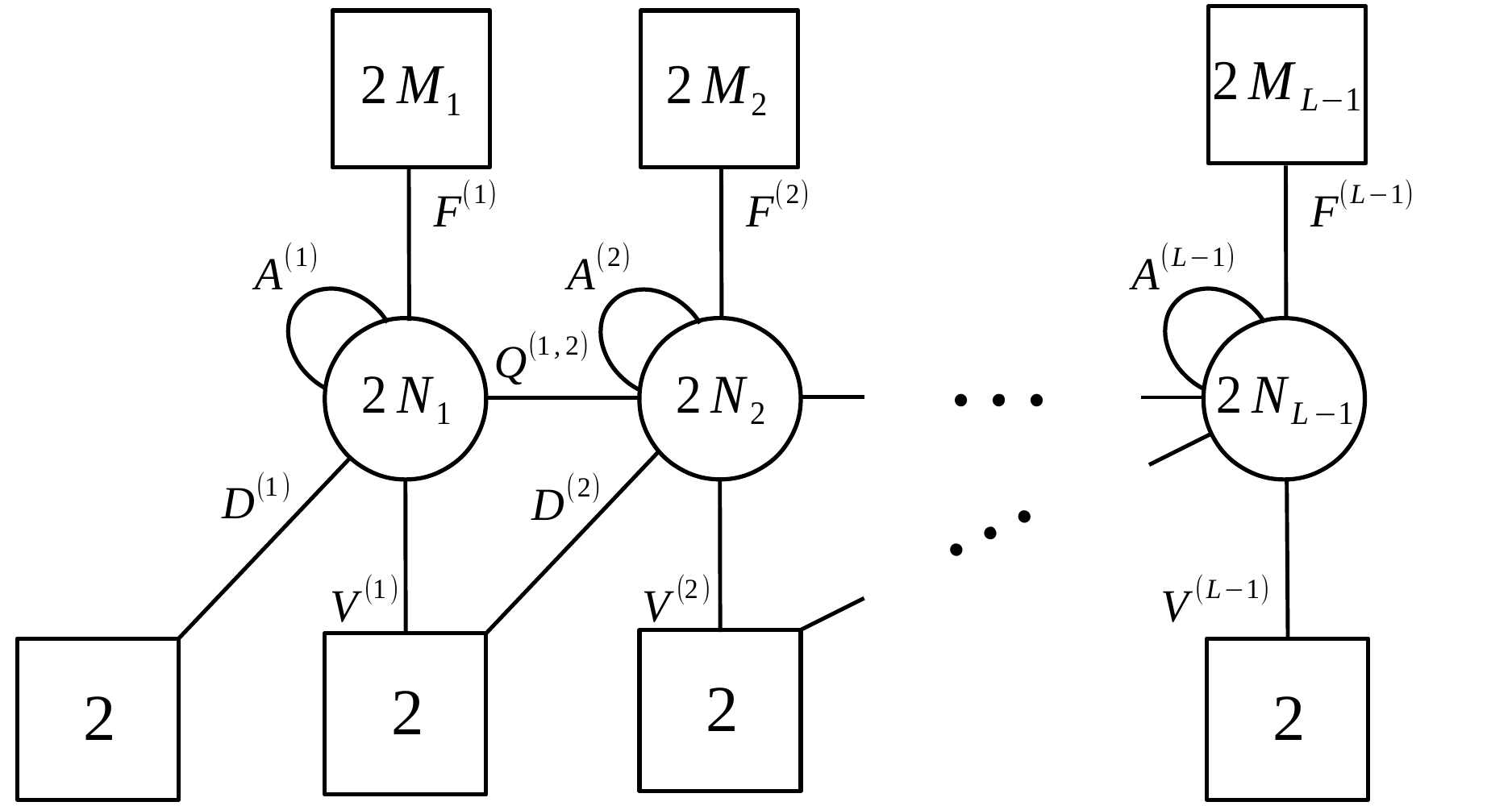}
\caption{The  $E_\rho^\sigma[USp(2 N)]$ quiver diagram. To avoid cluttering the drawing the gauge singlet $\gc_{nj}$ and $\pi^{(i,j)}$ are not shown in this diagram.}
\label{fig:T_C}
\end{figure}

At this point we can go from $\mathsf{T}$ to $E_\rho^\sigma[USp(2 N)]$ by iteratively applying Intriligator--Pouliot dualities
to move across the flip-flip dual frame.
The quiver diagram representation of the $E_\rho^\sigma[USp(2 N)]$ theory is shown in Figure \ref{fig:T_C}. 
There are also two sets of gauge singlets: the chirals $\gc_{nj}$ which are also singlets under the non-abelian global symmetries and  the chirals $\pi^{(i,j)}$  that transform non-trivially under the non-abelian symmetries. 
To avoid cluttering  Figure  \ref{fig:T_C}  we did not draw the gauge singlets (but we will do so in the examples we will present).
The flavor nodes in the top line and the gauge nodes in the middle line are $USp(2n)$ groups with ranks determined by the partitions $\gr$ and $\gs$ as for $T^\gs_\gr[SU(N)]$:
\begin{align}
\begin{aligned}
\label{eq:ranks}
M_{L-i} &= k_i \,, \\
N_{L-i} &= \sum_{j = i+1}^L \rho_j-\sum_{j = i+1}^N (j-i) k_j \,.
\end{aligned}
\end{align}
The $E_\rho^\sigma[USp(2 N)]$ superpotential is given by:
\begin{equation}
\makebox[\linewidth][c]{\scalebox{1}{$
\begin{split}
\mathcal W_{E_\rho^\sigma[USp(2 N)]} &= \sum_{n=1}^{L-1}\Tr_{n}\left[A^{(n)}\left(\Tr_{n+1}\mathbb{Q}^{(n,n+1)}-\Tr_{n-1}\mathbb{Q}^{(n-1,n)}+\Tr_{x^{(n)}}F^{(n)}F^{(n)}\right)\right]\\
&+\sum_{n=1}^{L-2}\Tr_{n}\Tr_{n+1}\left[V_{[1}^{(n)}Q^{(n,n+1)}D_{2]}^{(n+1)}\right]+\sum_{n=1}^{L-1}\sum_{j=1}^{N_n-N_{n-1}}\gc_{nj}\Tr_n\left[\left(A^{(n)}\right)^{j-1}D^{(n)}_{[1}D^{(n)}_{2]}\right]\\
&+\sum_{i=1}^{L-1}\sum_{j=i+1}^L\left(\prod_{k=1}^{j-1}\Tr_k\right)\Tr_{x^{(i)}}\left[F^{(i)} \left(\prod_{l=i}^{j-2}Q^{(l,l+1)}\right) V^{(j-1)}_{[1} \pi^{(i,j)}_{2]}\right]\, ,\\
\label{superpoterhosigma}
\end{split}$}}
\end{equation}
where we defined $N_0= 0$. We also recall that the $\Tr_n$ traces are taken over the $n$-th gauge node.
 Notice the interaction terms for the gauge singlets. In particular, the singlets $\gc_{nj}$ couple to the $n$-th diagonal meson dressed by the $(j-1)$-th power of the antisymmetric chiral $A^{(n)}$, with  $j=1,\cdots, N_n- N_{n-1}$. This means that the maximum power of the dressing is given by how much the rank of the
$n$-th gauge group jumps when compared to the $(n-1)$-th one. Moreover, we have singlets $\pi^{(i,j)}$  connecting the $i$-th  $USp(2M_i)$ flavor node to all the  $j$-th $SU(2)$ nodes of the saw sitting to its right, that is $j=i+1,\cdots, L$.
The $\pi^{(i,j)}$ singlets play a key role in  the enhancement of the nonabelian global symmetry since they enter the 
superpotential by flipping  gauge invariant operators which do not respect the enhanced symmetry.

The IR non-anomalous global symmetry of $E_\rho^\sigma[USp(2 N)]$ is
\be
\prod_{m = 1}^N USp(2 k_m)_{x^{(m)}} \times \prod_{n = 1}^N USp(2 l_n)_{y^{(n)}}\times U(1)_c\times U(1)_t\,.
\ee
Indeed, one can verify that the constraints coming from the superpotential \eqref{superpoterhosigma} and from the requirement that the NSVZ beta-functions vanish at each gauge node fix all the R-charges of the chiral fields up to two parameters, which correspond to the mixing coefficients $\mathfrak{c}$ and $\mathfrak{t}$ with $U(1)_c$ and $U(1)_t$.
For what concerns the non-abelian part, the global symmetry $USp(2 N)_x \times USp(2 N)_y$ of the original $E[USp(2 N)]$ theory is broken to
\begin{align}
USp(2 N)_x \times USp(2 N)_y \quad \longrightarrow \quad \prod_{m = 1}^N USp(2 k_m)_{x^{(m)}} \times \prod_{n = 1}^N USp(2 l_n)_{y^{(n)}}
\end{align}
where, like the original $E[USp(2 N)]$ theory, only $USp(2)^{l_n} \subset USp(2 l_n)_{y^{(n)}}$ is manifest in the quiver gauge theory description. 

Let's now  consider the mirror dual frame where,  because of the operator map \eqref{opmap4dmirror}, the deformation superpotential \eqref{eq:def} becomes
\begin{align}
\label{eq:mirror def}
\delta \mathcal W^\vee &=  \mathrm{Tr}_x \left[(J_\sigma+\mathsf S_\sigma) \cdot \mathsf C^\vee\right] + \mathrm{Tr}_y \left[(J_\rho+\mathsf T_\rho) \cdot \mathsf H^\vee\right]+
\sum_{\{(i,j) \neq (1,1)|1 \leq i \leq \sigma_j, \, 1 \leq j \leq \rho_i\}}
\mathsf{O}_{B}^{ij} \, B_{ji}^\vee\,.
\end{align}
This deformation triggers a flow from  \mirroreusp to  $\mathsf T^\vee$ which contains gauge singlets $\mathsf S_\sigma, \, \mathsf T_\rho$ and $\mathsf{O}_B$, which are mapped to the same gauge singlets in $\mathsf T$.

Next we take the flip-flip duality on $\mathsf T^\vee$. This leads to the mirror dual of $E_\rho^\sigma[USp(2 N)]$, denoted by $E_\sigma^\rho[USp(2 N)]$. Indeed, $E_\rho^\sigma[USp(2 N)]$ and $E_\sigma^\rho[USp(2 N)]$ have the same global symmetry as well as the same operator spectrum. In the following we will illustrate this construction in  various examples.
\\

\subsubsection{$\rho = [N]$ and $\sigma = [1^N]$}

\subsubsection*{Flow to $E_{[N]}[USp(2 N)]$}

In this case, the superpotential deformation triggering the flow to theory $\mathsf T$  is given by
\begin{align}
\delta \mathcal W &= \mathrm{Tr}_x \left[\mathsf S_{[1^N]} \cdot \mathsf H\right] + \mathrm{Tr}_y \left[\mathsf T_{[N]} \cdot \mathsf C\right] + \sum_{n = 1}^{N-1} \mathrm{Tr}_n \left[D^{(n)}_1 V^{(n)}_2\right] +\sum_{n = 2}^{N}  \mathsf{O}_{B}^{1n} \, \gb_{N-n+1}
\label{maxdef}
\end{align}
where $\mathsf S_{[1^N]}$ is an arbitrary $2 N \times 2 N$ antisymmetric matrix and $\mathsf T_{[N]}$ is determined requiring that it is traceless antisymmetric and that it commutes with $\mathsf{J}_{[N]}$
\begin{align}
\mathsf T_{[N]} =
\left(\begin{array}{cccc}
0 & -\mathsf T^{(2)}{}^T & \cdots & -\mathsf T^{(N)}{}^T \\
\mathsf T^{(2)} & 0 & \cdots & -\mathsf T^{(N-1)}{}^T \\
\vdots & \ddots & \ddots & \vdots \\
\mathsf T^{(N)} & \cdots &\mathsf T^{(2)} & 0 \\
\end{array}\right)
\end{align}
where each $\mathsf T^{(n)}$ is a $2 \times 2$ matrix with a single non-zero element:
\begin{align}
\mathsf T^{(n)} = \left(\begin{array}{ll}
0 & 0 \\
\mathsf t^{(n)} & 0 \\
\end{array}\right) \,.
\end{align}
Note that the flavor indices $1, \, 2$ of $D^{(n)}_1$ and $V^{(n)}_2$ do not belong to the same $SU(2)$; $D^{(n)}_1$ is charged under the $n$-th $SU(2)$ in the saw while $V^{(n)}_2$ is charged under the $(n+1)$-th $SU(2)$. It turns out that this deformation breaks the $USp(2 N)_y$ symmetry of the original $E[USp(2 N)]$ to $SU(2)_y$. The deformation also makes $D^{(n)}_1$ and $V^{(n)}_2$ massive for $n = 1, \dots, N-1$.

We obtain theory $\mathsf T$  by integrating out those massive fields. In theory $\mathsf T$ each gauge node except the last one now has only two fundamental chirals while the last gauge node has $2 N+2$ fundamental chirals in addition to the bifundamental and antisymmetric chirals which remain the same.\\

Now to reach  $E_{[N]}[USp(2 N)]$ we need to implement the flip-flip duality by  sequentially applying the Intriligator--Pouliot duality on each gauge node starting  from the left. The first gauge node is $USp(2)$ with a total of 6 fundamental chirals, the antisymmetric is a gauge singlet so we can apply directly the Intriligator--Pouliot duality. 
As the $USp(2)$ theory with 6 chirals is dual to a Wess-Zumino model with 15 chirals, the leftmost gauge node is confined once the duality is applied. Some of the 15 chirals make massive the traceless part of antisymmetric of the next $USp(4)$ gauge node, while the others partially cancel with the singlets $\mathsf{S}_{[1^N]}$, $\mathsf{T}_{[N]}$ and $\mathsf{O}_B^{1n}$. Now the $USp(4)$ node has 8 chirals and is also confined when we apply the Intriligator--Pouliot duality. Proceeding to the right we see that the entire chain of gauge nodes is sequentially confined leaving a set of chirals at the end. So the $E_{[N]}[USp(2 N)]$ theory will be a Wess-Zumino model. 
\\

This procedure of applying sequential Intriligator--Pouliot dualities can be realized at the level of the index. The mass deformation
$\sum_{n = 1}^{N-1} \mathrm{Tr}_n \left[D^{(n)}_1 V^{(n)}_2\right]$ in \eqref{maxdef} imposes the constraints on the fugacities of the saw
\be
\frac{y_{n+1}}{y_n}=t,\qquad n=1,\cdots,N-1\,,
\ee
which can be solved with
\be
y_n=t^{n-1}a,\qquad n=1,\cdots,N\,.
\ee
For our purpose, it is convenient to use $y = a t^{\frac{N-1}{2}} = y_n t^{\frac{N-2 n+1}{2}}$, which makes the unbroken $SU(2)_y \subset USp(2 N)_y$ manifest. The extra chirals we introduce give rise to the following index contributions:
\begin{align}
\mathsf S_{[1^N]} \quad &\longrightarrow \quad \Gpq{p q t^{-1}}^{N-1} \prod_{n < m}^N \Gpq{p q t^{-1} x^{\pm1}_n x^{\pm1}_m} \,, \\
\mathsf T_{[N]} \quad &\longrightarrow \quad \prod_{i = 2}^{N} \Gpq{t^i} \,, \\
 \mathsf{O}_{B}^{1n} \quad &\longrightarrow \quad \Gpq{t^{1-n} c^2} \,.
\end{align}
Hence, the complete index of theory  $\mathsf{T}$ is given by
\begin{align}
\mathcal I_{\mathsf T}(\vec x;y;t,c) &= \Gpq{p q t^{-1}}^{N-1} \prod_{n < m}^N \Gpq{p q t^{-1} x^{\pm1}_n x^{\pm1}_m} \prod_{i = 2}^{N} \Gpq{t^i} \prod_{n = 2}^{N} \Gpq{t^{1-n} c^2} \nn \\
&\times \mathcal{I}_{E[USp(2N)]}\left(\vec x;t^{\frac{1-N}{2}} y,t^{\frac{3-N}{2}} y,\cdots,t^{\frac{N-1}{2}}y;t;c\right) \,.
\end{align}
The  sequential confinement of the tail then translates in the identity
\begin{align}
&\mathcal{I}_{E[USp(2N)]}\left(\vec x;t^{N-1}u,t^{N-2}u,\cdots,u;t,c\right) \nn \\
&= \Gpq{c^2} \Gpq{t}^{N}\prod_{n<m}^N\Gpq{t\,x_n^{\pm1}x_m^{\pm1}}\prod_{i=1}^N\frac{\Gpq{u\,c\,x_i^{\pm1}}\Gpq{\frac{c}{u\,t^{N-1}}x_i^{\pm1}}}{\Gpq{t^{1-i}c^2}\Gpq{t^i}}\, ,
\label{maxdefindex}
\end{align}
which was proven by Rains in Corollary 2.8 of \citep{2014arXiv1408.0305R}. Putting this back into $\mathcal I_{\mathsf T}$ with $u =t^{\frac{1-N}{2}}y$\footnote{Notice that to apply \eqref{maxdefindex} we need to use  the $USp(2N)_y$ Weyl symmetry of $E[USp(2N)]$ to  reorder the  fugacities.}, we obtain the identity
\begin{align}
\label{eq:confinement}
\mathcal I_{\mathsf T}(\vec x;y;t,c) = \prod_{n = 1}^N \Gpq{y^{\pm1} t^{-\frac{N-1}{2}} c x_n^{\pm1}} = \mathcal I_{E_{[N]}[USp(2 N)]}(\vec x;y;t,c) \,.
\end{align}
As expected, $E_{[N]}[USp(2 N)]$ is a Wess-Zumino model with $2 N$ chirals, which are bifundamental betweeen $USp(2 N)_x \times SU(2)_y$. One can see that the new fugacity $y$ makes the $SU(2)_y$ symmetry manifest.
\\

\subsubsection*{Flow to $E^{[N]}[USp(2 N)]$}

Now let us examine this confinement on the mirror side. The superpotential deformation triggering the flow to theory $\mathsf T^\vee$  is given by

\begin{align}
\label{eq:mirror def}
\delta \mathcal W^\vee &= \mathrm{Tr}_x \left[\mathsf S_{[1^N]} \cdot \mathsf C^\vee\right] + \mathrm{Tr}_y \left[\mathsf T_{[N]} \cdot \mathsf H^\vee\right]  + \sum_{n = 1}^{N-1} \mathrm{Tr}_{N-1} \left[q^{(N-1,N)}_{2 n-1} q^{(N-1,N)}_{2 n+2}\right] \nn \\
&+\sum_{n = 2}^{N}  \mathsf{O}_{B}^{1n} \, \mathrm{Tr}_{N-1} \left[\left(A^{(N-1)}\right)^{n-2} v_{[1}^{(N-1)} v_{2]}^{(N-1)}\right] \,,
\end{align}
which makes $q^{(N-1,N)}$ massive except $q^{(N-1,N)}_2$ and $q^{(N-1,N)}_{2 N-1}$. Integrating out the massive $q^{(N-1,N)}$, we reach theory $\mathsf T^\vee$, which is mirror-like dual to theory $\mathsf T$.

 $\mathsf T^\vee$ differs from \eusp only by the fact that there are only two chirals attached to the last gauge node.
Now to reach  $E^{[N]}[USp(2 N)]$ we can  implement the flip-flip duality by sequentially applying the Intriligator--Pouliot duality on each gauge node starting from the leftmost $USp(2)$ node and proceeding along the tail. Since the first $N-2$ nodes are  $USp(2n)$
with $4n+4$ chirals, their rank does not change when we apply Intriligator--Pouliot duality. However when we act one the last gauge node
which is $USp(2(N-1))$ with $2n+2$ chirals it confines.
At the second iteration we start again from the leftmost $USp(2)$ node but when we reach the $USp(2(N-2))$ node it confines.
In this way the quiver is confined from the right until we are left with  the same gauge singlets as in \eqref{eq:confinement}, that is we reach  the $E^{[N]}[USp(2 N)]$ WZ model.
\\

\subsubsection{$\rho = [N-1,1]$ and $\sigma = [1^N]$}
\label{sec:next-to-maximal}

\subsubsection*{Flow to $E_{[N-1,1]}[USp(2 N)]$}

The deformation leading to theory $\mathsf{T}$ is given by:
\begin{align}
\delta \mathcal W &=  \mathrm{Tr}_x \left[\mathsf S_{[1^N]} \cdot \mathsf H\right] + \mathrm{Tr}_y \left[\mathsf T_{[N-1,1]} \cdot \mathsf C\right] + \sum_{n = 1}^{N-2} \mathrm{Tr}_n \left[D^{(n)}_1 V^{(n)}_2\right] +\sum_{n = 2}^{N-1}  \mathsf{O}_{B}^{1n} \, \gb_{N-n+1}
\label{eq:def_A}
\end{align}
where $\mathsf S_{[1^N]}$ is again an arbitrary $2 N \times 2 N$ skew-symmetric matrix and $\mathsf T_{[N-1,1]}$ is given by
\begin{align}
\mathsf{T}_{[N-1,1]}=\left(\begin{array}{lc}
\begin{array}{cccc}
\mathsf T^{(1)}_{11} & -\mathsf T^{(2)}_{11}{}^T & \cdots & -\mathsf T^{(N-1)}_{11}{}^T \\
\mathsf T^{(2)}_{11} & \mathsf T^{(1)}_{11} & \cdots & -\mathsf T^{(N-2)}_{11}{}^T \\
\vdots & \ddots & \ddots & \vdots \\
\mathsf T^{(N-1)}_{11} & \cdots &\mathsf T^{(2)}_{11} & \mathsf T^{(1)}_{11} \\
\end{array}
&
\begin{array}{c}
-\mathsf T^{(1)}_{N1}{}^T \\
\vdots \\
0 \\
\mathsf T^{(1)}_{1N} \\
\end{array}
\\
\begin{array}{cccc}
\,\,\,\, \mathsf T^{(1)}_{N1} & \qquad 0 & \,\,\,\,\, \cdots  & \quad -\mathsf T^{(1)}_{1N}{}^T \\
\end{array}
&
\, -(N-1) \mathsf T^{(1)}_{11}
\\
\end{array}\right)
\end{align}
where each $T^{(n)}_{ij}$ is a $2 \times 2$ matrix of the form:
\begin{align}
\mathsf T^{(1)}_{11} &=
\left(\begin{array}{ll}
0 & -\mathsf t^{(1)}_{11} \\
\mathsf t^{(1)}_{11} & 0 \\
\end{array}\right) \,, \\
\mathsf T^{(n)}_{ij} &=\left\{\begin{array}{ll}
 \left(\begin{array}{ll}
0 & 0 \\
\mathsf t^{(n)}_{ii} & 0 \\
\end{array}\right) \,, & \qquad i = j, \, n \neq 1 \,, \\
\\
\left(\begin{array}{ll}
\mathsf{r}^{(n)}_{ij} & 0 \\
\mathsf{s}^{(n)}_{ij} & 0 \\
\end{array}\right) \,, & \qquad i > j, \, n \neq 1 \,,\\
\\
\left(\begin{array}{ll}
0 & 0 \\
\mathsf{u}^{(n)}_{ij} & \mathsf{w}^{(n)}_{ij} \\
\end{array}\right) \,, & \qquad i < j, \, n \neq 1 \,.
\end{array}\right.
\end{align}

One can write down the superconformal index of theory $\mathsf{T}$ by constraining fugacities of the index of $E[USp(2 N)]$. 
The deformation \eqref{eq:def_A} demands the following conditions on the $USp(2 N)_y$ fugacities:
\begin{align}
\label{eq:fugacity conditions}
\frac{y_{n+1}}{y_n} = t \,, \qquad n = 1, \dots, N-2 \,,
\end{align}
which are satisfied by
\begin{align}
y_n = t^{n-1} a \,, \qquad n = 1, \dots, N-1 \,.
\label{eq:fugacity redef}
\end{align}
For later convenience, we introduce the new fugacities
\begin{align}
y_n &= t^{n-\frac{N}{2}} y^{(1)} \,, \qquad n = 1,\dots,N-1 \,, \nn \\
y_N &= y^{(2)}\,,
\end{align}
which will make the unbroken $USp(2)_{y^{(1)}} \times USp(2)_{y^{(2)}} \subset USp(2 N)_y$ manifest in the index.
The extra chiral singlets we introduce then give rise to the following index contributions:
\begin{align}
\begin{aligned}
\label{eq:singlets}
\mathsf S_{[1^N]} \quad &\longrightarrow \quad \Gpq{p q t^{-1}}^{N-1} \prod_{n < m}^N \Gpq{p q t^{-1} x^{\pm1}_n x^{\pm1}_m} \,, \\
\mathsf T_{[N-1,1]} \quad &\longrightarrow \quad \Gpq{t^\frac{N}{2} y^{(1)\pm1} y^{(2)\pm1}} \prod_{i = 1}^{N-1} \Gpq{t^i} \,, \\
\mathsf{O}_{B}^{1n} \quad &\longrightarrow \quad  \Gpq{t^{1-n} c^2} \,.
\end{aligned}
\end{align}
Substituting them into the recursive definition of the index of the $E[USp(2 N)]$ theory, we obtain the index of theory $\mathsf{T}$ as follows:
\begin{align}
&\mathcal I_{\mathsf T} \left(\vec x;y^{(1)},y^{(2)};t,c\right) \nn \\
&= \Gpq{p q t^{-1}}^{N-1} \prod_{n < m}^N \Gpq{p q t^{-1} x^{\pm1}_n x^{\pm1}_m} \Gpq{t^\frac{N}{2} y^{(1)\pm1} y^{(2)\pm1}} \prod_{i = 1}^{N-1} \Gpq{t^i} \nn \\
&\quad \times \prod_{n = 2}^{N-1} \Gpq{t^{1-n} c^2} \mathcal{I}_{E[USp(2N)]}\left(\vec x;t^{-\frac{N}{2}+1} y^{(1)},t^{-\frac{N}{2}+2}y^{(1)},\cdots,t^{\frac{N}{2}-1} y^{(1)},y^{(2)};t;c\right) \nn \\
&= \Gpq{p q t^{-1}}^{N-1} \prod_{n < m}^N \Gpq{p q t^{-1} x^{\pm1}_n x^{\pm1}_m} \Gpq{t^\frac{N}{2} y^{(1)\pm1} y^{(2)\pm1}} \prod_{i = 1}^{N-1} \Gpq{t^i} \nn \\
&\quad \times\prod_{n = 2}^{N-1} \Gpq{t^{1-n} c^2} \frac{\prod_{n=1}^N\Gpq{c\,y^{(2)\pm1}x_n^{\pm1}}}{\Gpq{t^{-1} c^2}}\oint\udl{\vec{z}^{(N-1)}_{N-1}} \Gpq{p q t^{-1}}^{N-1} \nn \\
&\quad \times \prod_{i < j}^{N-1} \Gpq{p q t^{-1} z^{(N-1)}_i{}^{\pm1} z^{(N-1)}_j{}^{\pm1}} \prod_{i=1}^{N-1}\frac{\prod_{n=1}^N\Gpq{t^{1/2}z^{(N-1)}_i{}^{\pm1}x_n^{\pm1}}}{\Gpq{t^{1/2}c\,y^{(2)\pm1}z^{(N-1)}_i{}^{\pm1}}} \nn\\
&\quad \times\mathcal{I}_{E[USp(2(N-1))]}\left(z^{(N-1)}_1,\cdots,z^{(N-1)}_{N-1};t^{-\frac{N}{2}+1} y^{(1)},t^{-\frac{N}{2}+2}y^{(1)},\cdots,t^{\frac{N}{2}-1} y^{(1)};t;t^{-1/2}c\right)\,.
\end{align}
At this stage, one can see that there is an $SU(2)$ symmetry for $y^{(2)}$ while it is not clear whether or not we have an enhanced $SU(2)$ symmetry for $y^{(1)}$.

To reach $E_{[N-1,1]}[USp(2 N)]$ we need to implement the flip-flip duality by applying iteratively the IP duality. We can recycle
some of the previous calculations noting that  the last factor of the integrand is the index of  $E[USp(2 N-2)]$
 with the specialisation of parameters leading to the evaluation formula  \eqref{maxdefindex} as we discussed in the previous subsection. 
Taking this into account, we obtain
\begin{align}
\label{eq:T_A^dual}
\mathcal{I}_{\mathsf{T}}&= \Gpq{p q t^{-1}}^{N-1} \prod_{n < m}^N \Gpq{p q t^{-1} x^{\pm1}_n x^{\pm1}_m} \Gpq{t^\frac{N}{2} y^{(1)\pm1} y^{(2)\pm1}} \nn\\
&\quad \times \frac{\prod_{n=1}^N \Gpq{c\,y^{(2)\pm1}x_n^{\pm1}}}{\Gpq{t^{-N+1} c^2}} \oint\udl{\vec{z}^{(N-1)}_{N-1}} \prod_{i=1}^{N-1}\Gpq{t^{-\frac{N-1}{2}} c y^{(1)\pm1} z^{(N-1)}_i{}^{\pm1}} \nn\\
&\quad \times \Gpq{pq t^{-1/2}c^{-1}y^{(2)\pm1} z^{(N-1)}_i{}^{\pm1}}\prod_{n=1}^N\Gpq{t^{1/2}z^{(N-1)}_i{}^{\pm1}x_n^{\pm1}}\,,
\end{align}
where the $SU(2)$ symmetry for $y^{(1)}$ is also manifest.

This  is the index of a $USp(2(N-1))$ theory with $2N+4$ favors and various flipping fields. To complete the derivation of the flip-flip duality we need to apply Intriligator--Pouliot  duality one more time and we obtain:
\begin{align}
\label{eq:T_C}
\mathcal{I}_{\mathsf{T}} &= \frac{\prod_{n=1}^N\Gpq{t^{-\frac{N}{2}+1} c y^{(1)\pm1} x_n^{\pm1}}}{\Gpq{p^{-1} q^{-1} t c^2}} \oint\udl{\vec z^{(1)}_{1}}\Gpq{t} \Gpq{p^{1/2} q^{1/2} t^{\frac{N-1}{2}} c^{-1} y^{(1)\pm1} z^{(1)}{}^{\pm1}} \nn\\
&\quad \times \Gpq{p^{-1/2} q^{-1/2} t^{1/2} c y^{(2)\pm1} z^{(1)}{}^{\pm1}}\prod_{n=1}^N\Gpq{p^{1/2} q^{1/2} t^{-1/2} x_n^{\pm1}z^{(1)}{}^{\pm1}} \nn \\
&= \mathcal I_{E_{[N-1,1]}[USp(2 N)]}\left(\vec x;y^{(2)},y^{(1)};t,c\right)
\end{align}
where
\begin{align}
\udl{\vec z^{(1)}_1} = \frac{(p;p) (q;q)}{2} \frac{\udl{z^{(1)}}}{2 \pi i z^{(1)}} \frac{1}{\Gpq{z^{(1)}{}^{\pm2}}} \,.
\end{align}

\begin{figure}[tbp]
\centering
\includegraphics[scale=0.6]{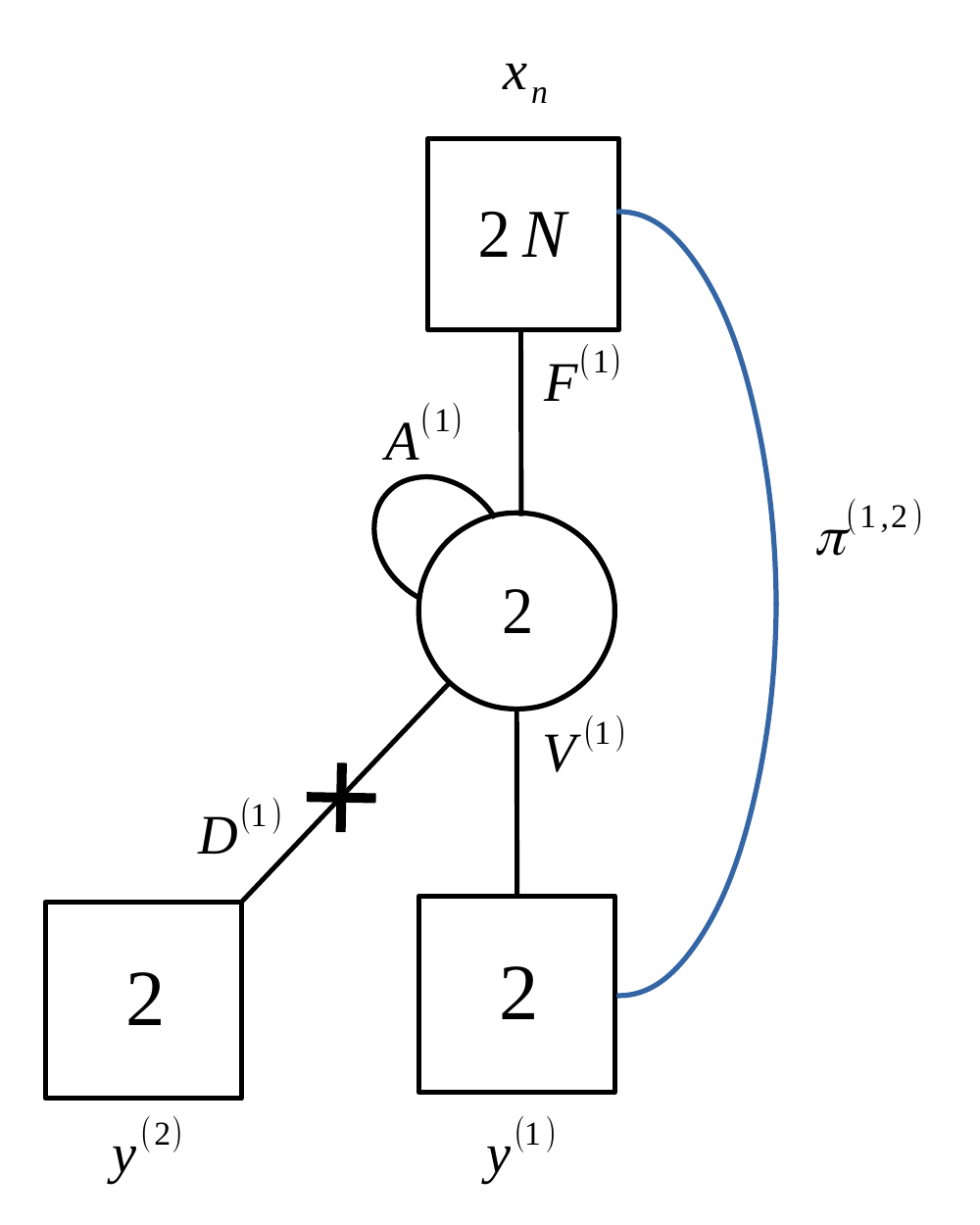}
\caption{\label{fig:EN-11} The quiver diagram representation of $E_{[N-1,1]}[USp(2N)]$.}
\end{figure}

The $E_{[N-1,1]}[USp(2 N)]$ theory is a $USp(2)$ theory with $2 N+4$ fundamental chirals and some additional singlets, which are shown in Figure \ref{fig:EN-11}.\footnote{Note that as, a consequence of the sequential application of the Intriligator--Pouliot duality, the fugacities are permuted and the two nodes in the saw are labeled by $y^{(2)}$ and $y^{(1)}$ respectively, from the left, which is  the opposite labelling compared to the definition of the original $E[USp(2 N)]$ index. For this reason, we call the index \eqref{eq:T_C} as  $\mathcal I_{E_{[N-1,1]}[USp(2 N)]}(\vec x;y^{(2)},y^{(1)};t,c)$ instead of $\mathcal I_{E_{[N-1,1]}[USp(2 N)]}(\vec x;y^{(1)},y^{(2)};t,c)$. Indeed we can't use the  $USp(2N)$ Weyl symmetry to reorder the two set of fugacities $y^{(1)}$ and $y^{(2)}$.}
From the index \eqref{eq:T_C}, one can read off the charges of each chiral multiplet and the available superpotential. For example, one can see that there is a singlet $\gc_{11}$, whose index contribution is $\Gpq{p^{-1} q^{-1} t c^2}^{-1}$, flipping the diagonal meson $\mathrm{Tr}_1 D^{(1)}_{[1} D^{(1)}_{2]}$ where $D^{(1)}$ contributes to the index by $\Gpq{p^{-1/2} q^{-1/2} t^{1/2} c y_2^{\pm1} z^{(1)}{}^{\pm1}}$.
The total superpotential of $E_{[N-1,1]}[USp(2 N)]$ is given by 
\begin{align}
\mathcal W_{E_{[N-1,1]}[USp(2 N)]} = \mathrm{Tr}_1 \mathrm{Tr}_{x} \left[A^{(1)} F^{(1)} F^{(1)}\right] + \mathrm{Tr}_1  \mathrm{Tr}_x \left[F^{(1)} V^{(1)}_{[1} \pi^{(1,2)}_{2]}\right] + \gc_{11} \mathrm{Tr}_2  \left[D^{(1)}_{[1} D^{(1)}_{2]}\right] \,.
\end{align}

We can work out some interesting gauge invariant operators:
\begin{align}
\begin{aligned}
\Pi^{(1)} &= \pi^{(1,2)} \,, \\
\Pi^{(2)} &= \mathrm{Tr}_1 \left[D^{(1)} F^{(1)}\right] \,, \\
\mathsf C^{(1)} &= \mathrm{Tr}_1 A^{(1)} \,, \\
\mathsf C^{(1,2)} &= \mathrm{Tr}_1 \left[D^{(1)} V^{(1)}\right] \,, \\
\mathsf H &= \mathrm{Tr}_1 \left[F^{(1)} F^{(1)}\right] \,. \\
\end{aligned}
\end{align}
Recall that the global symmetry of $E_{[N-1,1]}[USp(2 N)]$ includes $USp(2 N)_x \times USp(2)_{y^{(1)}} \times USp(2)_{y^{(2)}}$ rather than $USp(2 N)_x \times USp(4)_y$ unless $N = 2$. 
Indeed, we find that the would-be antisymmetric operator of $USp(4)_y$ is decomposed into one singlet operator and one bifundamental operator between $USp(2)_{y^{(1)}} \times USp(2)_{y^{(2)}}$, which are denoted by $\mathsf C^{(1)}$ and $\mathsf C^{(1,2)}$ respectively. Also each $\Pi^{(i)}$ is a bifundamental operator between $USp(2N)_x \times USp(2)_{y^{(i)}}$. As expected, $\mathsf C^{(1)}$ and $\mathsf C^{(1,2)}$ have different $U(1)$ global charges, and so do $\Pi^{(1)}$ and $\Pi^{(2)}$. Thus, only $USp(2)_{y^{(1)}} \times USp(2)_{y^{(2)}} \subset USp(4)_y$ is preserved. On the other hand, $\mathsf H$ is an antisymmetric operator respecting the entire $USp(2 N)_x$ symmetry.

Notice that $E_{[N-1,1]}[USp(2 N)]$ is asymptotically free only when $N < 4$. Among these three cases, $N = 1$ is the confining case while $N = 2$ is the self-dual case of Iintriligator--Pouliot duality. In the subsequent subsections, thus, we will mostly focus on the $N = 3$ case although the mathematical identities of the superconformal indices hold beyond $N = 3$.
\\

\subsubsection*{Flow to $E^{[N-1,1]}[USp(2 N)]$}

Now let us consider the mass deformation in the mirror dual frame. In this dual frame, the superpotential deformation \eqref{eq:def_A} is mapped to
\begin{align}
\label{eq:def_B}
\delta \mathcal W^\vee &=  \mathrm{Tr}_x \left[\mathsf S_{[1^N]} \cdot \mathsf C^\vee\right] + \mathrm{Tr}_y \left[\mathsf T_{[N-1,1]} \cdot \mathsf H^\vee\right] + \sum_{n = 1}^{N-2}\mathrm{Tr}_{N-1} q^{(N-1,N)}_{2 n-1} q^{(N-1,N)}_{2 n+2} \nn \\
&+\sum_{n = 2}^{N-1} \mathsf{O}_{B}^{1n} \, \mathrm{Tr}_{N-1} \left[\left(A^{(N-1)}\right)^{n-2} v_{[1}^{(N-1)} v_{2]}^{(N-1)}\right] \,,
\end{align}
which makes $q^{(N-1,N)}_n$ massive except $n = 2, \, 2 N-3, \, 2 N-1, \, 2 N$. The extra singlets we introduce are denoted by the same letters as in the original side.

The superconformal index of theory $\mathsf T^\vee$ can be obtained from that of $E[USp(2 N)]^\vee$ taking into account the extra singlet contributions \eqref{eq:singlets} and  by imposing the fugacity conditions \eqref{eq:fugacity conditions}-\eqref{eq:fugacity redef}:\begin{align}
\label{eq:T_B}
&\mathcal{I}_{\mathsf T^\vee}\left(\vec x;y^{(1)},y^{(2)};t,c\right)= \nn \\
&= \Gpq{p q t^{-1}}^{N-1} \prod_{n < m}^N \Gpq{p q t^{-1} x^{\pm1}_n x^{\pm1}_m} \Gpq{t^\frac{N}{2} y^{(1)\pm1} y^{(2)\pm1}} \prod_{i = 1}^{N-1} \Gpq{t^i} \prod_{n = 2}^{N-1} \Gpq{t^{1-n} c^2} \nn \\
&\quad \times \mathcal{I}_{E[USp(2N)]^\vee}\left(\vec x;t^{-\frac{N}{2}+1} y^{(1)},t^{-\frac{N}{2}+2}y^{(1)},\cdots,t^{\frac{N}{2}-1} y^{(1)},y^{(2)};t;c\right) \nn \\
&= \Gpq{p q t^{-1}}^{N-1} \prod_{n < m}^N \Gpq{p q t^{-1} x^{\pm1}_n x^{\pm1}_m} \Gpq{t^\frac{N}{2} y^{(1)\pm1} y^{(2)\pm1}} \prod_{i = 1}^{N-1} \Gpq{t^i} \prod_{n = 2}^{N-1} \Gpq{t^{1-n} c^2} \nn \\
&\quad \times \frac{\Gpq{c\,x_N^{\pm1} y^{(2)\pm1}} \prod_{n = 1}^{N-1} \Gpq{c\,x_N^{\pm1} \left(t^{n-\frac{N}{2}} y^{(1)}\right)^{\pm1}}}{\Gpq{p^{-1} q^{-1} t c^2}} \nn \\
&\quad \times \oint\udl{\vec{z}^{(N-1)}_{N-1}} \Gpq{t}^{N-1} \prod_{i < j}^{N-1} \Gpq{t z^{(N-1)}_i{}^{\pm1} z^{(N-1)}_j{}^{\pm1}} \nn\\
&\quad \times \frac{\prod_{i=1}^{N-1} \Gpq{p^{1/2} q^{1/2} t^{-1/2}z^{(N-1)}_i{}^{\pm1}y^{(2)\pm1}} \Gpq{p^{1/2} q^{1/2} t^{-\frac{N-1}{2}} z^{(N-1)}_i{}^{\pm1} y^{(1)\pm1}}}{\prod_{i=1}^{N-1}\Gpq{p^{1/2} q^{1/2} t^{-1/2}c\,x_N^{\pm1}z^{(N-1)}_i{}^{\pm1}}} \nn \\
&\quad \times \mathcal{I}_{E[USp(2(N-1))]}\left(z^{(N-1)}_1,\cdots,z^{(N-1)}_{N-1};x_1,\cdots,x_{N-1};p q/t,p^{-1/2} q^{-1/2} t^{1/2}c\right) \,.
\end{align}
We see that theory $\mathsf T^\vee$ is basically the same quiver theory as $E[USp(2N)]^\vee$ but there are only 4  fundamental chirals attached to the $(N-1)$-th gauge node on top of those of the saw.  Two of these 4 chirals couple to $A^{(N-1)}$, while the other two couple to $\left(A^{(N-1)}\right)^{N-1}$.

\begin{figure}[tbp]
\centering
\includegraphics[scale=0.6]{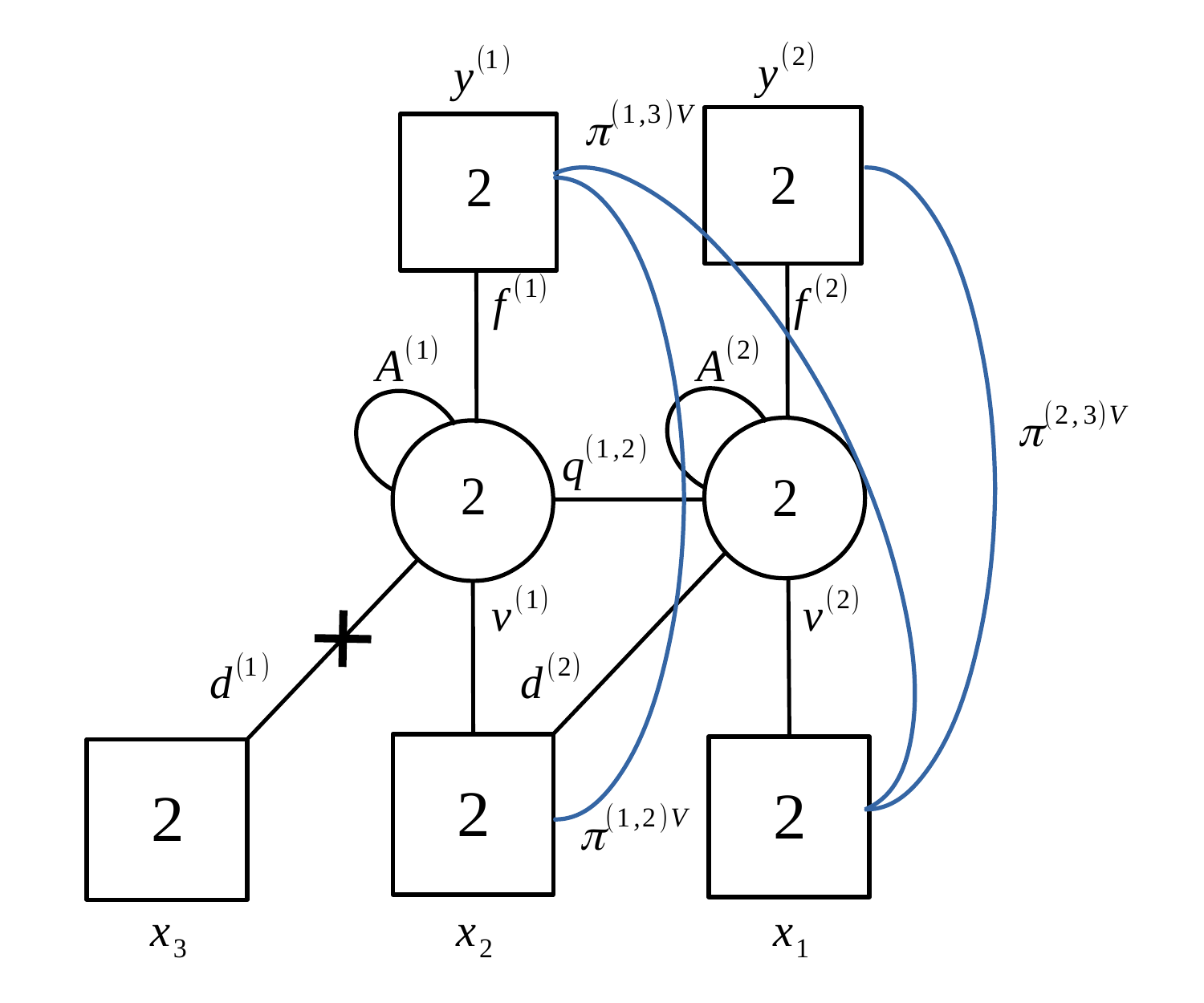}
\caption{\label{fig:T_D} The quiver diagram representation of $E^{[2,1]}[USp(6)]$. The fugacity corresponding to each gauge/flavor node is also indicated.}
\end{figure}

Now we need to implement the flip-flip duality as a chain of sequential Intriligator--Pouliot dualities. In Appendix \ref{app:next-to-maximal} we do this at the level of the superconformal index for the $N = 3$ case obtaining\footnote{{ Again, the labelling of the saw by the  $x_n$ fugaicties is in the opposite order compared to the original $E[USp(2 N)]^\vee$ index. This time, however, the permutations of $x_n$ belong to the Weyl symmetry of the $USp(6)_x$ global symmetry. Thus, the index is invariant under such permutations, so we just call the index $\mathcal{I}_{E^{[2,1]}[USp(6)]}(y^{(1)},y^{(2)};\vec{x};p q/t,c)$ without specifying a particular order of $x_n$.}}:
\begin{align}
\label{eq:T_D}
&\mathcal{I}_{\mathsf{T}^\vee}=\Gamma_e(t^{-1/2} c x_1^{\pm1} y^{(1)\pm1}) \Gamma_e(t^{-1/2} c x_2^{\pm1} y^{(1)\pm1}) \Gamma_e(c x_1^{\pm1} y^{(2)\pm1})  \Gamma_e(p q t^2 c^{-2}) \nonumber \\
&\times \oint \udl{\vec z^{(1)}_1} \udl{\vec z^{(2)}_1} \Gpq{p q t^{-1}}^{2} \Gamma_e(t^{1/2} z^{(1)}{}^{\pm1} y^{(1)\pm1})  \nonumber \\
&\times \Gamma_e(p q c^{-1} x_2^{\pm1} z^{(1)}{}^{\pm1}) \Gamma_e(t^{-1} c x_3^{\pm1} z^{(1)}{}^{\pm1}) \Gamma_e(t^{1/2} z^{(2)}{}^{\pm1} y^{(2)\pm1}) \nonumber \\
&\times \Gamma_e(p q t^{-1/2} c^{-1} x_1^{\pm1} z^{(2)}{}^{\pm1}) \Gamma_e(t^{-1/2} c x_2^{\pm1} z^{(2)}{}^{\pm1}) \Gamma_e(t^{1/2} z^{(1)}{}^{\pm1} z^{(2)}{}^{\pm1}) \nonumber \\
&= \mathcal{I}_{E^{[2,1]}[USp(6)]}\left(y^{(1)},y^{(2)};\vec x;p q/t,c\right) \,.
\end{align}
One can read off the matter content of $E^{[2,1]}[USp(6)]$ from the index \eqref{eq:T_D}, which is shown in Figure \ref{fig:T_D}.
In particular, there is a single flipping field $\gc_{11}^\vee$, denoted by a cross in Figure \ref{fig:T_D}, which flips the diagonal meson $\mathrm{Tr}_1 d^{(1)}_{[1} d^{(1)}_{2]}$.
The total superpotential is given by 
\begin{align}
\label{eq:sup21}
&\mathcal W_{E^{[2,1]}[USp(6)]} \nn \\
&= \mathrm{Tr}_1 \left[A^{(1)} \left(\mathrm{Tr}_2 \, q^{(1,2)} q^{(1,2)}+\mathrm{Tr}_{y_1} f^{(1)} f^{(1)}\right)\right] + \mathrm{Tr}_2 \left[A^{(2)} \left(\mathrm{Tr}_1 \, q^{(1,2)} q^{(1,2)}+\mathrm{Tr}_{y_2} f^{(2)} f^{(2)}\right)\right] \nn \\
&\quad + \mathrm{Tr}_1 \mathrm{Tr}_2 \left[v^{(1)}_{[1} q^{(1,2)} d^{(2)}_{2]}\right] + \mathrm{Tr}_1 \mathrm{Tr}_{y_2} \left[f^{(1)} v^{(1)}_{[1} \pi^{(1,2)\vee}_{2]}\right]  + \mathrm{Tr}_1 \mathrm{Tr}_2 \mathrm{Tr}_{y_2} \left[f^{(1)} q^{(1,2)} v^{(2)}_{[1} \pi^{(1,3)\vee}_{2]}\right] \nn \\
&\quad + \mathrm{Tr}_2 \mathrm{Tr}_{y_2} \left[f^{(2)} v^{(2)}_{[1} \pi^{(2,3)\vee}_{2]}\right] + \gc_{11}^\vee \, \mathrm{Tr}_1 \left[d^{(1)}_{[1} d^{(1)}_{2]}\right] \,.
\end{align}

Some examples of gauge invariant operators are as follows:
\begin{align}
\begin{aligned}
\Pi^{(1)}{}^\vee &= \left(\pi^{(1,3)\vee}, \pi^{(1,2)\vee}, \mathrm{Tr}_1 \left[d^{(1)} f^{(1)}\right]\right) \,, \\
\Pi^{(2)}{}^\vee &= \left(\pi^{(2,3)\vee}, \mathrm{Tr}_2 \left[d^{(2)} f^{(2)}\right], \mathrm{Tr}_1 \mathrm{Tr}_2 \left[d^{(1)} q^{(1,2)} f^{(2)}\right]\right) \,, \\
\mathsf H^{(1)}{}^\vee &= \mathrm{Tr}_1 \left[f^{(1)} f^{(1)}\right] = \mathsf H^{(2)}{}^\vee = \mathrm{Tr}_2 \left[f^{(2)} f^{(2)}\right] \,, \\
\mathsf H^{(1,2)}{}^\vee &= \mathrm{Tr}_1 \mathrm{Tr}_2 \left[f^{(1)} q^{(1,2)} f^{(2)}\right] \,, \\
\mathsf C{}^\vee &= \left(\begin{array}{ccc}
i \sigma_2 \mathrm{Tr}_1 A^{(1)} & \mathrm{Tr}_1 d^{(1)} v^{(1)} & \mathrm{Tr}_1 \mathrm{Tr}_2 d^{(1)} q^{(1,2)} v^{(2)} \\
-\mathrm{Tr}_1 d^{(1)} v^{(1)} & i \sigma_2 \mathrm{Tr}_2 A^{(2)} & \mathrm{Tr}_2 d^{(2)} v^{(2)} \\
-\mathrm{Tr}_1 \mathrm{Tr}_2 d^{(1)} q^{(1,2)} v^{(2)} & -\mathrm{Tr}_2 d^{(2)} v^{(2)} & -i \sigma_2 \mathrm{Tr}_1 A^{(1)}-i \sigma_2 \mathrm{Tr}_2 A^{(2)}
\end{array}\right) \,.
\end{aligned}
\end{align}
Each $\Pi^{(i)}{}^\vee$ is a bifundamental operator between $USp(6)_x \times USp(2)_{y^{(i)}}$. Note that the superpotential \eqref{eq:sup21} is crucial to realize the nonabelian part of the global symmetry, $USp(6)_x \times USp(2)_{y^{(1)}} \times USp(2)_{y^{(2)}}$, because other bifundamental operators $\mathrm{Tr}_1 f^{(1)} v^{(1)}, \, \mathrm{Tr}_1 f^{(1)} q^{(1,2)} v^{(2)}$ and $\mathrm{Tr}_2 f^{(2)} v^{(2)}$, which do not respect this symmetry, are flipped by $\pi^{(1,2)\vee}, \, \pi^{(1,3)\vee}$ and $\pi^{(2,3)\vee}$ respectively and thus are trivial in the chiral ring. Each $\mathsf H^{(i)}{}^\vee$ is an $USp(2)_{y^{(i)}}$ antisymmetric, i.e. a singlet operator. Note that $\mathsf H^{(1)^\vee}$ and $\mathsf H^{(2)^\vee}$ are identified due to the superpotential, which implies that
\begin{align}
\mathrm{Tr}_1  \left[f^{(1)}_{[1} f^{(1)}_{2]}\right] \sim \left[\mathrm{Tr}_1 \mathrm{Tr}_2 q^{(1,2)} q^{(1,2)}\right] \sim \mathrm{Tr}_2 \left[f^{(2)}_{[1} f^{(2)}_{2]}\right] \,.
\end{align}
$\mathsf H^{(1,2)}{}^\vee$ is a bifundamental operator between $USp(2)_{y^{(1)}} \times USp(2)_{y^{(2)}}$.
Lastly $\mathsf C{}^\vee$ is an $USp(6)_x$ antisymmetric operator.
 
We also find the map of these operators between $E_{[2,1]}[USp(2 N)]$ and $E^{[2,1]}[USp(2 N)]$:
\begin{align}
\begin{aligned}
\Pi^{(1)} \quad &\longleftrightarrow \quad \Pi^{(1)}{}^\vee \,, \\
\Pi^{(2)} \quad &\longleftrightarrow \quad \Pi^{(2)}{}^\vee \,, \\
\mathsf C^{(1)} \quad &\longleftrightarrow \quad \mathsf H^{(1)}{}^\vee = \mathsf H^{(2)}{}^\vee\,, \\
\mathsf C^{(1,2)} \quad &\longleftrightarrow \quad \mathsf H^{(1,2)}{}^\vee \,, \\
\mathsf H \quad &\longleftrightarrow \quad \mathsf C{}^\vee \,.
\end{aligned}
\end{align}
This shows that $E_{[2,1]}[USp(2 N)]$ and $E^{[2,1]}[USp(2 N)]$ have the same low-lying operator spectrum, which respects the same global symmetry.\\

\begin{figure}[tbp]
\centering
\includegraphics[scale=0.45]{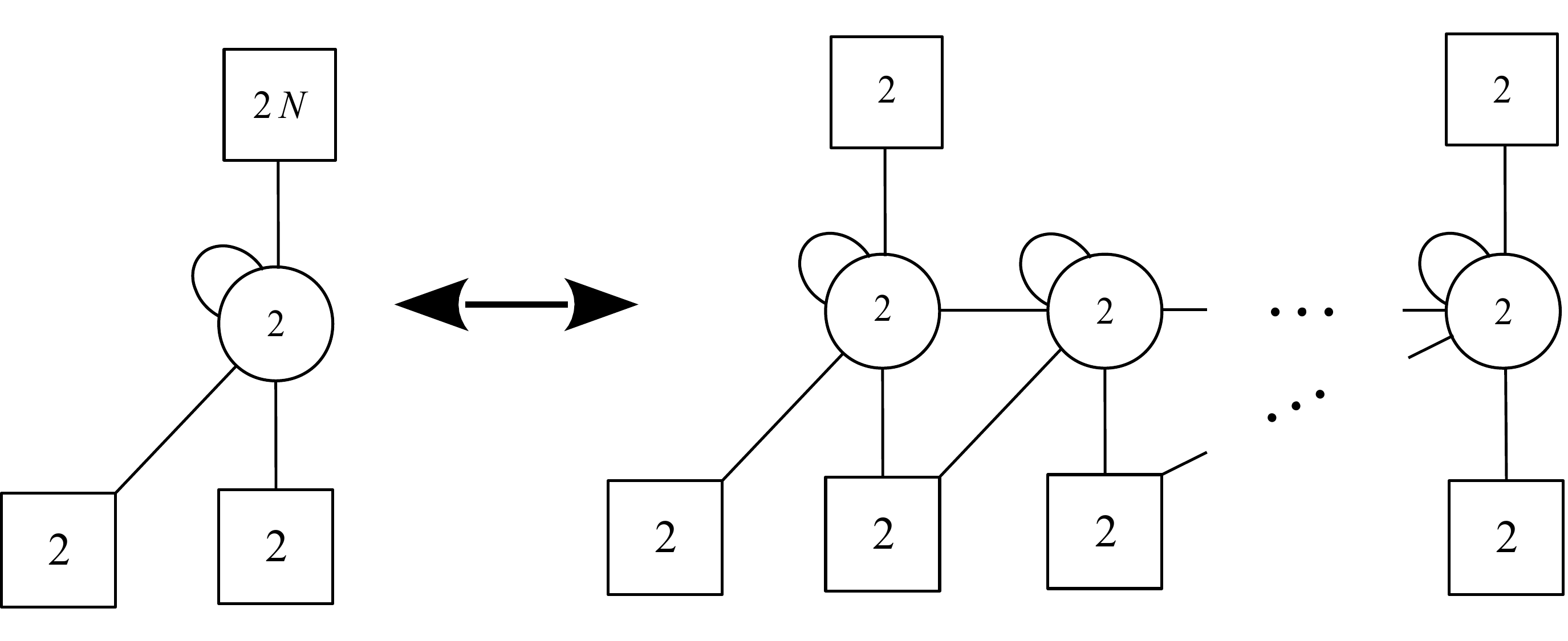}
\caption{\label{fig:T_D2} Duality between $E_{[N-1,1]}[USp(2 N)]$ and $E^{[N-1,1]}[USp(2 N)]$.}
\end{figure}

Although here we only considered the $N=3$ case, we checked that the superconformal index identity holds for  higher $N$ as well.
The mirror duality between $E^{[N-1,1]}[USp(2N)]$ and $E_{[N-1,1]}[USp(2N)]$ for arbitrary $N$ is represented in Figure \ref{fig:T_D2} in a simplified version where we omit gauge singlets. This is the $4d$ analogue of the $3d$ abelian mirror duality\footnote{See \cite{Sacchi:2020pet} for the $2d$ $\mathcal{N}=(0,2)$ reduction of this $4d$ $\mathcal{N}=1$ duality and for an analogue of the piecewise derivation in that context.}. As shown in \cite{Kapustin:1999ha}, the abelian  $3d$ Mirror Symmetry for SQED with $N$ flavors can be derived from the basic duality between SQED with one flavor and the XYZ model with a piecewise procedure. Interestingly, we can do the same  in $4d$ and derive the duality \ref{fig:T_D2} with a similar piecewise procedure, where the role of the basic duality is now played by the Intriligator--Pouliot duality in the confining case of $USp(2)$ with 6 chirals dual to a WZ model of 15 chiral fields. We show this at the level of the index in the $N=3$ case in Appendix \ref{4dabelian}.
\\

\subsubsection{$\gr=[2^2]$ and $\gs=[1^4]$}

\subsubsection*{Flow to $E_{[2^2]}[USp(8)]$}

\begin{figure}[tbp]
\centering
\includegraphics[scale=0.6]{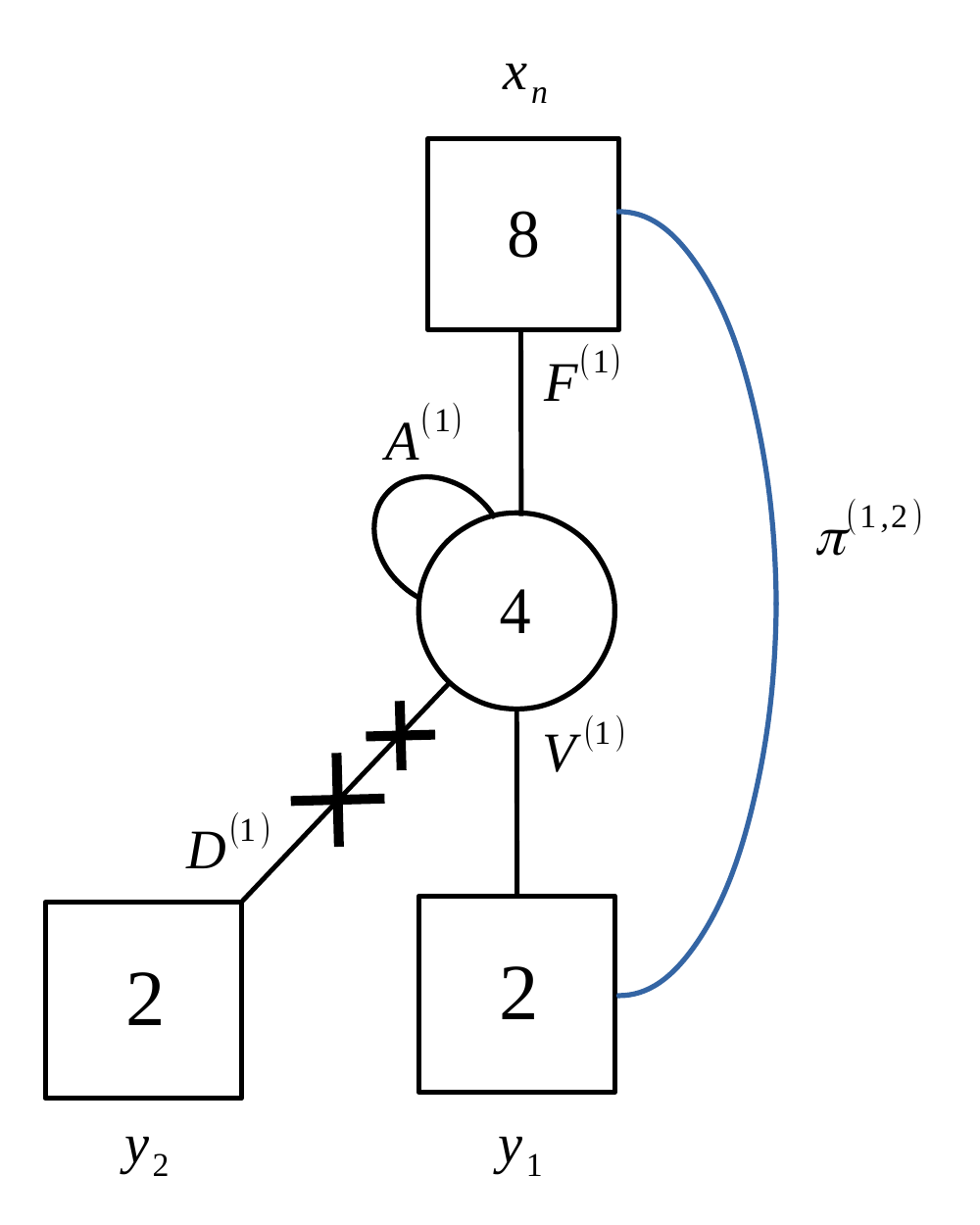}
\caption{\label{fig:E_22} The quiver diagram representation of $E_{[2^2]}[USp(8)]$. Two crosses with different sizes on top of the diagonal line denote the singlets $\gc_{11}$ and $\gc_{12}$, which flip $\mathrm{Tr}_1 \left[D^{(1)}_{[1} D^{(1)}_{2]}\right]$ and $\mathrm{Tr}_1 \left[A^{(1)} D^{(1)}_{[1} D^{(1)}_{2]}\right]$ respectively.}
\end{figure}

Starting from $E[USp(8)]$ we introduce the superpotential \eqref{eq:def} with $\rho = [2^2]$ and $\sigma = [1^4]$, which includes the mass terms
\begin{align}
\label{eq:mass22}
\delta \mathcal W = \dots + \mathrm{Tr}_1 D^{(1)}_{[1} V^{(1)}_{2]}+\mathrm{Tr}_3 D^{(3)}_{[1} V^{(3)}_{2]} +\dots \,,
\end{align}
which lead to the following constraints on fugacities:
\begin{align}
y_1 = t^{-\frac12} y^{(1)}_1 \,, \qquad y_2  = t^\frac12 y^{(1)}_1 \,, \qquad y_3 = t^{-\frac12} y^{(1)}_2 \,, \qquad y_4 = t^\frac12 y^{(1)}_2 \,.
\end{align}
For simplicity, we will omit the superscript $(1)$ of the new variables $y^{(1)}_i$, which should not be confused with the original variables $y_i$.
We also introduce a set of extra flipping fields, which contribute to the index as follows:
\begin{align}
\begin{aligned}
\label{eq:singlets_22}
\mathsf S_{[1^4]} \quad &\longrightarrow \quad \Gpq{p q t^{-1}}^{3} \prod_{n < m}^4 \Gpq{p q t^{-1} x^{\pm1}_n x^{\pm1}_m} \,, \\
\mathsf T_{[2,2]} \quad &\longrightarrow \quad \Gpq{t} \Gpq{t^2}^2 \prod_{i = 1}^2 \Gpq{t^i y_1^{\pm1} y_2^{\pm1}} \,, \\
\mathsf{O}_B^{12} \quad &\longrightarrow \quad \Gpq{t^{-1} c^2} \,.
\end{aligned}
\end{align}
After integrating out the massive fields and applying sequentially the Intriligator--Pouliot duality
we obtain the index of the $E_{[2^2]}[USp(8)]$ theory is as follows:
\begin{align}
\label{eq:SCI_22}
&\mathcal  I_{\mathsf{T}}=  \Gpq{p q t^{-1}}^{2} \prod_{n < m}^4 \Gpq{p q t^{-1} x^{\pm1}_n x^{\pm1}_m} \Gpq{t^2}^2 \prod_{i = 1}^2 \Gpq{t^i y_1^{\pm1} y_2^{\pm1}} \Gpq{t^{-1} c^2} \nn \\
&\quad \times \left.\mathcal I_{E[USp(8)]}(\vec x,\vec y,t,c)\right|_{y_1 \rightarrow t^{-\frac12} y_1, \, y_2  \rightarrow t^\frac12 y_1, \, y_3 \rightarrow t^{-\frac12} y_2, \, y_4 \rightarrow t^\frac12 y_2} \nn \\
&= \Gpq{p^2 q^2 c^{-2}} \Gpq{p^2 q^2 t^{-1} c^{-2}} \prod_{m = 1}^4 \Gpq{t^{-1/2} c y_1^{\pm1} x_m^{\pm1}} \nn \\
&\quad \times \oint \udl{\vec{z}^{(1)}_2} \Gpq{t}^2 \prod_{i < j}^2 \Gpq{t z^{(1)}_i{}^{\pm1} z^{(1)}_j{}^{\pm1}} \prod_{j = 1}^2 \Gpq{p^{-1/2} q^{-1/2} c y_2^{\pm1} z^{(1)}_j{}^{\pm1}}  \nn\\
&\quad \times \prod_{i = 1}^2 \prod_{m = 1}^4 \Gpq{p^{1/2} q^{1/2} t^{-1/2} z^{(1)}_i{}^{\pm1} x_m{}^{\pm1}} \prod_{j = 1}^2 \Gpq{p^{1/2} q^{1/2} t c^{-1} y_1^{\pm1} z^{(1)}_j{}^{\pm1}}=\nn\\
&=\mathcal  I_{E_{[2^2]}[USp(8)]}(\vec x ;\vec{y};t,c)  \,.
\end{align}
From the superconformal index \eqref{eq:SCI_22}, one can read off the matter content and the superpotential of the $E_{[2^2]}[USp(8)]$ theory, which we represent using the quiver diagram of Figure \ref{fig:E_22}.
Furthermore, the total superpotential of $E_{[2^2]}[USp(8)]$ is given by
\begin{align}
\mathcal W_{E_{[2^2]}[USp(8)]} &= \mathrm{Tr}_1 \mathrm{Tr}_x \left[A^{(1)} F^{(1)} F^{(1)}\right] + \mathrm{Tr}_1 \mathrm{Tr}_x \left[F^{(1)} v^{(1)}_{[1} \pi^{(1,2)}_{2]}\right] \nn \\
&\quad + \gc_{11} \, \mathrm{Tr}_1 D^{(1)}_{[1} D^{(1)}_{2]} + \gc_{12} \, \mathrm{Tr}_1 \left[A^{(1)} D^{(1)}_{[1} D^{(1)}_{2]}\right] \,.
\end{align}
One can see that the superpotential involves a set of gauge singlet operators, which contribute to the resulting index \eqref{eq:SCI_22} by
\begin{align}
\begin{aligned}
\pi^{(1,2)} \quad &\longrightarrow \quad \prod_{m = 1}^4 \Gpq{t^{-1/2} c y_1^{\pm1} x_m^{\pm1}} \,, \\
\gc_{11} \quad &\longrightarrow \quad\Gpq{p^2 q^2 c^{-2}} \,, \\
\gc_{12} \quad &\longrightarrow \quad\Gpq{p^2 q^2 t^{-1} c^{-2}} \,.
\end{aligned}
\end{align}

The nonabelian global symmetry of $E_{[2^2]}[USp(8)]$ is $USp(8)_x \times USp(4)_y$. A few examples of gauge invarint operators respecting this symmetry are as follows:
\begin{align}
\begin{aligned}
\Pi &= \left(\pi^{(1,2)},\mathrm{Tr}_1 \left[D^{(1)} F^{(1)}\right]\right) \,, \\
\mathsf H &= \mathrm{Tr}_1 \left[F^{(1)} F^{(1)}\right] \,, \\
\mathsf C &= \left(\begin{array}{cc}
i \sigma_2 \mathrm{Tr}_1 A^{(1)} & \mathrm{Tr}_1 \left[D^{(1)} V^{(1)}\right] \\
-\mathrm{Tr}_1 \left[D^{(1)} V^{(1)}\right] & -i \sigma_2 \mathrm{Tr}_1 A^{(1)} \\
\end{array}\right)
\end{aligned}
\end{align}
where $\Pi$ is a bifundamental between $USp(8)_x \times USp(4)_y$, while $\mathsf H$ and $\mathsf C$ are antisymmetrics of $USp(8)_x$ and $USp(4)_y$ respectively.
\\

\subsubsection*{Flow to $E^{[2^2]}[USp(8)]$}

\begin{figure}[tbp]
\centering
\includegraphics[scale=0.6]{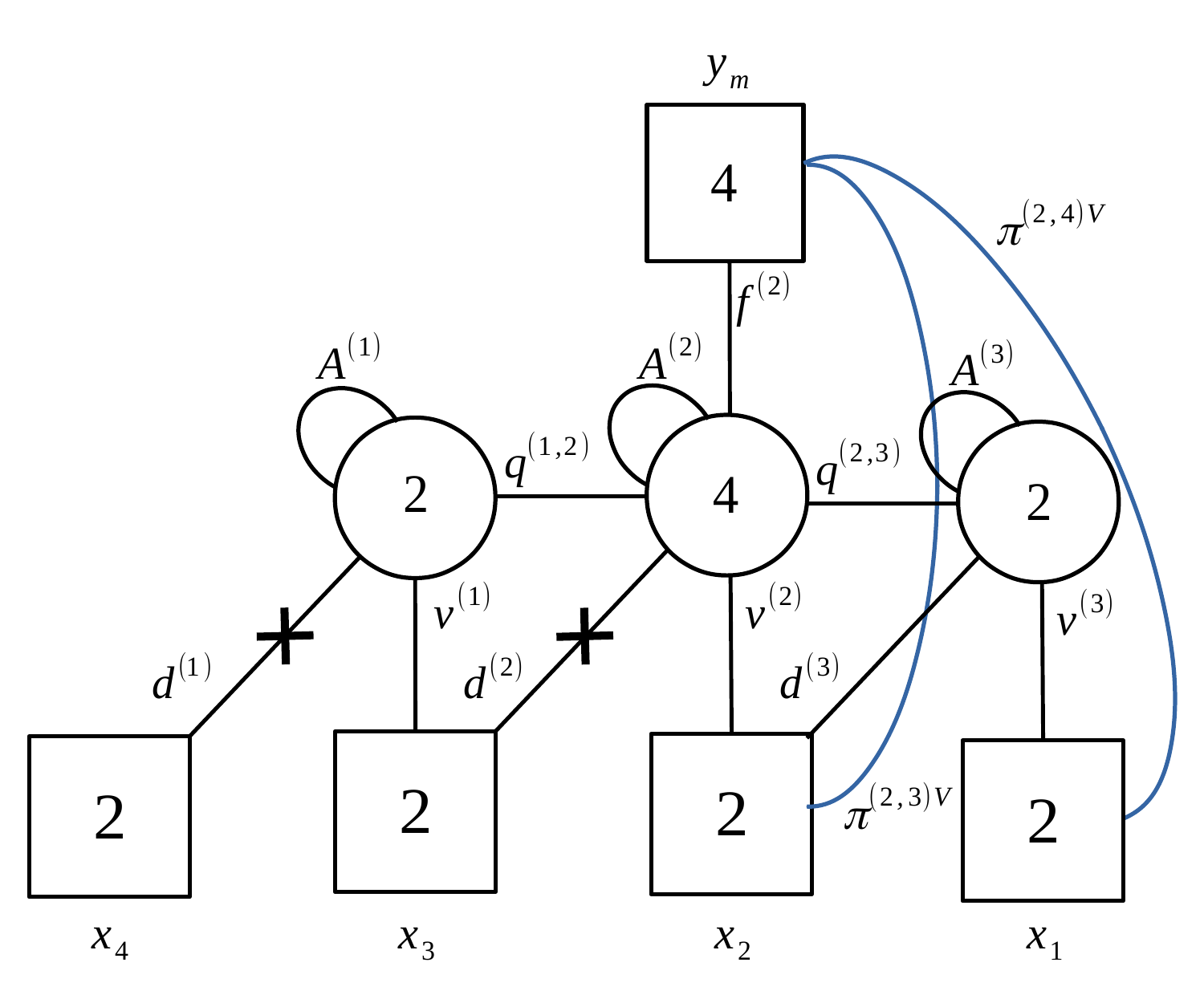}
\caption{\label{fig:E^22} The quiver diagram representation of $E^{[2^2]}[USp(8)]$. Two flipping fields $\gc_{11}^\vee$ and $\gc_{11}^\vee$, denoted by crosses, flip $\mathrm{Tr}_1 \left[d^{(1)}_{[1} d^{(1)}_{2]}\right]$ and $\mathrm{Tr}_2 \left[d^{(2)}_{[1} d^{(2)}_{2]}\right]$ respectively.}
\end{figure}

Let's now look at the mirror side. The  deformation \eqref{eq:mass22} is mapped to a deformation of $E[USp(8)]^\vee$ which includes the mass terms
\begin{align}
\delta \mathcal W = \dots+ q^{(3,4)}_1 q^{(3,4)}_4 + q^{(3,4)}_5 q^{(3,4)}_8 +\dots \,,
\end{align}
implying the constraints on fugacities
\begin{align}
y_1 = t^{-1/2} y^{(1)}_1 \,, \qquad y_2  = t^{1/2} y^{(1)}_1 \,, \qquad y_3 = t^{-1/2} y^{(1)}_2 \,, \qquad y_4 = t^{1/2} y^{(1)}_2 \,.
\end{align}
As before we will omit the superscript $(1)$ of $y^{(1)}_i$.
Taking into account the contributions of the extra flipping fields \eqref{eq:singlets_22}
and applying sequentially the Intriligator--Pouliot duality
we obtain the superconformal index of $E^{[2^2]}[USp(8)]$:
\begin{align}
\label{eq:EUSp^22}
& \mathcal I_{E^{[2^2]}[USp(8)]}(\vec{y};\vec x;p q/t,c) \nn \\
&= \prod_{m = 1}^2 \Gpq{t^{-1/2} c x_1^{\pm1} y_m^{\pm1}} \prod_{m = 1}^2 \Gpq{t^{-1/2} c x_2^{\pm1} y_m^{\pm1}} \Gpq{p q t^3 c^{-2}} \Gpq{p q t^2 c^{-2}} \nn \\
&\quad \times \oint \udl{\vec z^{(1)}_1} \udl{\vec z^{(2)}_2} \udl{\vec z^{(3)}_1} \Gpq{p q t^{-1}}^4 \prod_{i < j}^2 \Gpq{p q t^{-1} z^{(2)}_i{}^{\pm1} z^{(2)}_j{}^{\pm1}}\nn \\
&\quad \times \Gpq{t^{-3/2} c x_4^{\pm1} z^{(1)}{}^{\pm1}} \prod_{j = 1}^2 \Gpq{t^{-1} c x_3^{\pm1} z^{(2)}_j{}^{\pm1}} \Gpq{t^{-1/2} c x_2^{\pm1} z^{(3)}{}^{\pm1}} \nn \\
&\quad \times \prod_{j = 1}^2 \Gpq{t^{1/2} z^{(1)}{}^{\pm1} z^{(2)}_j{}^{\pm1}} \prod_{i = 1}^2 \Gpq{t^{1/2} z^{(2)}_i{}^{\pm1} z^{(3)}{}^{\pm1}} \prod_{i = 1}^2 \prod_{m = 1}^2 \Gpq{t^{1/2} z^{(2)}_i{}^{\pm1} y_m^{\pm1}} \nn \\
&\quad \times \Gpq{p q t^{1/2} c^{-1} x_3^{\pm1} z^{(1)}{}^{\pm1}} \prod_{j = 1}^2 \Gpq{p q c^{-1} x_2^{\pm1} z^{(2)}_j{}^{\pm1}} \Gpq{p q t^{-1/2} c^{-1} x_1^{\pm1} z^{(3)}{}^{\pm1}} \,.
\end{align}
Starting from the identity for the mirror-like duality of $E[USp(8)]$
we have derived a new identity for the duality between $E_{[2^2]}[USp(8)]$ and $E^{[2^2]}[USp(8)]$:
\begin{align}
\mathcal I_{E_{[2^2]}[USp(8)]}(\vec x;\vec{y};t,c) = \mathcal I_{E^{[2^2]}[USp(8)]}(\vec{y};\vec x;p q/t,c) \,.
\end{align}
The quiver diagram of $E^{[2^2]}[USp(8)]$ can be read from \eqref{eq:EUSp^22} and it's represented in Figure \ref{fig:E^22}.
The superpotential of $E^{[2^2]}[USp(8)]$ is given by
\begin{align}
\label{eq:sup_22}
& \mathcal W_{E^{[2^2]}[USp(8)]} \nn \\
&= \mathrm{Tr}_1 \mathrm{Tr}_2 \left[A^{(1)} q^{(1,2)} q^{(1,2)}\right] + \mathrm{Tr}_2 \left[A^{(2)} \left(\mathrm{Tr}_1 \, q^{(1,2)} q^{(1,2)}+\mathrm{Tr}_{y} f^{(2)} f^{(2)}+\mathrm{Tr}_3 q^{(2,3)} q^{(2,3)}\right)\right] \nn \\
&\quad + \mathrm{Tr}_2 \mathrm{Tr}_3 \left[A^{(3)} q^{(2,3)} q^{(2,3)}\right] + \mathrm{Tr}_1 \mathrm{Tr}_2 \left[v^{(1)}_{[1} q^{(1,2)} d^{(2)}_{2]}\right] + \mathrm{Tr}_2 \mathrm{Tr}_3 \left[v^{(2)}_{[1} q^{(2,3)} d^{(3)}_{2]}\right] \nn \\
&\quad + \mathrm{Tr}_2 \mathrm{Tr}_{y} \left[f^{(2)} v^{(2)}_{[1} \pi^{(2,3)\vee}_{2]}\right] + \mathrm{Tr}_2 \mathrm{Tr}_3 \mathrm{Tr}_{y} \left[f^{(2)} q^{(2,3)} v^{(3)}_{[1} \pi^{(2,4)\vee}_{2]}\right] + \sum_{i = 1}^2 \gc_{i1}^\vee \, \mathrm{Tr}_i \left[d^{(i)}_{[1} d^{(i)}_{2]}\right] \,,
\end{align}
which involves the gauge singlet operators whose index contributions are as follows:
\begin{align}
\begin{aligned}
\pi^{(2,3)\vee} \quad &\longrightarrow \quad \prod_{m = 1}^2 \Gpq{t^{-1/2} c x_2^{\pm1} y_m^{\pm1}} \,, \\
\pi^{(2,4)\vee} \quad &\longrightarrow \quad \prod_{m = 1}^2 \Gpq{t^{-1/2} c x_1^{\pm1} y_m^{\pm1}} \,, \\
\gc_{11}^\vee \quad &\longrightarrow \quad \Gpq{p q t^3 c^{-2}} \,, \\
\gc_{21}^\vee \quad &\longrightarrow \quad \Gpq{p q t^2 c^{-2}} \,.
\end{aligned}
\end{align}

One can also construct gauge invariant operators. For example,
\begin{equation}
\makebox[\linewidth][c]{\scalebox{0.9}{$
\begin{split}
\Pi^\vee &= \left(\pi^{(2,4)\vee},\pi^{(2,3)\vee},\mathrm{Tr}_2 \left[d^{(2)} f^{(2)}\right],\mathrm{Tr}_1 \mathrm{Tr}_2 \left[d^{(1)} q^{(1,2)} f^{(2)}\right]\right) \,, \\
\mathsf H^\vee &= \mathrm{Tr}_2 \left[f^{(2)} f^{(2)}\right] \,, \\
\mathsf C^\vee &= \left(\begin{array}{cccc}
i \sigma_2 \mathrm{Tr}_1 A^{(1)} & \mathrm{Tr}_1 d^{(1)} v^{(1)} & \mathrm{Tr}_1 \mathrm{Tr}_2 d^{(1)} q^{(1,2)} v^{(2)} & \mathrm{Tr}_1 \mathrm{Tr}_2 \mathrm{Tr}_3 d^{(1)} q^{(1,2)} q^{(2,3)} v^{(3)} \\
-\mathrm{Tr}_1 d^{(1)} v^{(1)} & i \sigma_2 \mathrm{Tr}_2 A^{(2)} & \mathrm{Tr}_2 d^{(2)} v^{(2)} & \mathrm{Tr}_2 \mathrm{Tr}_3 d^{(2)} q^{(2,3)} v^{(3)} \\
-\mathrm{Tr}_1 \mathrm{Tr}_2 d^{(1)} q^{(1,2)} v^{(2)} & -\mathrm{Tr}_2 d^{(2)} v^{(2)} & i \sigma_2 \mathrm{Tr}_3 A^{(3)} & \mathrm{Tr}_3 d^{(3)} v^{(3)} \\
-\mathrm{Tr}_1 \mathrm{Tr}_2 \mathrm{Tr}_3 d^{(1)} q^{(1,2)} q^{(2,3)} v^{(3)} & -\mathrm{Tr}_2 \mathrm{Tr}_3 d^{(2)} q^{(2,3)} v^{(3)} & -\mathrm{Tr}_3 d^{(3)} v^{(3)} & -i \sigma_2 \sum_{i = 1}^3 \mathrm{Tr}_i A^{(i)}\\
\end{array}\right) \,,
\end{split}$}}
\end{equation}
which are mapped to operators of $E_{[2^2]}[USp(8)]$ as follows:
\begin{align}
\begin{aligned}
\Pi \quad &\longleftrightarrow \quad \Pi^\vee \,, \\
\mathsf H \quad &\longleftrightarrow \quad \mathsf C^\vee \,, \\
\mathsf C \quad &\longleftrightarrow \quad \mathsf H^\vee \,.
\end{aligned}
\end{align}
Note that $\Pi^\vee$ is a bifundamental between $USp(8)_x \times USp(4)_y$, while $\mathsf H^\vee$ and $\mathsf C^\vee$ are antisymmetrics of $USp(4)_y$ and $USp(8)_x$ respectively.
\\

\subsubsection{$\gr=[2,1^2]$ and $\gs=[1^4]$}

\subsubsection*{Flow to $E_{[2,1^2]}[USp(8)]$}

\begin{figure}[tbp]
\centering
\includegraphics[scale=0.6]{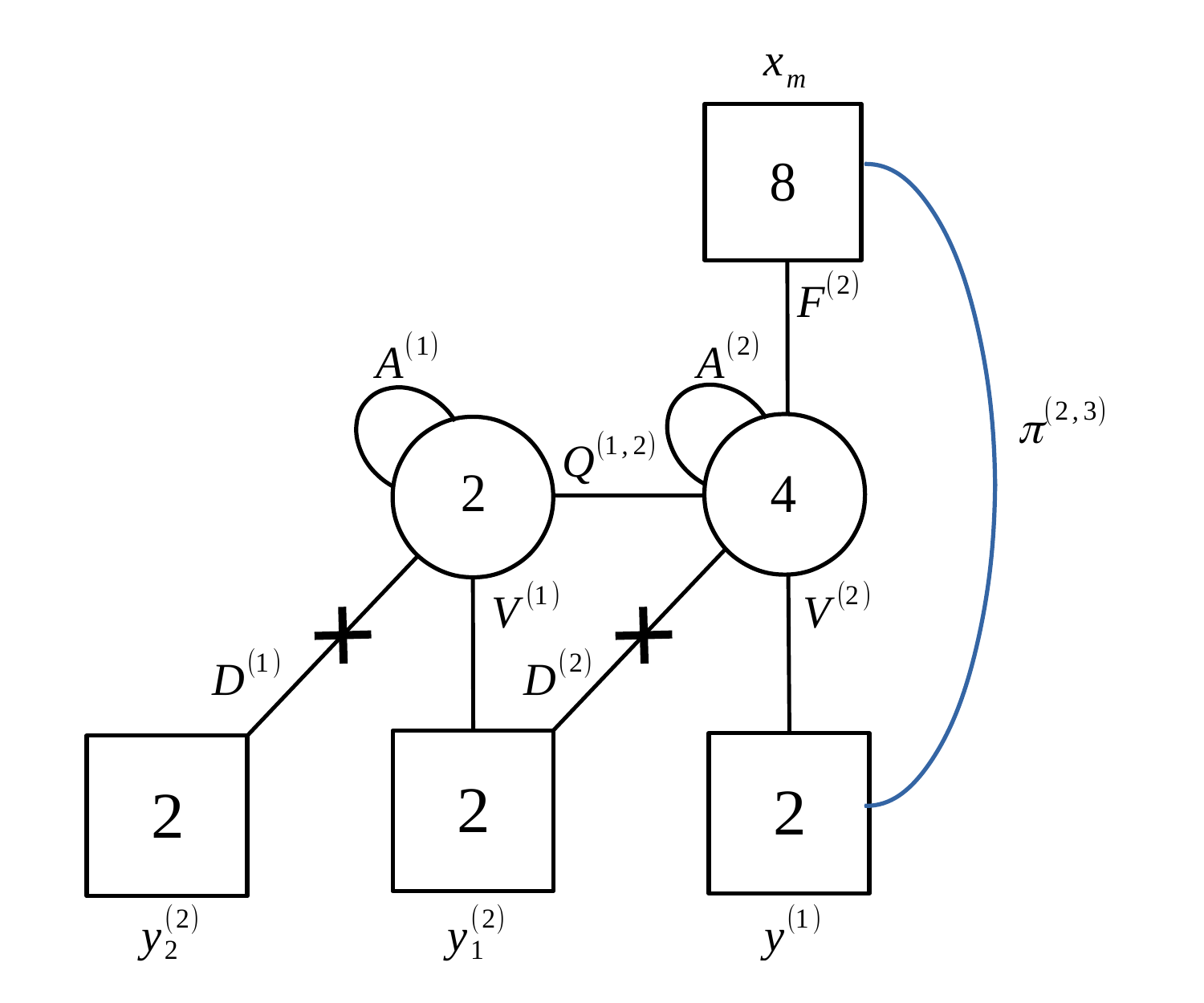}
\caption{\label{fig:E_211} The quiver diagram representation of $E_{[2,1^2]}[USp(8)]$. Two flipping fields $\gc_{11}$ and $\gc_{21}$, denoted by crosses, flip $\mathrm{Tr}_1 \left[D^{(1)}_{[1} D^{(1)}_{2]}\right]$ and $\mathrm{Tr}_2 \left[D^{(2)}_{[1} D^{(2)}_{2]}\right]$ respectively.}
\end{figure}

We now consider a deformation of $E[USp(8)]$ corresponding to $\rho = [2,1^2]$ and $\sigma = [1^4]$, which includes a mass term
\begin{align}
\delta \mathcal W = \dots + \mathrm{Tr}_1 D^{(1)}_{[1} V^{(1)}_{2]} + \dots
\end{align}
which relates $y_1$ and $y_2$ as follows:
\begin{align}
y_1 = t^{-\frac12} y^{(1)} \,, \qquad y_2  = t^\frac12 y^{(1)}.
\end{align}
For later convenience, we also rename $y_3$ and $y_4$ as
\begin{align}
y_3 = y^{(2)}_1 \,, \qquad y_4 = y^{(2)}_2 \,.
\end{align}
The extra flipping fields we introduce in this case are
\begin{align}
\begin{aligned}
\label{eq:singlets_211}
\mathsf S_{[1^4]} \quad &\longrightarrow \quad \Gpq{p q t^{-1}}^{3} \prod_{n < m}^4 \Gpq{p q t^{-1} x^{\pm1}_n x^{\pm1}_m} \,, \\
\mathsf T_{[2,1^2]} \quad &\longrightarrow \quad \Gpq{t}^2 \Gpq{t^2} \prod_{i = 1}^2 \Gpq{t^\frac32 y^{(1)}{}^{\pm1} y^{(2)}_i{}^{\pm1}} \Gpq{t y^{(2)}_1{}^{\pm1} y^{(2)}_2{}^{\pm1}} \,, \\
\mathsf{O}_B^{12} \quad &\longrightarrow \quad \Gpq{t^{-1} c^2} \,.
\end{aligned}
\end{align}
After applying sequentially the Intriligator--Pouliot duality, we obtain the superconformal index of $E_{[2,1^2]}[USp(8)]$:\begin{align}
\label{eq:SCI_211}
&\mathcal I_{E_{[2,1^2]}[USp(8)]}\left(\vec x; \vec y^{(2)},y^{(1)};t,c\right) \nn \\
&= \Gpq{p^3 q^3 t^{-2} c^{-2}} \Gpq{p^2 q^2 t^{-1} c^{-2}} \prod_{m = 1}^4 \Gpq{t^{-1/2} c y^{(1)}{}^{\pm1} x_m{}^{\pm1}} \nn \\
&\quad \times \oint \udl{\vec z^{(1)}_1} \udl{\vec{z}^{(2)}_2} \Gpq{t}^3 \prod_{i < j}^2 \Gpq{t z^{(2)}_i{}^{\pm1} z^{(2)}_j{}^{\pm1}} \nn \\
&\quad \times \Gpq{p^{-1} q^{-1} t c z^{(1)}{}^{\pm1} y^{(2)}_2{}^{\pm1}} \prod_{j = 1}^2 \Gpq{p^{-1/2} q^{-1/2} t^{1/2} c y^{(2)}_1{}^{\pm1} z^{(2)}_j{}^{\pm1}} \nn \\
&\quad \times  \prod_{j = 1}^2 \Gpq{p^{1/2} q^{1/2} t^{-1/2} z^{(1)}{}^{\pm1} z^{(2)}_j{}^{\pm1}} \prod_{i = 1}^2 \prod_{m = 1}^4 \Gpq{p^{1/2} q^{1/2} t^{-1/2} z^{(2)}_i{}^{\pm1} x_m{}^{\pm1}} \nn \\
&\quad \times \Gpq{p q c^{-1} y^{(2)}_1{}^{\pm1} z^{(1)}{}^{\pm1}} \prod_{j = 1}^2 \Gpq{p^{1/2} q^{1/2} t c^{-1} y^{(1)}{}^{\pm1} z^{(2)}_j{}^{\pm1}} \,.
\end{align}
The quiver diagram of $E_{[2,1^2]}[USp(8)]$ is drawn in Figure \ref{fig:E_211}, which can be worked out from the superconformal index \eqref{eq:SCI_211}.
The total superpotential of $E_{[2,1^2]}[USp(8)]$ is given by
\begin{align}
\mathcal W_{E_{[2,1^2]}[USp(8)]} &= \mathrm{Tr}_1 \mathrm{Tr}_2 \left[A^{(1)} Q^{(1,2)} Q^{(1,2)}\right] + \mathrm{Tr}_2 \left[A^{(2)} \left(\mathrm{Tr}_1 Q^{(1,2)} Q^{(1,2)}+\mathrm{Tr}_x F^{(2)} F^{(2)}\right)\right] \nn \\
&+ \mathrm{Tr}_1 \mathrm{Tr}_2 \left[V^{(1)}_{[1} Q^{(1,2)} D^{(2)}_{2]}\right]+\mathrm{Tr}_2 \mathrm{Tr}_x \left[F^{(2)} V^{(2)}_{[1} \pi^{(2,3)}_{2]}\right] + \sum_{i = 1}^2 \gc_{i1} \, \mathrm{Tr}_i D^{(i)}_{[1} D^{(i)}_{2]} \,.
\end{align}
One can see that the superpotential involves a set of gauge singlet operators, which contribute to the index \eqref{eq:SCI_22} by
\begin{align}
\begin{aligned}
\pi^{(2,3)} \quad &\longrightarrow \quad \prod_{m = 1}^4 \Gpq{t^{-1/2} c y_1^{\pm1} x_m^{\pm1}} \,, \\
\gc_{11} \quad &\longrightarrow \quad \Gpq{p^3 q^3 t^{-2} c^{-2}} \,, \\
\gc_{21} \quad &\longrightarrow \quad \Gpq{p^2 q^2 t^{-1} c^{-2}} \,.
\end{aligned}
\end{align}

The nonabelian global symmetry of $E_{[2,1^2]}[USp(8)]$ is $USp(8)_x \times USp(2)_{y^{(1)}} \times USp(4)_{y^{(2)}}$. Some interesting examples of gauge invariant operators, which respect this symmetry, are
\begin{align}
\begin{aligned}
\Pi^{(1)} &= \pi^{(2,3)} \,, \\
\Pi^{(2)} &= \left(\mathrm{Tr}_2 \left[D^{(2)} F^{(2)}\right], \mathrm{Tr}_1 \mathrm{Tr}_2 \left[D^{(1)} Q^{(1,2)} F^{(2)}\right]\right) \,, \\
\mathsf H &= \mathrm{Tr}_2 \left[F^{(2)} F^{(2)}\right] \,, \\
\mathsf C^{(1)} &= A^{(2)} \,, \\
\mathsf C^{(2)} &= \left(\begin{array}{cc}
i \sigma_2 \mathrm{Tr}_1 A^{(1)} & \mathrm{Tr}_1 D^{(1)} V^{(1)} \\
-\mathrm{Tr}_1 D^{(1)} V^{(1)} & -i \sigma_2 \mathrm{Tr}_1 A^{(1)} \\
\end{array}\right) \,, \\
\mathsf C^{(1,2)} &= \mathrm{Tr}_1 \mathrm{Tr}_2 \left[D^{(1)} Q^{(1,2)} V^{(2)}\right] \,,
\end{aligned}
\end{align}
where $\Pi^{(i)}$ is a bifundamental between $USp(8)_x \times USp(2l_i)_{y^{(i)}}$ with $l_1 = 1$ and $l_2 = 2$, $\mathsf H$ and $\mathsf C^{(i)}$ are antisymmetrics of $USp(8)_x$ and $USp(2l_i)_{y^{(i)}}$ respectively, and lastly $C^{(1,2)}$ is a bifundamental between $USp(2)_{y^{(1)}} \times USp(4)_{y^{(2)}}$.
\\

\subsubsection*{Flow to $E^{[2,1^2]}[USp(8)]$}

\begin{figure}[tbp]
\centering
\includegraphics[scale=0.6]{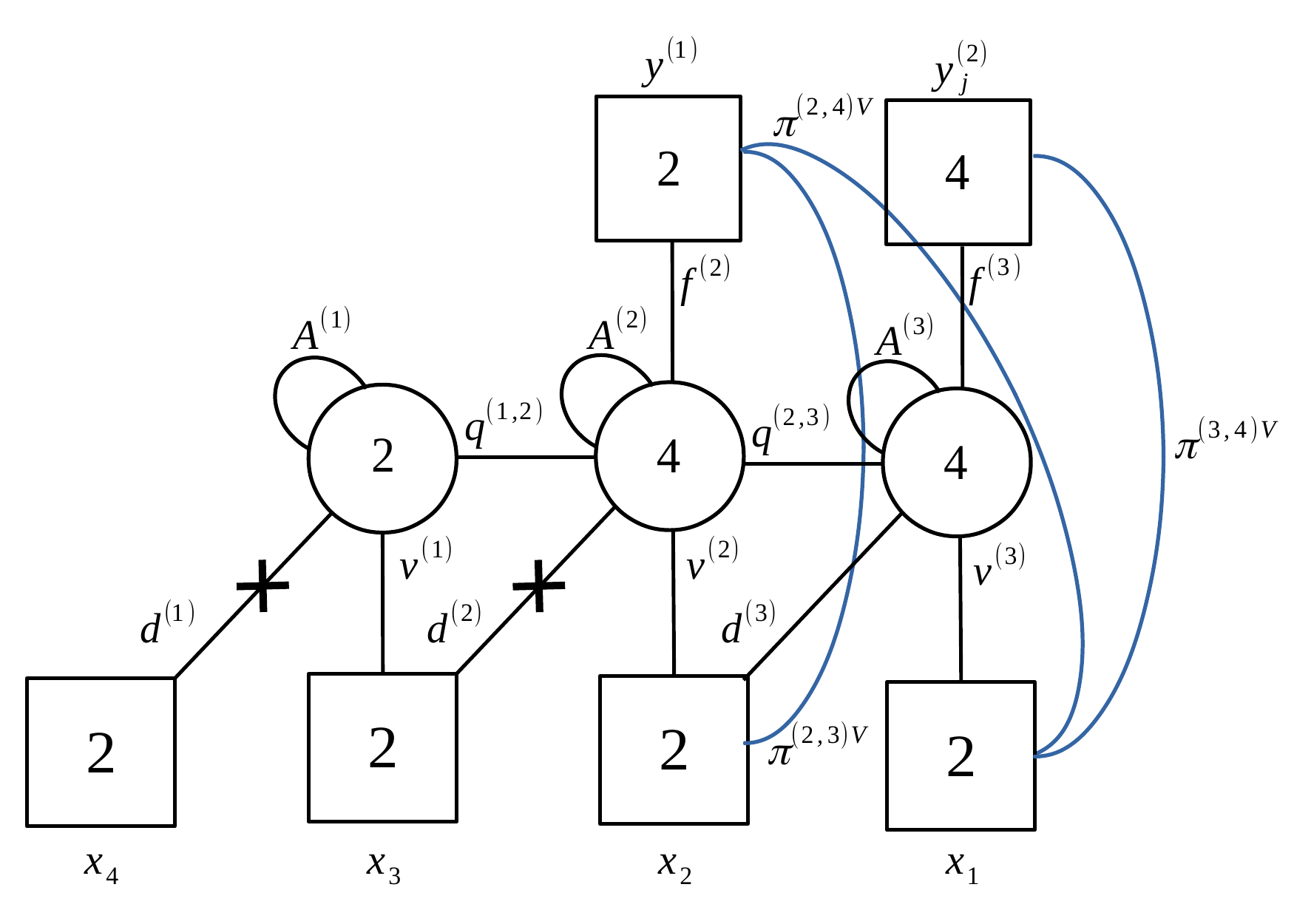}
\caption{\label{fig:E^211} The quiver diagram representation of $E^{[2,1^2]}[USp(8)]$. Two flipping fields $\gc_{11}^\vee$ and $\gc_{21}^\vee$, denoted by crosses, flip $\mathrm{Tr}_1 \left[d^{(1)}_{[1} d^{(1)}_{2]}\right]$ and $\mathrm{Tr}_2 \left[d^{(2)}_{[1} d^{(2)}_{2]}\right]$ respectively. }
\end{figure}

On the mirror side we start from the index of $E[USp(8)]^\vee$ and impose the fugacity constraints
\begin{align}
y_1 = t^{-\frac12} y^{(1)} \,, \qquad y_2  = t^\frac12 y^{(1)} \,, \quad y_3 = y^{(2)}_1 \,, \qquad y_4 = y^{(2)}_2 \,,
\end{align}
which is due to the mirror deformation superpotential 
\begin{align}
\delta \mathcal W = \dots+ q^{(3,4)}_1 q^{(3,4)}_4 +\dots \,,
\end{align}
We also introduce the extra flipping fields given in \eqref{eq:singlets_211}. 
After sequentially applying Intriligator--Pouliot duality we obtain  the
index of the $E^{[2,1^2]}[USp(8)]$ theory:
\begin{align}
\label{eq:SCI^211}
& \mathcal I_{E^{[2,1^2]}[USp(8)]}\left(y^{(1)},\vec y^{(2)};\vec x;p q t^{-1},c\right) \nn \\
&= \Gpq{t^{-1/2} c x_1^{\pm1} y^{(1)}{}^{\pm1}} \prod_{j = 1}^2 \Gpq{c x_1^{\pm1} y^{(2)}_j{}^{\pm1}} \Gpq{t^{-1/2} c x_2^{\pm1} y^{(1)}{}^{\pm1}} \Gpq{p q t^3 c^{-2}} \Gpq{p q t^2 c^{-2}} \nn \\
&\quad \times \oint \udl{\vec z^{(1)}_1} \udl{\vec z^{(2)}_2} \udl{\vec z^{(3)}_2} \Gpq{p q t^{-1}}^5 \prod_{i < j}^2 \Gpq{p q t^{-1} z^{(2)}_i{}^{\pm1} z^{(2)}_j{}^{\pm1}} \prod_{i < j}^2 \Gpq{p q t^{-1} z^{(3)}_i{}^{\pm1} z^{(3)}_j{}^{\pm1}} \nn \\
&\quad \times \Gpq{t^{-3/2} c x_4^{\pm1} z^{(1)}{}^{\pm1}} \prod_{j = 1}^2 \Gpq{t^{-1} c x_3^{\pm1} z^{(2)}_j{}^{\pm1}} \prod_{j = 1}^2 \Gpq{t^{-1/2} c x_2^{\pm1} z^{(3)}_j{}^{\pm1}} \nn \\
&\quad \times \prod_{j = 1}^2 \Gpq{t^{1/2} z^{(1)}{}^{\pm1} z^{(2)}_j{}^{\pm1}} \prod_{i = 1}^2 \prod_{j = 1}^2 \Gpq{t^{1/2} z^{(2)}_i{}^{\pm1} z^{(3)}_j{}^{\pm1}} \nn \\
&\quad \times \prod_{i = 1}^2 \Gpq{t^{1/2} z^{(2)}_i{}^{\pm1} y^{(1)}{}^{\pm1}} \prod_{i = 1}^2 \prod_{j = 1}^2 \Gpq{t^{1/2} z^{(3)}_i{}^{\pm1} y^{(2)}_j{}^{\pm1}} \nn \\
&\quad \times \Gpq{p q t^{1/2} c^{-1} x_3^{\pm1} z^{(1)}{}^{\pm1}} \prod_{j = 1}^2 \Gpq{p q c^{-1} x_2^{\pm1} z^{(2)}_j{}^{\pm1}} \prod_{j = 1}^2 \Gpq{p q t^{-1/2} c^{-1} x_1^{\pm1} z^{(3)}_j{}^{\pm1}} \,,
\end{align}
We then have shown the equality of indices
\begin{align}
\mathcal I_{E_{[2,1^2]}[USp(8)]}(\vec x;\vec y^{(2)},y^{(1)};t,c) = \mathcal I_{E^{[2,1^2]}[USp(8)]}(y^{(1)},\vec y^{(2)};\vec x;p q/t,c) \,.
\end{align}
The quiver diagram read from the index \eqref{eq:SCI^211} is shown in Figure \ref{fig:E^211}.
The superpotential of $E^{[2,1^2]}[USp(8)]$ is
\begin{align}
\label{eq:sup_211}
& \mathcal W_{E^{[2,1^2]}[USp(8)]} = \nn \\
& \mathrm{Tr}_1 \mathrm{Tr}_2 \left[A^{(1)} q^{(1,2)} q^{(1,2)}\right] + \mathrm{Tr}_2 \left[A^{(2)} \left(\mathrm{Tr}_1 \, q^{(1,2)} q^{(1,2)}+\mathrm{Tr}_{y^{(1)}} f^{(2)} f^{(2)}+\mathrm{Tr}_3 q^{(2,3)} q^{(2,3)}\right)\right] \nn \\
& + \mathrm{Tr}_3 \left[A^{(3)} \left(\mathrm{Tr}_2 \, q^{(2,3)} q^{(2,3)}+\mathrm{Tr}_{y^{(2)}} f^{(3)} f^{(3)}\right)\right] + \mathrm{Tr}_1 \mathrm{Tr}_2 \left[v^{(1)}_{[1} q^{(1,2)} d^{(2)}_{2]}\right] + \mathrm{Tr}_2 \mathrm{Tr}_3 \left[v^{(2)}_{[1} q^{(2,3)} d^{(3)}_{2]}\right] \nn \\
& + \mathrm{Tr}_2 \mathrm{Tr}_{y^{(1)}} \left[f^{(2)} v^{(2)}_{[1} \pi^{(2,3)\vee}_{2]}\right] + \mathrm{Tr}_2 \mathrm{Tr}_3 \mathrm{Tr}_{y^{(1)}} \left[f^{(2)} q^{(2,3)} v^{(3)}_{[1} \pi^{(2,4)\vee}_{2]}\right] + \mathrm{Tr}_3 \mathrm{Tr}_{y^{(2)}} \left[f^{(3)} v^{(3)}_{[1} \pi^{(3,4)\vee}_{2]}\right] \nn \\
&+ \sum_{i = 1}^2 \gc_{i1}^\vee \, \mathrm{Tr}_i \left[d^{(i)}_{[1} d^{(i)}_{2]}\right] \,,
\end{align}
which involves a set of gauge singlet operators, which contribute to the index \eqref{eq:SCI^211} by
\begin{align}
\begin{aligned}
\pi^{(2,3)\vee} \quad &\longrightarrow \quad \Gpq{t^{-1/2} c x_2^{\pm1} y^{(1)}{}^{\pm1}} \,, \\
\pi^{(2,4)\vee} \quad &\longrightarrow \quad \Gpq{t^{-1/2} c x_1^{\pm1} y^{(1)}{}^{\pm1}} \,, \\
\pi^{(3,4)\vee} \quad &\longrightarrow \quad \prod_{j = 1}^2 \Gpq{c x_1^{\pm1} y^{(2)}_j{}^{\pm1}} \,, \\
\gc_{11}^\vee \quad &\longrightarrow \quad \Gpq{p q t^3 c^{-2}} \,, \\
\gc_{21}^\vee \quad &\longrightarrow \quad \Gpq{p q t^2 c^{-2}} \,.
\end{aligned}
\end{align}

We also exhibit some gauge invariant operators:
\begin{equation}
\makebox[\linewidth][c]{\scalebox{1}{$
\begin{split}
\label{eq:ops^211}
&\Pi^{(1)}{}^\vee = \left(\pi^{(2,4)\vee},\pi^{(2,3)\vee},\mathrm{Tr}_2 \left[d^{(2)} f^{(2)}\right],\mathrm{Tr}_1 \mathrm{Tr}_2 \left[d^{(1)} q^{(1,2)} f^{(2)}\right]\right) \,, \\
&\Pi^{(2)}{}^\vee = \left(\pi^{(3,4)\vee},\mathrm{Tr}_3 \left[d^{(3)} f^{(3)}\right], \mathrm{Tr}_2 \mathrm{Tr}_3 \left[d^{(2)} q^{(2,3)} f^{(3)}\right], \mathrm{Tr}_1 \mathrm{Tr}_2 \mathrm{Tr}_3 \left[d^{(1)} q^{(1,2)} q^{(2,3)} f^{(3)}\right]\right) \,, \\
&\mathsf H^{(1)}{}^\vee = \mathrm{Tr}_2 \left[f^{(2)} f^{(2)}\right] \,, \\
&\mathsf H^{(2)}{}^\vee = \mathrm{Tr}_3 \left[f^{(3)} f^{(3)}\right] \,, \\
&\mathsf H^{(1,2)}{}^\vee = \mathrm{Tr}_2 \mathrm{Tr}_3 \left[f^{(2)} q^{(2,3)} f^{(3)}\right]
\end{split}$}}
\end{equation}
and
\begin{equation}
\makebox[\linewidth][c]{\scalebox{0.9}{$
\begin{split}
\mathsf C^\vee &= \left(\begin{array}{cccc}
i \sigma_2 \mathrm{Tr}_1 A^{(1)} & \mathrm{Tr}_1 d^{(1)} v^{(1)} & \mathrm{Tr}_1 \mathrm{Tr}_2 d^{(1)} q^{(1,2)} v^{(2)} & \mathrm{Tr}_1 \mathrm{Tr}_2 \mathrm{Tr}_3 d^{(1)} q^{(1,2)} q^{(2,3)} v^{(3)} \\
-\mathrm{Tr}_1 d^{(1)} v^{(1)} & i \sigma_2 \mathrm{Tr}_2 A^{(2)} & \mathrm{Tr}_2 d^{(2)} v^{(2)} & \mathrm{Tr}_2 \mathrm{Tr}_3 d^{(2)} q^{(2,3)} v^{(3)} \\
-\mathrm{Tr}_1 \mathrm{Tr}_2 d^{(1)} q^{(1,2)} v^{(2)} & -\mathrm{Tr}_2 d^{(2)} v^{(2)} & i \sigma_2 \mathrm{Tr}_3 A^{(3)} & \mathrm{Tr}_3 d^{(3)} v^{(3)} \\
-\mathrm{Tr}_1 \mathrm{Tr}_2 \mathrm{Tr}_3 d^{(1)} q^{(1,2)} q^{(2,3)} v^{(3)} & -\mathrm{Tr}_2 \mathrm{Tr}_3 d^{(2)} q^{(2,3)} v^{(3)} & -\mathrm{Tr}_3 d^{(3)} v^{(3)} & -i \sigma_2 \sum_{i = 1}^3 \mathrm{Tr}_i A^{(i)} \\
\end{array}\right) \,,
\end{split}$}}
\end{equation}
where $\Pi^{(i)}$ is a bifundamental between $USp(8)_x \times USp(2l_i)_{y^{(i)}}$ with $l_1 = 1$ and $l_2 = 2$, $\mathsf C^\vee$ and $\mathsf H^{(i)}{}^\vee$ are antisymmetrics of $USp(8)_x$ and $USp(2l_i)_{y^{(i)}}$ respectively and lastly $H^{(1,2)}{}^\vee$ is a bifundamental between $USp(2)_{y^{(1)}} \times USp(4)_{y^{(2)}}$. Note that the nonabelian global symmetry of $E^{[2,1^2]}[USp(8)]$ is $USp(8)_x \times USp(2)_{y^{(1)}} \times USp(4)_{y^{(2)}}$. The operators in \eqref{eq:ops^211} are mapped to those of $E_{[2,1^2]}[USp(8)]$ as follows:
\begin{align}
\begin{aligned}
\Pi^{(1)} \quad &\longleftrightarrow \quad \Pi^{(1)}{}^\vee \,, \\
\Pi^{(2)} \quad &\longleftrightarrow \quad \Pi^{(2)}{}^\vee \,, \\
\mathsf H \quad &\longleftrightarrow \quad \mathsf C^\vee \,, \\
\mathsf C^{(1)} \quad &\longleftrightarrow \quad \mathsf H^{(1)}{}^\vee \,, \\
\mathsf C^{(2)} \quad &\longleftrightarrow \quad \mathsf H^{(2)}{}^\vee \,, \\
\mathsf C^{(1,2)} \quad &\longleftrightarrow \quad \mathsf H^{(1,2)}{}^\vee \,.
\end{aligned}
\end{align}
\\

\subsubsection{$\rho = \sigma = [2^3,1]$}
So far we focused on cases with one non-trivial partitions, however we checked that our construction consistently produces
mirror pairs of theories  also when both  $\rho$ and $\sigma$ are non-trivial (we checked  this for all partitions up to $N = 14$).  
Here we exhibit one particular example with  $N=7$ and  $\rho = \sigma = [2^3,1]$, which corresponds to a self-duality. This example exhibits diverse increments of the gauge rank along the tail, so one can see how such different rank increments affect the number of the flipping fields in the resulting $E_\rho^\sigma[SU(N)]$ theory.

We  start with the $E[USp(14)]$ theory and introduce the deformation \eqref{eq:def} for $\rho = \sigma = [2^3,1]$. This deformation requires the following specialization of fugacities, now both for $\vec x$ and for $\vec y$:
\begin{gather}
\begin{gathered}
x_1 = t^{-\frac12} x^{(1)}_1 \,, \quad x_2 = t^\frac12 x^{(1)}_1 \,, \quad  x_3 = t^{-\frac12} x^{(1)}_2 \,, \quad x_4 = t^\frac12 x^{(1)}_2 \,, \quad x_5 = t^{-\frac12} x^{(1)}_3 \,, \quad x_6 = t^\frac12 x^{(1)}_3 \,, \\
y_1 = t^{-\frac12} y^{(1)}_1 \,, \quad y_2 = t^\frac12 y^{(1)}_1 \,, \quad y_3 = t^{-\frac12} y^{(1)}_2 \,, \quad y_4 = t^\frac12 y^{(1)}_2 \,, \quad y_5 = t^{-\frac12} y^{(1)}_3 \,, \quad y_6 = t^\frac12 y^{(1)}_3 \,.
\end{gathered}
\end{gather}
We also rename $x_7$ and $y_7$ as follows:
\begin{align}
x_7 = x^{(2)}_1 \,, \qquad y_7 = y^{(2)}_1 \,.
\end{align}
Then those new variables will be the fugacities for the enhanced non-abelian global symmetry in the IR, which is $USp(6)_{x^{(1)}} \times USp(2)_{x^{(2)}}\times  USp(6)_{y^{(1)}} \times USp(2)_{y^{(2)}}$ for $\rho = \sigma = [2^3,1]$.

\begin{figure}[tbp]
\centering
\includegraphics[scale=0.6]{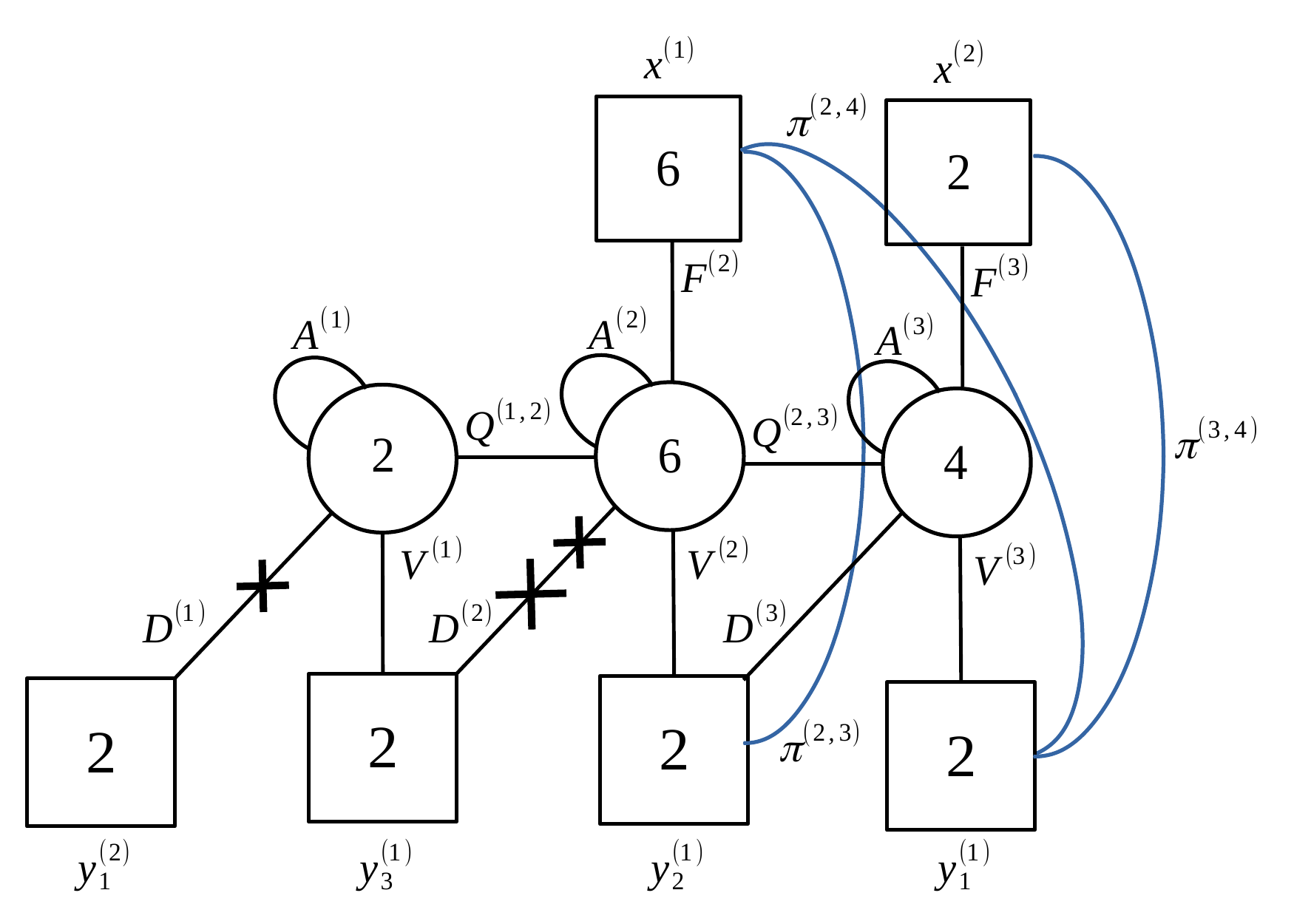}
\caption{\label{fig:E2221} The quiver diagram representation of $E_{[2^3,1]}^{[2^3,1]}[USp(14)]$. Three flipping fields $\gamma_{11}, \, \gamma_{21}$ and $\gamma_{22}$, denoted by crosses with two different sizes, flip $\mathrm{Tr}_1 D^{(1)}_{[1} D^{(1)}_{2]}, \, \mathrm{Tr}_2 D^{(2)}_{[1} D^{(2)}_{2]}$ and $\mathrm{Tr}_2 A^{(2)} D^{(2)}_{[1} D^{(2)}_{2]}$ respectively. }
\end{figure}

In addition, we introduce the extra singlets, which contribute to the index as follows:
\begin{align}
\begin{aligned}
\mathsf S_{[2^3,1]} \quad \longrightarrow \quad &\Gpq{p^2 q^2 t^{-2}}^{3} \Gpq{p q t^{-1}}^2 \prod_{m < n}^3 \Gpq{p^2 q^2 t^{-2} x^{(1)}_m{}^{\pm1} x^{(1)}_n{}^{\pm1}} \\
&\times \prod_{m < n}^3 \Gpq{p q t^{-1} x^{(1)}_m{}^{\pm1} x^{(1)}_n{}^{\pm1}} \prod_{n = 1}^3 \Gpq{p^{3/2} q^{3/2} t^{-\frac32} x^{(1)}_n{}^{\pm1} x^{(2)}_1{}^{\pm1}} \,, \\
\mathsf T_{[2^3,1]} \quad \longrightarrow \quad &\Gpq{t^2}^3 \Gpq{t}^2 \prod_{m < n}^3 \Gpq{t^2 y^{(1)}_m{}^{\pm1} y^{(1)}_n{}^{\pm1}} \\
&\times \prod_{n < m}^3 \Gpq{t y^{(1)}_m{}^{\pm1} y^{(1)}_n{}^{\pm1}}  \prod_{n = 1}^3 \Gpq{t^{3/2} y^{(1)}_n{}^{\pm1} y^{(2)}_1{}^{\pm1}} \,, \\
\mathsf O_B^{12} \quad \longrightarrow \quad &\Gpq{t^{-1} c^2} \,, \\
\mathsf O_B^{21} \quad \longrightarrow \quad &\Gpq{p^{-1} q^{-1} t c^2} \,, \\
\mathsf O_B^{22} \quad \longrightarrow \quad &\Gpq{p^{-1} q^{-1}  c^2} \,.
\end{aligned}
\end{align}
Adding the singlets and applying sequentially the Intriligator--Pouliot duality we obtain the index of the $E_{[2^3,1]}^{[2^3,1]}[USp(14)]$ theory:
\begin{equation}
\makebox[\linewidth][c]{\scalebox{0.95}{$
\begin{split}
&I_{E_{[2^3,1]}[USp(14)]} \left(\vec x^{(1)},x^{(2)};y^{(2)}, \vec y^{(1)};t,c\right)\\
&= \Gpq{p^4 q^4 t^{-3} c^{-2}} \Gpq{p^3 q^3 t^{-2} c^{-2}} \Gpq{p^3 q^3 t^{-1} c^{-2}}  \\
&\quad \times  \prod_{m = 1}^3 \Gpq{p^{-1/2} q^{-1/2} c y^{(1)}_2{}^{\pm1} x^{(1)}_m{}^{\pm1}} \prod_{m = 1}^3 \Gpq{p^{-1/2} q^{-1/2} c y^{(1)}_1{}^{\pm1} x^{(1)}_m{}^{\pm1}} \Gpq{t^{-1/2} c y^{(1)}_1{}^{\pm1} x^{(2)}_1{}^{\pm1}} \\
&\quad \times \oint \udl{\vec z^{(1)}_1} \udl{\vec z^{(2)}_3} \udl{\vec z^{(3)}_2} \Gpq{t}^6 \prod_{i < j}^3 \Gpq{t z^{(2)}_i{}^{\pm1} z^{(2)}_j{}^{\pm1}} \prod_{i < j}^2 \Gpq{t z^{(3)}_i{}^{\pm1} z^{(3)}_j{}^{\pm1}}  \\
&\quad \times \Gpq{p^{-3/2} q^{-3/2} t^{3/2} c z^{(1)}{}^{\pm1} y^{(2)}_1{}^{\pm1}} \prod_{j = 1}^3 \Gpq{p^{-1} q^{-1} t^{1/2} c y^{(1)}_3{}^{\pm1} z^{(2)}_j{}^{\pm1}} \prod_{j = 1}^2 \Gpq{p^{-1/2} q^{-1/2} c y^{(1)}_2{}^{\pm1} z^{(3)}_j{}^{\pm1}} \\
&\quad \times \prod_{j = 1}^3 \Gpq{p^{1/2} q^{1/2} t^{-1/2} z^{(1)}{}^{\pm1} z^{(2)}_j{}^{\pm1}} \prod_{i = 1}^3 \prod_{j = 1}^2 \Gpq{p^{1/2} q^{1/2} t^{-1/2} z^{(2)}_i{}^{\pm1} z^{(3)}_j{}^{\pm1}} \\
&\quad \times \Gpq{p^{3/2} q^{3/2} c^{-1} y^{(1)}_3{}^{\pm1} z^{(1)}{}^{\pm1}} \prod_{j = 1}^3 \Gpq{p q t^{1/2} c^{-1} y^{(1)}_2{}^{\pm1} z^{(2)}_j{}^{\pm1}} \prod_{j = 1}^2 \Gpq{p^{1/2} q^{1/2} t c^{-1} y^{(1)}_1{}^{\pm1} z^{(3)}_j{}^{\pm1}} \\
&\quad \times \prod_{i = 1}^3 \prod_{n = 1}^3 \Gpq{p^{1/2} q^{1/2} t^{-1/2} z^{(2)}_i{}^{\pm1} x^{(1)}_n{}^{\pm1}} \prod_{i = 1}^2 \Gpq{p^{1/2} q^{1/2} t^{-1/2} z^{(3)}_i{}^{\pm1} x^{(2)}_1{}^{\pm1}} \,,
\end{split}$}}
\end{equation}
from which one can read off the matter content and the superpotential. The matter content is conveniently represented using the quiver diagram, which is drawn in Figure \ref{fig:E2221}.
In particular we find the gauge singlets
\begin{align}
\begin{aligned}
\pi^{(2,3)} \quad &\longrightarrow \quad \prod_{m = 1}^3 \Gpq{p^{-1/2} q^{-1/2} c y^{(1)}_2{}^{\pm1} x^{(1)}_m{}^{\pm1}} \,, \\
\pi^{(2,4)} \quad &\longrightarrow \quad \prod_{m = 1}^3 \Gpq{p^{-1/2} q^{-1/2} c y^{(1)}_1{}^{\pm1} x^{(1)}_m{}^{\pm1}} \,, \\
\pi^{(3,4)} \quad &\longrightarrow \quad \Gpq{t^{-1/2} c y^{(1)}_1{}^{\pm1} x^{(2)}_1{}^{\pm1}} \,, \\
\gamma_{11} \quad &\longrightarrow \quad \Gpq{p^4 q^4 t^{-3} c^{-2}} \,, \\
\gamma_{21} \quad &\longrightarrow \quad \Gpq{p^3 q^3 t^{-2} c^{-2}} \,, \\
\gamma_{22} \quad &\longrightarrow \quad \Gpq{p^3 q^3 t^{-1} c^{-2}} \,,
\end{aligned}
\end{align}
and the superpotential
\begin{align}
&\mathcal W_{E_{[2^3,1]}[USp(14)]} = \nn \\
&\mathrm{Tr}_1 \mathrm{Tr}_2 \left[A^{(1)} Q^{(1,2)} Q^{(1,2)}\right] + \mathrm{Tr}_2 \left[A^{(2)} \left(\mathrm{Tr}_1 Q^{(1,2)} Q^{(1,2)}+\mathrm{Tr}_{x^{(1)}} F^{(2)} F^{(2)}+\mathrm{Tr}_3 Q^{(2,3)} Q^{(2,3)}\right)\right] \nn \\
&+ \mathrm{Tr}_3 \left[A^{(3)} \left(\mathrm{Tr}_2 Q^{(2,3)} Q^{(2,3)}+\mathrm{Tr}_{x^{(2)}} F^{(3)} F^{(3)}\right)\right] + \mathrm{Tr}_1 \mathrm{Tr}_2 \left[V^{(1)}_{[1} Q^{(1,2)} D^{(2)}_{2]}\right] + \mathrm{Tr}_2 \mathrm{Tr}_3 \left[V^{(2)}_{[1} Q^{(2,3)} D^{(3)}_{2]}\right] \nn \\
&+ \mathrm{Tr}_2 \mathrm{Tr}_{x^{(1)}} \left[F^{(2)} V^{(2)}_{[1} \pi^{(2,3)}_{2]}\right] + \mathrm{Tr}_2 \mathrm{Tr}_3 \mathrm{Tr}_{x^{(1)}} \left[F^{(2)} Q^{(2,3)} V^{(3)}_{[1} \pi^{(2,4)}_{2]}\right] + \mathrm{Tr}_3 \mathrm{Tr}_{x^{(2)}} \left[F^{(3)} V^{(3)}_{[1} \pi^{(3,4)}_{2]}\right] \nn \\
&+ \sum_{i = 1}^2 \sum_{j = 1}^i \gamma_{ij} \, \mathrm{Tr}_i \left[(A^{(i)})^{j-1} D^{(i)}_{[1} D^{(i)}_{2]}\right] \,.
\end{align}
where, as before, subscripts 1, 2 denote the flavor indices for the corresponding $SU(2)$ in the saw. This superpotential is perfectly consistent with the general form of the $E_\rho^\sigma[USp(2 N)]$ theory given by \eqref{superpoterhosigma}.

\section*{Acknowledgements}
We would like to thank  A.~Amariti, F.~Aprile, S.~Benvenuti, N.~Mekareeya, S.~Razamat and  A.~Zaffaroni  for helpful comments and discussions.
C.H. and S.P. are partially supported by the ERC-STG grant 637844-HBQFTNCER and  by  the  INFN.   M.S. is partially supported by the ERC-STG grant 637844-HBQFTNCER, by the University of Milano-Bicocca grant 2016-ATESP0586, by the MIUR-PRIN contract 2017CC72MK003 and by the INFN.

\appendix

\section{Partition function computations in \boldmath$3d$}

\subsection{Basic $3d$ dualities}
\label{funddualities}

In the main text we used intessively two basic $3d$ $\mathcal{N}=2$ dualities: Aharony duality and a variant with one monopole turned on in the superpotential. Both of these dualities can be derived from a more fundamental one, which was first proposed in \cite{Benini:2017dud}:

\medskip
\noindent \textbf{Theory A}: $U(N_c)$ gauge theory with $N_f$ flavors and superpotential $\mathcal{W}=\mathfrak{M}^++\mathfrak{M}^-$.

\medskip
\noindent \textbf{Theory B}: $U(N_f-N_c-2)$ gauge theory with $N_f$ flavors, $N_f^2$ singlets (collected in a matrix $M_{ab}$) and superpotential $\hat{\mathcal{W}}=\sum_{a,b=1}^{N_f}M_{ab}\tilde{q}_aq_b+\hat{\mathfrak{M}}^++\hat{\mathfrak{M}}^-$.

\medskip
\noindent 
The global symmetry of these theories is $SU(N_f)_m\times SU(N_f)_s$. Indeed, the monopole superpotential breaks both the axial and the topological symmetry. Moreover, requiring that the two fundamental monopoles of $U(N_c)$ are marginal we can fix the R-charges of all the chiral fields to $\frac{N_f-N_c-1}{N_f}$. At the level of $S^3_b$ partition functions, this duality translates into the following integral identity:
\be
\mathcal{Z}_{\mathcal{T}_A}&=&\frac{1}{N_c!}\int\prod_{i=1}^{N_c}\udl{x_i}\frac{\prod_{i=1}^{N_c}\prod_{a=1}^{N_f}s_b\left(i\frac{Q}{2}\pm(x_i+m_a)-s_a\right)}{\prod_{i<j}^{N_c}s_b\left(i\frac{Q}{2}\pm(x_i-x_j)\right)}\nn\\
&=&\frac{1}{(N_f-N_c-2)!}\prod_{a,b=1}^{N_f}s_b\left(i\frac{Q}{2}-(s_a+s_b-m_a+m_b)\right)\nn\\
&\times&\int\prod_{i=1}^{N_f-N_c-2}\udl{x_i}\frac{\prod_{i=1}^{N_f-N_c-2}\prod_{a=1}^{N_f}s_b\left(\pm(x_i-m_a)+s_a\right)}{\prod_{i<j}^{N_f-N_c-2}s_b\left(i\frac{Q}{2}\pm(x_i-x_j)\right)}=\mathcal{Z}_{\mathcal{T}_B}\, ,
\label{twomonopoles}
\ee
where $m_a$, $s_a$ are real masses in the Cartan subalgebra of the two $SU(N_f)$ flavor symmetries. Hence, the vector masses sum to zero $\sum m_a=0$, while the axial masses have to satisfy the following constraint due to the monopole superpotential:
\be
2\sum_{a=1}^{N_f}s_a=iQ(N_f-N_c-1)\, .
\label{constraintCONF2}
\ee

From this duality, we can derive the two that we actually need by performing suitable real mass deformations. The first one involves theories with only one monopole linearly turned on in the superpotential \cite{Benini:2017dud}:

\medskip
\noindent \textbf{Theory A}: $U(N_c)$ gauge theory with $N_f$ fundamental flavors and superpotential $\mathcal{W}=\mathfrak{M}^-$.

\medskip
\noindent \textbf{Theory B}: $U(N_f-N_c-1)$ gauge theory with $N_f$ fundamental flavors, $N_f^2$ singlets (collected in a matrix $M_{ab}$), an extra singlet $S^+$ and superpotential $\hat{\mathcal{W}}=\sum_{a,b=1}^{N_f}M_{ab}\tilde{q}_aq_b+\hat{\mathfrak{M}}^++S^+\hat{\mathfrak{M}}^-$.

\medskip
\noindent Implementing the real mass deformation on the partition functions, we get the following integral identity:
\be
\mathcal{Z}_{\mathcal{T}_A}&=&\frac{1}{N_c!}\int\prod_{i=1}^{N_c}\udl{x_i}\e^{i\pi\left(\sum_{i=1}^{N_c}x_i\right)(\eta-iQ)}\frac{\prod_{i=1}^{N_c}\prod_{a=1}^{N_f}s_b\left(i\frac{Q}{2}\pm(x_i+m_a)-s_a\right)}{\prod_{i<j}^{N_c}s_b\left(i\frac{Q}{2}\pm(x_j-x_i)\right)}\nn\\
&=&\frac{1}{(N_f-N_c-1)!}\e^{-i\pi\left(2\sum_{a=1}^{N_f}m_as_a+(\eta-iQ)\sum_{a=1}^{N_f}m_a\right)}s_b\left(i\frac{Q}{2}-\eta\right)\nn\\
&&\quad\times\prod_{a,b=1}^{N_f}s_b\left(i\frac{Q}{2}-(s_a+s_b-m_a+m_b)\right)\nn\\
&&\quad\times\int\prod_{i=1}^{N_f-N_c-1}\udl{x_i}\e^{i\pi\eta\sum_{i=1}^{N_c}x_i}\frac{\prod_{i=1}^{N_f-N_c-1}\prod_{a=1}^{N_f}s_b\left(\pm(x_i-m_a)+s_a\right)}{\prod_{i<j}^{N_f-N_c-1}s_b\left(i\frac{Q}{2}\pm(x_j-x_i)\right)}=\mathcal{Z}_{\mathcal{T}_B}\, ,\nn\\
\label{onemonopole}
\ee
where $\eta$ is the real mass for a restored combination of the topological and the axial symmetry. The condition of the monopole superpotential is now
\be
\eta+2\sum_{a=1}^{N_f}s_a=iQ(N_f-N_c)\, .
\label{constraintM-}
\ee
Finally, we can perform a further real mass deformation that leads to Aharony duality \cite{Aharony:1997gp}:

\medskip
\noindent \textbf{Theory A}: $U(N_c)$ gauge theory with $N_f$ flavors and superpotential $\mathcal{W}=0$.

\medskip
\noindent \textbf{Theory B}: $U(N_f-N_c)$ gauge theory with $N_f$ flavors, $N_f^2$ singlets (collected in a matrix $M_{ab}$), two extra singlets $S^{\pm}$ and superpotential $\hat{\mathcal{W}}=\sum_{a,b=1}^{N_f}M_{ab}\tilde{q}_aq_b+S^-\hat{\mathfrak{M}}^++S^+\hat{\mathfrak{M}}^-$.

\medskip
\noindent The equality of partition functions of the dual theories is
\be
\mathcal{Z}_{\mathcal{T}_A}&=&\frac{1}{N_c!}\int\prod_{i=1}^{N_c}\udl{x_i}\e^{i\pi\xi\left(\sum_{i=1}^{N_c}x_i\right)}\frac{\prod_{i=1}^{N_c}\prod_{a=1}^{N_f}s_b\left(i\frac{Q}{2}\pm(x_i+m_a)-s_a\right)}{\prod_{i<j}^{N_c}s_b\left(i\frac{Q}{2}\pm(x_j-x_i)\right)}\nn\\
&=&\e^{-i\pi\xi\sum_{a=1}^{N_f}m_a}s_b\left(i\frac{Q}{2}-\frac{iQ(N_f-N_c+1)-2\sum_{a=1}^{N_f}s_a\pm\xi}{2}\right)\nn\\
&\times&\prod_{a,b=1}^{N_f}s_b\left(i\frac{Q}{2}-(s_a+s_b-m_a+m_b)\right)\nn\\
&\times&\frac{1}{(N_f-N_c)!}\int\prod_{i=1}^{N_f-N_c}\udl{x_i}\e^{i\pi\xi\sum_{i=1}^{N_c}x_i}\frac{\prod_{i=1}^{N_f-N_c}\prod_{a=1}^{N_f}s_b\left(\pm(x_i-m_a)+s_a\right)}{\prod_{i<j}^{N_f-N_c}s_b\left(i\frac{Q}{2}\pm(x_j-x_i)\right)}=\mathcal{Z}_{\mathcal{T}_B}\, ,\nn\\
\label{aha}
\ee
where $\xi$ is the FI parameter for the restored topological symmetry, while $\sum_as_a=s$ with $s$ being the axial mass.

\subsection{$3d$ flip-flip duality as repeated Aharony duality}
\label{appflipflipaha}

In this appendix we explicitly show that, when the theory has no monopole superpotential, the flip-flip duality is equivalent to sequentially applying the Aharony duality. At the level of  the $S^3_b$ partition function, we sequentially apply the integral identity \eqref{aha} for Aharony duality. We first consider the flip-flip duality between $T[SU(3)]$ and $FFT[SU(3)]$ and then the deformation of $T[SU(3)]^\vee$ labelled by the partition $\gr=[2,1]$ which leads to $T^{[2,1]}[SU(3)]$.

\subsubsection{Derivation of $T[SU(3)]\leftrightarrow FFT[SU(3)]$}\label{ff3}

Let us consider the partition function of $T[SU(3)]$
\be
\mathcal{Z}_{T[SU(3)]}&=&\int\udl{\vec{z}_2^{(2)}}\e^{2\pi i(Y_2-Y_3)\sum_{i=1}^2z^{(2)}_i}\prod_{i,j=1}^2\sbfunc{-i\frac{Q}{2}+(z^{(2)}_i-z^{(2)}_j)+2m_A}\nn\\
&\times&\prod_{i=1}^2\prod_{n=1}^3\sbfunc{i\frac{Q}{2}\pm(z^{(2)}_i-X_n)-m_A}\int\udl{z_1^{(1)}}\e^{2\pi i(Y_1-Y_2)z^{(1)}}\sbfunc{-i\frac{Q}{2}+2m_A}\nn\\
&\times&\prod_{i=1}^2\sbfunc{i\frac{Q}{2}\pm(z^{(1)}-z^{(2)}_i)-m_A}\, .
\ee
In order to get the partition function of $FFT[SU(3)]$ we have to apply Aharony duality $2+1=3$ times. At the first iteration, we first apply it to the $U(1)$ gauge node associated to the $z^{(1)}$ integration variable
\begin{equation}
\makebox[\linewidth][c]{\scalebox{0.95}{$
\begin{split}
\mathcal{Z}_{T[SU(3)]}&=\sbfunc{-i\frac{Q}{2}+2m_A}\sbfunc{-i\frac{Q}{2}\pm(Y_1-Y_2)+2m_A}\int\udl{\vec{z}_2^{(2)}}\e^{2\pi i(Y_1-Y_3)\sum_{i=1}^2z^{(2)}_i}\\
&\times\prod_{i=1}^2\prod_{n=1}^3\sbfunc{i\frac{Q}{2}\pm(z^{(2)}_i-X_n)-m_A}\int\udl{z_1^{(1)}}\e^{2\pi i(Y_1-Y_2)z^{(1)}}\prod_{i=1}^2\sbfunc{\pm(z^{(1)}+z^{(2)}_i)+m_A}
\end{split}$}}
\end{equation}
and then to the $U(2)$ gauge node associated to the $z^{(2)}_i$ integration variable
\be
\mathcal{Z}_{T[SU(3)]}&=&\e^{2\pi i(Y_1-Y_3)\sum_{n=1}^3X_n}\sbfunc{-i\frac{Q}{2}+2m_A}^2\sbfunc{-i\frac{Q}{2}\pm(Y_1-Y_2)+2m_A}\nn\\
&\times&\sbfunc{-i\frac{Q}{2}\pm(Y_1-Y_3)+2m_A}\prod_{n,m=1}^3\sbfunc{i\frac{Q}{2}+(X_n-X_m)-2m_A}\nn\\
&\times&\int\udl{\vec{z}_2^{(2)}}\e^{2\pi i(Y_1-Y_3)\sum_{i=1}^2z^{(2)}_i}\prod_{i=1}^2\prod_{n=1}^3\sbfunc{\pm(z^{(2)}_i+X_n)+m_A}\nn\\
&\times&\int\udl{z_1^{(1)}}\e^{2\pi i(Y_3-Y_2)z^{(1)}}\prod_{i=1}^2\sbfunc{i\frac{Q}{2}\pm(z^{(1)}-z^{(2)}_i)-m_A}\,.
\ee
The second iteration only consists of applying Aharony duality once, again to the $U(1)$ gauge node
\be
\mathcal{Z}_{T[SU(3)]}&=&\e^{2\pi i(Y_1-Y_3)\sum_{n=1}^3X_n}\sbfunc{-i\frac{Q}{2}+2m_A}^2\sbfunc{-i\frac{Q}{2}\pm(Y_1-Y_2)+2m_A}\nn\\
&\times&\sbfunc{-i\frac{Q}{2}\pm(Y_1-Y_3)+2m_A}\sbfunc{-i\frac{Q}{2}\pm(Y_2-Y_3)+2m_A}\nn\\
&\times&\prod_{n,m=1}^3\sbfunc{i\frac{Q}{2}+(X_n-X_m)-2m_A}\int\udl{\vec{z}_2^{(2)}}\e^{2\pi i(Y_1-Y_2)\sum_{i=1}^2z^{(2)}_i}\nn\\
&\times&\prod_{i,j=1}^2\sbfunc{i\frac{Q}{2}\pm(z^{(2)}_i-z^{(2)}_j)-2m_A}\prod_{i=1}^2\prod_{n=1}^3\sbfunc{\pm(z^{(2)}_i+X_n)+m_A}\nn\\
&\times&\int\udl{z_1^{(1)}}\e^{2\pi i(Y_3-Y_2)z^{(1)}}\prod_{i=1}^2\sbfunc{\pm(z^{(1)}+z^{(2)}_i)-m_A}\,.
\ee
Re-arranging the contribution of the singlets in the prefactor, imposing the tracelessness condition $\sum_{n=1}^3X_n=0$ and performing the change of variables $z^{(2)}_i\to-z^{(2)}_i$, we get
\be
\mathcal{Z}_{T[SU(3)]}&=&\prod_{n,m=1}^3\sbfunc{i\frac{Q}{2}+(X_n-X_m)-2m_A}\sbfunc{-i\frac{Q}{2}+(Y_n-Y_m)+2m_A}\nn\\
&\times&\int\udl{\vec{z}_2^{(2)}}\e^{2\pi i(Y_2-Y_1)\sum_{i=1}^2z^{(2)}_i}\prod_{i,j=1}^2\sbfunc{i\frac{Q}{2}\pm(z^{(2)}_i-z^{(2)}_j)-2m_A}\nn\\
&\times&\prod_{i=1}^2\prod_{n=1}^3\sbfunc{\pm(z^{(2)}_i-X_n)+m_A}\int\udl{z_1^{(1)}}\e^{2\pi i(Y_3-Y_2)z^{(1)}}\nn\\
&\times&\prod_{i=1}^2\sbfunc{\pm(z^{(1)}-z^{(2)}_i)-m_A}\,.
\ee
This is precisely the partition function of $FFT[SU(3)]$
up to the exchange $Y_1\leftrightarrow Y_3$, which is just an element of the Weyl group of $SU(3)_Y$ that acts trivially on the partition function. Hence, we get \eqref{idflipflip} in the particular case $N=3$
\be
\mathcal{Z}_{T[SU(3)]}(\vec X;\vec Y;m_A)&=&\prod_{n,m=1}^3\sbfunc{i\frac{Q}{2}+(X_n-X_m)-2m_A}\sbfunc{-i\frac{Q}{2}+(Y_n-Y_m)+2m_A}\nn\\
&\times&\mathcal{Z}_{T[SU(3)]}\left(\vec X;\vec Y;i\frac{Q}{2}-m_A\right)\,.
\ee
With the same strategy, one can derive flip-flip duality for \tsu with arbitrary rank $N$.

\subsubsection{The case $\gr=[2,1]$}

The starting point of the computation is the partition function of $T[SU(3)]$, to which we have to impose the constraint on the real masses \eqref{solconstrN-11} due to the superpotential deformation \eqref{deltaWN-11}
\be
Y_2=Y_1+2m_A\, .
\ee
We know that the effect of the massive deformation \eqref{deltaWN-11} is of making some of the flavors at the end of the tail of $T[SU(3)]^\vee$ massive. This is realized at the level of the partition function using the identity $\sbfunc{x}\sbfunc{-x}=1$. Denoting with $z_i^{(2)}$ the integration variables of the $U(2)$ gauge node, we have
\begin{equation}
\makebox[\linewidth][c]{\scalebox{0.95}{$
\begin{split}
\prod_{n=1}^3\sbfunc{\pm(z_i^{(2)}-Y_n)+m_A}&=\sbfunc{z_i^{(2)}-Y_1+m_A}\sbfunc{-z_i^{(2)}+Y_1+3m_A}\sbfunc{\pm(z_i^{(2)}-Y_3)+m_A}\\
&\rightarrow\sbfunc{\pm(z_i^{(2)}-Y_1)+2m_A}\sbfunc{\pm(z_i^{(2)}-Y_3)+m_A}\, ,
\end{split}$}}
\end{equation}
where at the last step we redefined
\be
Y_1\rightarrow Y_1-m_A\, .
\label{Y1shift2}
\ee
Hence, the partition function of theory $\mathcal{T}^\vee$ is
\begin{equation}
\makebox[\linewidth][c]{\scalebox{0.95}{$
\begin{split}
\mathcal{Z}_{\mathcal{T}^\vee}&=\mathcal{B}\int\udl{\vec z_2^{(2)}}\e^{2\pi i(X_2-X_3)\sum_{i=1}^2z_i^{(2)}}\prod_{i,j=1}^2\sbfunc{i\frac{Q}{2}+(z_i^{(2)}-z_j^{(2)})-2m_A}\prod_{i=1}^2\sbfunc{\pm(z_i^{(2)}-Y_1)+2m_A}\\
&\times\sbfunc{\pm(z_i^{(2)}-Y_3)+m_A}\int\udl{z_1^{(1)}}\e^{2\pi i(X_1-X_2)z^{(1)}}\sbfunc{i\frac{Q}{2}-2m_A}\prod_{i=1}^2\sbfunc{\pm(z^{(1)}-z_i^{(2)})+m_A}\, ,
\end{split}$}}
\end{equation}
where $\mathcal{B}$ denotes the contribution of the flipping fields $\mathcal{S}_{[1^3]}$ and $\mathcal{T}_i,\mathcal{T},\tilde{\mathcal{T}}$ contained in $\mathcal{T}_{[2,1]}$
\be
\mathcal{B}&=&\sbfunc{i\frac{Q}{2}-2m_A}^2\sbfunc{i\frac{Q}{2}-4m_A}\sbfunc{i\frac{Q}{2}\pm(Y_1-Y_3)-3m_A}\nn\\
&\times&\prod_{n,m=1}^3\sbfunc{-i\frac{Q}{2}+(X_n-X_m)+2m_A}
\ee
Since the adjoint chiral field at the $U(1)$ node is just a singlet, we can apply Aharony duality at this node. Using \eqref{aha} we find
\be
\mathcal{Z}_{\mathcal{T}^\vee}&=&\mathcal{B}\sbfunc{i\frac{Q}{2}-2m_A}\sbfunc{i\frac{Q}{2}\pm(X_1-X_2)-2m_A}\int\udl{\vec z_2^{(2)}}\e^{2\pi i(X_1-X_3)\sum_{i=1}^2z_i^{(2)}}\nn\\
&\times&\prod_{i=1}^2\sbfunc{\pm(z_i^{(2)}-Y_1)+2m_A}\sbfunc{\pm(z_i^{(2)}-Y_3)+m_A}\int\udl{z_1^{(1)}}\e^{2\pi i(X_1-X_2)z^{(1)}}\nn\\
&\times&\prod_{i=1}^2\sbfunc{i\frac{Q}{2}\pm(z^{(1)}+z_i^{(2)})-m_A}\, .\nn\\
\ee
This had the effect of removing the adjoint chiral of the adjacent $U(2)$ gauge node, so now we can apply Aharony duality to it. In this case the rank of the group gets lowered by one unit
\be
\mathcal{Z}_{\mathcal{T}^\vee}&=&\mathcal{B}\,\e^{2\pi i(X_1-X_3)(Y_1+Y_3)}\sbfunc{i\frac{Q}{2}-2m_A}\sbfunc{-i\frac{Q}{2}+4m_A}\sbfunc{i\frac{Q}{2}\pm(X_1-X_2)-2m_A}\nn\\
&\times&\sbfunc{i\frac{Q}{2}\pm(X_1-X_3)-2m_A}\sbfunc{-i\frac{Q}{2}\pm(Y_1-Y_3)+3m_A}\int\udl{z_1^{(2)}}\e^{2\pi i(X_1-X_3)z^{(2)}}\nn\\
&\times&\sbfunc{i\frac{Q}{2}\pm(z^{(2)}+Y_1)-2m_A}\sbfunc{i\frac{Q}{2}\pm(z^{(2)}+Y_3)-m_A}\int\udl{z_1^{(1)}}\e^{2\pi i(X_3-X_2)z^{(1)}}\nn\\
&\times&\sbfunc{\pm(u+Y_1)+m_A}\sbfunc{\pm(z^{(1)}-z_i^{(2)})+m_A}\, .\nn\\
\ee
The last step of the computation consists of applying Aharony duality on the first $U(1)$ node once again. The various flipping fields produced in the derivation perfectly cancel with those contained in the prefactor $\mathcal{B}$ and we get
\be
\mathcal{Z}_{\mathcal{T}^\vee}&=&\e^{2\pi i(X_1+X_2-2X_3)Y_1}\e^{2\pi i(X_1-X_3)Y_3}\int\udl{z_1^{(2)}}\e^{2\pi i(X_2-X_1)z^{(2)}}\sbfunc{-i\frac{Q}{2}+2m_A}\nn\\
&\times&\sbfunc{i\frac{Q}{2}\pm(z^{(2)}-Y_3)-m_A}\int\udl{z_1^{(1)}}\e^{2\pi i(X_3-X_2)z^{(1)}}\sbfunc{-i\frac{Q}{2}+2m_A}\nn\\
&\times&\sbfunc{i\frac{Q}{2}\pm(u-Y_1)-m_A}\sbfunc{i\frac{Q}{2}\pm(z^{(1)}-z_i^{(2)})-m_A}\, .
\ee
At this point we recall that $Y_1$ and $Y_3$ are not independent variables because of the original tracelessness condition $\sum_{n=1}^3Y_n=0$, which after the constraint \eqref{solconstrN-11} and the shift \eqref{Y1shift} becomes
\be
2Y_1+Y_3=0\,.
\ee
We parametrize the residual $U(1)_{Y^{(1)}}$ symmetry with
\be
Y^{(1)}=Y_1-Y_3\,
\ee
and we also perform the change of variables $z^{(i)}\to z^{(i)}+Y^{(1)}/3$, so that
\be
\mathcal{Z}_{\mathcal{T}^\vee}&=&\e^{-2\pi i X_1 Y^{(1)}}\int\udl{z_1^{(2)}}\e^{2\pi i(X_2-X_1)z^{(2)}}\sbfunc{-i\frac{Q}{2}+2m_A}\nn\\
&\times&\sbfunc{i\frac{Q}{2}\pm\left(z^{(2)}-\frac{Y^{(1)}}{3}\right)-m_A}\int\udl{z_1^{(1)}}\e^{2\pi i(X_3-X_2)z^{(1)}}\sbfunc{-i\frac{Q}{2}+2m_A}\nn\\
&\times&\sbfunc{i\frac{Q}{2} \pm z^{(1)}-m_A}\sbfunc{i\frac{Q}{2}\pm(z^{(1)}-z_i^{(2)})-m_A}=\mathcal{Z}_{T^{[2,1]}[SU(3)]}\, .\nn\\
\ee
This coincides with the partition function $T^{[2,1]}[SU(3)]$ which, from the deformation of the duality web of $T[SU(N)]$, we expect to be flip-flip dual to theory $\mathcal{T}$. The real masses $X_n$ correspond to the $SU(3)_X$ global symmetry of $T^{[2,1]}[SU(3)]$ that enhances from the $U(1)^2$ topological symmetry that is manifest in the UV. Instead, the flavor symmetry of $T^{[2,1]}[SU(3)]$ is $U(1)_{Y^{(1)}}$.
Hence, we showed that flip-flip duality is equivalent to sequentially applying Aharony duality. 

\subsection{Derivation of the partition functions of $T_{[2,1^2]}[SU(4)]$ and its mirror dual}
\label{appA.2}

\subsubsection*{Flow to $T_{[2,1^2]}[SU(N)]$}
As discussed in Section \ref{trhosigmaweb}, the vev for the CB moment map of $T[SU(4)]$ can be studied as a linear superpotential in  $FFT[SU(4)]$ or, using flip-flip duality, as a monopole deformation of $T[SU(4)]$ with the addition of extra singlet fields flipping the components of the HB and CB moment maps that remain free after the vev. Hence, in our computation we start from the partition function \eqref{tsunPF} of $T[SU(4)]$, impose the constraint on the fugacities
\be
Y_2=Y_1+2m_A
\ee
due to the monopole deformation \eqref{superpotT211CB}, as well as the redefinition
\be
Y_1\rightarrow Y_1-m_A
\ee
and add the contribution of the flipping fields
\be
&&\mathcal{Z}_{\mathcal{T}}=\mathcal{B}\int\udl{\vec{z}_3^{(3)}}\e^{2\pi i(Y_3-Y_4)\sum_{i=1}^3z_i^{(3)}}\prod_{i,j=1}^3\sbfunc{-i\frac{Q}{2}+(z_i^{(3)}-z_j^{(3)})+2m_A}\nn\\
&&\qquad\times\prod_{i=1}^3\prod_{n=1}^4\sbfunc{i\frac{Q}{2}\pm(z_i^{(3)}-X_n)-m_A}\int\udl{\vec{z}_2^{(2)}}\e^{2\pi i(Y_1-Y_3+m_A)\sum_{a=1}^2z_a^{(2)}}\nn\\
&&\qquad\times\prod_{a,b=1}^2\sbfunc{-i\frac{Q}{2}+(z_a^{(2)}-z_b^{(2)})+2m_A}\prod_{a=1}^2\prod_{i=1}^3\sbfunc{i\frac{Q}{2}\pm(z_a^{(2)}-z_i^{(3)})-m_A}\nn\\
&&\qquad\times\int\udl{z_1^{(1)}}\e^{-4\pi im_Az^{(1)}}\sbfunc{-i\frac{Q}{2}+2m_A}\prod_{a=1}^2\sbfunc{i\frac{Q}{2}\pm(z^{(1)}-z_a^{(2)})-m_A}\, ,
\ee
where $\mathcal{B}$ is the contribution of the singlets
\be
\mathcal{B}&=&\prod_{n,m=1}^4\sbfunc{-i\frac{Q}{2}+(X_n-X_m)+2m_A}\sbfunc{i\frac{Q}{2}-2m_A}\sbfunc{i\frac{Q}{2}-4m_A}\nn\\
&\times&\prod_{\ga,\gb=3}^4\sbfunc{i\frac{Q}{2}+(Y_\ga-Y_\gb)-2m_A}\prod_{\ga=3}^4\sbfunc{i\frac{Q}{2}\pm(Y_1-Y_\ga)-3m_A}\, .
\ee
As we explained in Section \ref{sec2.4.1} we first apply the integral identity for the one-monopole duality \eqref{onemonopole} to the $U(1)$ gauge node where the monopole superpotential is turned on. In this way, this node confines and we get the partition function of a dual frame of theory $\mathcal{T}$ where we have no monopole superpotential
\be
\mathcal{Z}_{\mathcal{T}}&=&\mathcal{B}\,\sbfunc{-i\frac{Q}{2}+2m_A}\sbfunc{-i\frac{Q}{2}+4m_A}\int\udl{\vec{z}_3^{(3)}}\e^{2\pi i(Y_3-Y_4)\sum_{i=1}^3z_i^{(3)}}\nn\\
&\times&\prod_{i,j=1}^3\sbfunc{-i\frac{Q}{2}+(z_i^{(3)}-z_j^{(3)})+2m_A}\prod_{i=1}^3\prod_{n=1}^4\sbfunc{i\frac{Q}{2}\pm(z_i^{(3)}-X_n)-m_A}\nn\\
&\times&\int\udl{\vec{z}_2^{(2)}}\e^{2\pi i(Y_1-Y_3)\sum_{a=1}^2z_a^{(2)}}\prod_{a=1}^2\prod_{i=1}^3\sbfunc{i\frac{Q}{2}\pm(z_a^{(2)}-z_i^{(3)})-m_A}\, .
\ee
In order to find the flip-flip dual of $\mathcal{T}$,  we now have to sequentially apply the integral identity for Aharony duality \eqref{aha}. First we apply the duality to the $U(2)$ gauge node, whose rank decreases by one since we confined the previous node
\be
&&\mathcal{Z}_{\mathcal{T}}=\mathcal{B}\,\sbfunc{-i\frac{Q}{2}+2m_A}\sbfunc{-i\frac{Q}{2}+4m_A}\sbfunc{-i\frac{Q}{2}\pm(Y_1-Y_3)+3m_A}\nn\\
&&\qquad\quad\times\int\udl{\vec{z}_3^{(3)}}\e^{2\pi i(Y_1-Y_4)\sum_{i=1}^3z_i^{(3)}}\prod_{i=1}^3\prod_{n=1}^4\sbfunc{i\frac{Q}{2}\pm(z_i^{(3)}-X_n)-m_A}\nn\\
&&\qquad\quad\times\int\udl{z_1^{(2)}}\e^{2\pi i(Y_1-Y_3)z^{(2)}}\prod_{i=1}^3\sbfunc{\pm(z^{(2)}+z_i^{(3)})+m_A}\, .
\ee
Now we can apply Aharony duality on the $U(3)$ gauge node since its adjoint chiral became massive and was integrated out. The rank of the node decreases to two and we get
\begin{equation}
\makebox[\linewidth][c]{\scalebox{0.95}{$
\begin{split}
&\mathcal{Z}_{\mathcal{T}}=\mathcal{B}\,\e^{2\pi i(Y_1-Y_4)\sum_{n=1}X_n}\sbfunc{-i\frac{Q}{2}+2m_A}^2\sbfunc{-i\frac{Q}{2}+4m_A}\sbfunc{-i\frac{Q}{2}\pm(Y_1-Y_3)+3m_A}\\
&\qquad\times\sbfunc{-i\frac{Q}{2}\pm(Y_1-Y_4)+3m_A}\prod_{n,m=1}^4\sbfunc{i\frac{Q}{2}+(X_n-X_m)-2m_A}\int\udl{\vec{z}_2^{(3)}}\e^{2\pi i(Y_1-Y_4)\sum_{i=1}^2z_i^{(3)}}\\
&\qquad\times\prod_{i=1}^3\prod_{n=1}^4\sbfunc{\pm(z_i^{(3)}+X_n)+m_A}\int\udl{z_1^{(2)}}\e^{2\pi i(Y_4-Y_3)z^{(2)}}\prod_{i=1}^3\sbfunc{i\frac{Q}{2}\pm(z^{(2)}-z_i^{(3)})-m_A}\, .
\end{split}$}}
\end{equation}
Finally, we apply Aharony duality to the $U(1)$ gauge node. Simplifying the contributions of the singlets we produced in the derivation of the flip-flip dual with those contained in the prefactor $\mathcal{B}$ and performing the change of variable $z^{(2)}\rightarrow -z^{(2)}$ we get
\begin{equation}
\makebox[\linewidth][c]{\scalebox{0.95}{$
\begin{split}
&\mathcal{Z}_{\mathcal{T}}=\e^{2\pi i (Y_1-Y_4)\sum_{n=1}^4X_n}\int\udl{\vec{z}_2^{(3)}}\e^{2\pi i(Y_1-Y_3)\sum_{i=1}^2z_i^{(3)}}\prod_{i,j=1}^2\sbfunc{i\frac{Q}{2}+(z_i^{(3)}-z_j^{(3)})-m_A}\\
&\quad\times\prod_{i=1}^2\prod_{n=1}^4\sbfunc{\pm(z_i^{(3)}+X_n)+m_A}\int\udl{z^{(2)}_1}\e^{2\pi i(Y_3-Y_4)z^{(2)}}\sbfunc{i\frac{Q}{2}-2m_A}\prod_{i=1}^2\sbfunc{\pm(z^{(2)}-z_i^{(3)})+m_A}\, .
\end{split}$}}
\end{equation}
Notice that the contact term is actually trivial, since the $X_n$ parameters still parametrize the Cartan of the $SU(4)_X$ HB global symmetry. 
Moreover, we should recall that the original $Y_n$ real masses were parametrizing the $SU(4)_Y$ CB global symmetry of $T[SU(4)]$, meaning that $\sum_{n=1}^4Y_n=0$. After imposing the condition $Y_2=Y_1+2m_A$ and redefining $Y_1\rightarrow Y_1-m_A$, this translates into a condition for the real masses $Y_1$, $Y_\ga$ of the remaining $U(1)\times SU(2)$ CB global symmetry
\be
2Y_1+\sum_{\ga=3}^4Y_\ga=0\, .
\ee
This means that the proper $U(1)_{Y^{(1)}}\times SU(2)_{Y^{(2)}}$ fugacities are
\be
&&Y^{(1)}=Y_1\nn\\
&&Y^{(2)}_1=Y_3+Y_1\nn\\
&&Y^{(2)}_2=Y_4+Y_1\,,
\label{newfug211}
\ee
so that $\sum_{\ga=1}^2Y^{(2)}_\ga=0$. After this shift, we get
\be
&&\mathcal{Z}_{\mathcal{T}}=\int\udl{\vec{z}_2^{(3)}}\e^{2\pi i(2Y^{(1)}-Y^{(2)}_1)\sum_{i=1}^2z_i^{(3)}}\prod_{i,j=1}^2\sbfunc{i\frac{Q}{2}+(z_i^{(3)}-z_j^{(3)})-m_A}\nn\\
&&\quad\times\prod_{i=1}^2\prod_{n=1}^4\sbfunc{\pm(z_i^{(3)}+X_n)+m_A}\int\udl{z_1^{(2)}}\e^{2\pi i(Y^{(2)}_1-Y^{(2)}_2)z^{(2)}}\sbfunc{i\frac{Q}{2}-2m_A}\nn\\
&&\quad\times\prod_{i=1}^2\sbfunc{\pm(z^{(2)}-z_i^{(3)})+m_A}=\mathcal{Z}_{T_{[2,1^2]}[SU(4)]}(\vec X;\vec Y^{(2)},Y^{(1)};m_A)\, ,
\label{pf211}
\ee
where the contact term disappeared because of the tracelessness condition $\sum_{n=1}^4X_n=0$. This is precisely the partition function of $T_{[2,1^2]}[SU(4)]$, whose global symmetry is indeed $SU(4)_X\times U(1)_{Y^{(1)}}\times SU(2)_{Y^{(2)}}$ with the CB factor $U(1)_{Y^{(1)}}\times SU(2)_{Y^{(2)}}$ being enhanced at low energies.

\subsubsection*{Flow to $T^{[2,1^2]}[SU(4)]$}

As discussed in Section \ref{trhosigmaweb}, on the mirror dual side we should consider the vev for the HB moment map of $T[SU(4)]^\vee$, which can be studied as a linear superpotential in  $FFT[SU(4)]^\vee$ or, using flip-flip duality, as a mass deformation of $T[SU(4)]^\vee$ with the addition of extra singlet fields flipping the components of the HB and CB moment maps that remain free after the vev. Hence, in our computation we start from the partition function $T[SU(4)]^\vee$ and impose the constraint on the fugacities $Y_2=Y_1+2m_A$ due to the mass deformation. Using the relation $\sbfunc{x}\sbfunc{-x}=1$, we have that the contribution of some of the chiral fields attached to the last $U(3)$ gauge node cancel each other, meaning that they have become massive fields. Denoting with $z_i^{(3)}$ the integration variables of the $U(3)$ gauge node, we have
\be
\prod_{n=3}^4\sbfunc{\pm(z_i^{(3)}-Y_n)+m_A}&=&\prod_{\ga=3}^4\sbfunc{\pm(z_i^{(3)}-Y_n)+m_A}\sbfunc{z_i^{(3)}-Y_1+m_A}\nn\\
&\times&\sbfunc{-z_i^{(3)}+Y_1+3m_A}\nn\\
&\rightarrow&\prod_{\ga=3}^4\sbfunc{\pm(z_i^{(3)}-Y_n)+m_A}\sbfunc{\pm(z_i^{(3)}-Y_1)+2m_A}\, ,\nn\\
\ee
where at the last step we redefined
\be
Y_1\rightarrow Y_1-m_A\,.
\ee
Thus, the starting point of our computation is the partition function of theory $\mathcal{T}^\vee$
\begin{equation}
\makebox[\linewidth][c]{\scalebox{1}{$
\begin{split}
&\mathcal{Z}_{\mathcal{T}^\vee}=\mathcal{B}\int\udl{\vec{z}_3^{(3)}}\e^{2\pi i(X_3-X_4)\sum_{i=1}^3z_i^{(3)}}\prod_{i,j=1}^3\sbfunc{i\frac{Q}{2}+(z_i^{(3)}-z_j^{(3)})-2m_A}\\
&\qquad\times\prod_{i=1}^3\sbfunc{\pm(z_i^{(3)}-Y_1)+2m_A}\prod_{\ga=3}^4\sbfunc{\pm(z_i^{(3)}-Y_\ga)+m_A}\int\udl{\vec{z}_2^{(2)}}\e^{2\pi i(X_2-X_3)\sum_{a=1}^2z_a^{(2)}}\\
&\qquad\times\prod_{a,b=1}^2\sbfunc{i\frac{Q}{2}+(z_a^{(2)}-z_b^{(2)})-2m_A}\prod_{a=1}^2\prod_{i=1}^3\sbfunc{\pm(z_a^{(2)}-z_i^{(3)})+m_A}\int\udl{z_1^{(1)}}\e^{2\pi i(X_1-X_2)z^{(1)}}\\
&\qquad\times\sbfunc{i\frac{Q}{2}-2m_A}\prod_{a=1}^2\sbfunc{\pm(z^{(1)}-z_a^{(2)})+m_A}\, .
\end{split}$}}
\end{equation}
Again we claim that in order to reach the flip-flip dual frame which corresponds to $T^{[2,1^2]}[SU(4)]$, we can iteratively apply the integral identity for Aharony duality \eqref{aha}. We start from the $U(1)$ gauge node since its adjoint chiral field is just a singlet. This node has two flavors attached to it, so it remains a $U(1)$ node and we get
\begin{equation}
\makebox[\linewidth][c]{\scalebox{1}{$
\begin{split}
&\mathcal{Z}_{\mathcal{T}^\vee}=\mathcal{B}\,\sbfunc{i\frac{Q}{2}-2m_A}\sbfunc{i\frac{Q}{2}\pm(X_1-X_2)-2m_A}\int\udl{\vec{z}_3^{(3)}}\e^{2\pi i(X_3-X_4)\sum_{i=1}^3z^{(3)}_i}\\
&\qquad\times\prod_{i,j=1}^3\sbfunc{i\frac{Q}{2}+(z_i^{(3)}-z_j^{(3)})-2m_A}\prod_{i=1}^3\sbfunc{\pm(z_i^{(3)}-Y_1)+2m_A}\prod_{\ga=3}^4\sbfunc{\pm(z_i^{(3)}-Y_\ga)+m_A}\\
&\qquad\times\int\udl{\vec{z}_2^{(2)}}\e^{2\pi i(X_1-X_3)\sum_{a=1}^2z_a^{(2)}}\prod_{a=1}^2\prod_{i=1}^3\sbfunc{\pm(z_a^{(2)}-z_i^{(3)})+m_A}\int\udl{z_1^{(1)}}\e^{2\pi i(X_1-X_2)z^{(1)}}\\
&\qquad\times\prod_{a=1}^2\sbfunc{i\frac{Q}{2}\pm(z^{(1)}+z_a^{(2)})-m_A}\, .
\end{split}$}}
\end{equation}
Notice that this application of Aharony duality had the effect of removing the adjoint chiral field for the next $U(2)$ gauge node, which allows us to apply the duality again on this second node.
This is a $U(2)$ gauge node with four flavors, so its rank doesn't change
\begin{equation}
\makebox[\linewidth][c]{\scalebox{1}{$
\begin{split}
&\mathcal{Z}_{\mathcal{T}^\vee}=\mathcal{B}\,\sbfunc{i\frac{Q}{2}-2m_A}^2\sbfunc{i\frac{Q}{2}\pm(X_1-X_2)-2m_A}\sbfunc{i\frac{Q}{2}\pm(X_1-X_3)-2m_A}\\
&\qquad\times\int\udl{\vec{z}_3^{(3)}}\e^{2\pi i(X_1-X_4)\sum_{i=1}^3z_i^{(3)}}\prod_{i=1}^3\sbfunc{\pm(z_i^{(3)}-Y_1)+2m_A}\prod_{\ga=3}^4\sbfunc{\pm(z_i^{(3)}-Y_\ga)+m_A}\\
&\qquad\times\int\udl{\vec{z}_2^{(2)}}\e^{2\pi i(X_1-X_3)\sum_{a=1}^2z_a^{(2)}}\prod_{a=1}^2\prod_{i=1}^3\sbfunc{i\frac{Q}{2}\pm(z_a^{(3)}+z_i^{(3)})-m_A}\int\udl{z_1^{(1)}}\e^{2\pi i(X_3-X_2)z^{(1)}}\\
&\qquad\times\prod_{a=1}^2\sbfunc{\pm(z^{(1)}-z_a^{(2)})+m_A}\, .
\end{split}$}}
\end{equation}
Again, since we removed the adjoint chiral field from the $U(3)$ node we can apply Aharony duality to it. In this case the rank of the group decreases, since some of the flavors that used to be attached to it became massive, so this node is not balanced anymore. Hence, we get
\begin{equation}
\makebox[\linewidth][c]{\scalebox{0.95}{$
\begin{split}
&\mathcal{Z}_{\mathcal{T}^\vee}=\mathcal{B}\,\e^{2\pi i(X_1-X_4)(Y_1+\sum_{\ga=1}^2Y_\ga)}\sbfunc{i\frac{Q}{2}-2m_A}^2\sbfunc{-i\frac{Q}{2}+4m_A}\sbfunc{i\frac{Q}{2}\pm(X_1-X_2)-2m_A}\\
&\qquad\times\sbfunc{i\frac{Q}{2}\pm(X_1-X_3)-2m_A}\sbfunc{i\frac{Q}{2}\pm(X_1-X_4)-2m_A}\prod_{\ga,\gb=3}^4\sbfunc{-i\frac{Q}{2}+(Y_\ga-Y_\gb)+2m_A}\\
&\qquad\times\prod_{\ga=3}^4\sbfunc{-i\frac{Q}{2}\pm(Y_1-Y_\ga)+3m_A}\int\udl{\vec{z}_2^{(3)}}\e^{2\pi i(X_1-X_4)\sum_{i=1}^2z_i^{(3)}}\prod_{i=1}^2\sbfunc{i\frac{Q}{2}\pm(z_i^{(3)}+Y_1)-2m_A}\\
&\qquad\times\prod_{\ga=3}^4\sbfunc{i\frac{Q}{2}\pm(z_i^{(3)}+Y_\ga)-m_A}\int\udl{\vec{z}_2^{(2)}}\e^{2\pi i(X_4-X_3)\sum_{a=1}^2z_a^{(2)}}\prod_{a,b=1}^2\sbfunc{i\frac{Q}{2}+(z_a^{(2)}-z^{(2)}_b)-2m_A}\\
&\qquad\times\prod_{a=1}^2\prod_{i=1}^2\sbfunc{\pm(z^{(2)}_a-z_i^{(3)})+m_A}\int\udl{z_1^{(1)}}\e^{2\pi i(X_3-X_2)z^{(1)}}\prod_{a=1}^2\sbfunc{\pm(z^{(1)}-z^{(2)}_a)+m_A}\, .
\end{split}$}}
\end{equation}
This concludes the first iteration of the sequential application of Aharony duality along the whole tail. In the second iteration, we again sequentially apply the duality starting from the left $U(1)$ gauge node, but stopping at the second last node in order to restore the adjoint chiral at the $U(2)$ gauge node labelled by $\vec{z}^{(3)}$. From the first application of Aharony duality we get
\begin{equation}
\makebox[\linewidth][c]{\scalebox{0.95}{$
\begin{split}
&\mathcal{Z}_{\mathcal{T}^\vee}=\mathcal{B}\,\e^{2\pi i(X_1-X_4)(Y_1+\sum_{\ga=1}^2Y_\ga)}\sbfunc{i\frac{Q}{2}-2m_A}^2\sbfunc{-i\frac{Q}{2}+4m_A}\sbfunc{i\frac{Q}{2}\pm(X_1-X_2)-2m_A}\\
&\qquad\times\sbfunc{i\frac{Q}{2}\pm(X_1-X_3)-2m_A}\sbfunc{i\frac{Q}{2}\pm(X_1-X_4)-2m_A}\sbfunc{i\frac{Q}{2}\pm(X_2-X_3)-2m_A}\\
&\qquad\times\prod_{\ga,\gb=3}^4\sbfunc{-i\frac{Q}{2}+(Y_\ga-Y_\gb)+2m_A}\prod_{\ga=3}^4\sbfunc{-i\frac{Q}{2}\pm(Y_1-Y_\ga)+3m_A}\\
&\qquad\times\int\udl{\vec{z}_2^{(3)}}\e^{2\pi i(X_1-X_4)\sum_{i=1}^2z_i^{(3)}}\prod_{i=1}^2\sbfunc{i\frac{Q}{2}\pm(z_i^{(3)}+Y_1)-2m_A}\prod_{\ga=3}^4\sbfunc{i\frac{Q}{2}\pm(z_i^{(3)}+Y_\ga)-m_A}\\
&\qquad\times\int\udl{\vec{z}_2^{(2)}}\e^{2\pi i(X_4-X_2)\sum_{a=1}^2z^{(2)}_a}\prod_{a=1}^2\sbfunc{\pm(z^{(2)}_a+Y_1)+m_A}\prod_{i=1}^2\sbfunc{\pm(z^{(2)}_a-z_i^{(3)})+m_A}\\
&\qquad\times\int\udl{z^{(1)}_1}\e^{2\pi i(X_3-X_2)z^{(1)}}\prod_{a=1}^2\sbfunc{i\frac{Q}{2}\pm(z^{(1)}+z^{(2)}_a)-m_A}\, .
\end{split}$}}
\end{equation}
Now we apply Aharony duality to the $U(2)$ gauge node labelled by $\vec{z}^{(2)}$
\begin{equation}
\makebox[\linewidth][c]{\scalebox{0.95}{$
\begin{split}
&\mathcal{Z}_{\mathcal{T}^\vee}=\mathcal{B}\,\e^{2\pi i(X_1+X_2-2X_4)Y_1}\e^{2\pi i(X_1-X_4)\sum_{\ga=1}^2Y_\ga}\sbfunc{i\frac{Q}{2}-2m_A}^2\sbfunc{-i\frac{Q}{2}+4m_A}\\
&\qquad\times\sbfunc{i\frac{Q}{2}\pm(X_1-X_2)-2m_A}\sbfunc{i\frac{Q}{2}\pm(X_1-X_3)-2m_A}\sbfunc{i\frac{Q}{2}\pm(X_1-X_4)-2m_A}\\
&\qquad\times\sbfunc{i\frac{Q}{2}\pm(X_2-X_3)-2m_A}\sbfunc{i\frac{Q}{2}\pm(X_2-X_4)-2m_A}\prod_{\ga,\gb=3}^4\sbfunc{-i\frac{Q}{2}+(Y_\ga-Y_\gb)+2m_A}\\
&\qquad\times\prod_{\ga=3}^4\sbfunc{-i\frac{Q}{2}\pm(Y_1-Y_\ga)+3m_A}\int\udl{\vec{z}_2^{(3)}}\e^{2\pi i(X_1-X_2)\sum_{i=1}^2z_i^{(3)}}\prod_{i,j=1}^2\sbfunc{-i\frac{Q}{2}+(z_i^{(3)}-z_j^{(3)})+2m_A}\\
&\qquad\times\prod_{i=1}^2\prod_{\ga=3}^4\sbfunc{i\frac{Q}{2}\pm(z_i^{(3)}+Y_\ga)-m_A}\int\udl{\vec{z}_2^{(2)}}\e^{2\pi i(X_4-X_2)\sum_{a=1}^2z^{(2)}_a}\prod_{a=1}^2\sbfunc{i\frac{Q}{2}\pm(z^{(2)}_a-Y_1)-m_A}\\
&\qquad\times\prod_{i=1}^2\sbfunc{i\frac{Q}{2}\pm(z^{(2)}_a+z_i^{(3)})-m_A}\int\udl{z^{(1)}_1}\e^{2\pi i(X_3-X_4)z^{(1)}}\prod_{a=1}^2\sbfunc{\pm(z^{(1)}-z^{(2)}_a)+m_A}\, .
\end{split}$}}
\end{equation}
This concludes also the second iteration. The last iteration only consists of applying Aharony duality on the original $U(1)$ node, so to restore the adjoint chiral also at the $U(2)$ node labelled by $\vec{z}^{(2)}$. Simplifying the contributions of the many singlets we produced by the sequential application of Aharony duality with those contained in the prefactor $\mathcal{B}$ and performing the change of variables $z^{(2)}_a\rightarrow-z^{(2)}_a$ we get
\be
\mathcal{Z}_{\mathcal{T}^\vee}&=&\e^{2\pi i(X_1+X_2-2X_4)Y_1}\e^{2\pi i(X_1-X_4)\sum_{\ga=3}^4Y_\ga}\int\udl{\vec{z}_2^{(3)}}\e^{2\pi i(X_1-X_2)\sum_{i=1}^2z_i^{(3)}}\nn\\
&\times&\prod_{i,j=1}^2\sbfunc{-i\frac{Q}{2}+(z_i^{(3)}-z_j^{(3)})+2m_A}\prod_{i=1}^2\prod_{\ga=3}^4\sbfunc{i\frac{Q}{2}\pm(z_i^{(3)}+Y_\ga)-m_A}\nn\\
&\times&\int\udl{\vec{z}_2^{(2)}}\e^{2\pi i(X_2-X_3)\sum_{a=1}^2z^{(2)}_a}\prod_{a,b=1}^2\sbfunc{-i\frac{Q}{2}+(z^{(2)}_a-z^{(2)}_b)+2m_A}\nn\\
&\times&\prod_{a=1}^2\sbfunc{i\frac{Q}{2}\pm(z^{(2)}_a+Y_1)-m_A}\prod_{i=1}^2\sbfunc{i\frac{Q}{2}\pm(z^{(2)}_a-z^{(3)}_i)-m_A}\nn\\
&\times&\sbfunc{-i\frac{Q}{2}+2m_A}\int\udl{z^{(2)}_1}\e^{2\pi i(X_3-X_4)z^{(1)}}\prod_{a=1}^2\sbfunc{i\frac{Q}{2}\pm(z^{(1)}-z^{(2)}_a)-m_A}\, .\nn\\
\ee
At this point we implement the redefinition of the fugacities \eqref{newfug211} and we also perform the change of variables $z^{(i)}\to z^{(i)}+Y^{(1)}$. By taking into account the tracelessness conditions $\sum_{n=1}^4X_n=\sum_{\ga=1}^2Y^{(2)}_\ga=0$, we get
\be
\mathcal{Z}_{\mathcal{T}^\vee}&=&\e^{4\pi i(X_1+X_2)Y^{(1)}}\int\udl{\vec{z}_2^{(3)}}\e^{2\pi i(X_1-X_2)\sum_{i=1}^2z_i^{(3)}}\prod_{i,j=1}^2\sbfunc{-i\frac{Q}{2}+(z_i^{(3)}-z_j^{(3)})+2m_A}\nn\\
&\times&\prod_{i=1}^2\prod_{\ga=1}^2\sbfunc{i\frac{Q}{2}\pm(z_i^{(3)}+Y^{(2)}_\ga)-m_A}\int\udl{\vec{z}_2^{(2)}}\e^{2\pi i(X_2-X_3)\sum_{a=1}^2z^{(2)}_a}\nn\\
&\times&\prod_{a,b=1}^2\sbfunc{-i\frac{Q}{2}+(z^{(2)}_a-z^{(2)}_b)+2m_A}\prod_{a=1}^2\sbfunc{i\frac{Q}{2}\pm(z^{(2)}_a+Y^{(1)})-m_A}\nn\\
&\times&\prod_{i=1}^2\sbfunc{i\frac{Q}{2}\pm(z^{(2)}_a-z_i)-m_A}\sbfunc{-i\frac{Q}{2}+2m_A}\int\udl{z^{(1)}_1}\e^{2\pi i(X_3-X_4)z^{(1)}}\nn\\
&\times&\prod_{a=1}^2\sbfunc{i\frac{Q}{2}\pm(z^{(1)}-z^{(2)}_a)-m_A}=\mathcal{Z}_{T^{[2,1^2]}[SU(4)]}(Y^{(1)},\vec Y^{(2)};\vec X;i\frac{Q}{2}-m_A)\, .
\label{pfmirror211}
\ee
This is precisely the partition function of $T^{[2,1^2]}[SU(4)]$, whose global symmetry is indeed $U(1)_{Y^{(1)}}\times SU(2)_{Y^{(2)}}\times SU(4)_X$ with the CB factor $SU(4)_X$ being enhanced at low energies.

Combining the results \eqref{pf211} and \eqref{pfmirror211} with the integral identity for the mirror self-duality of $T[SU(4)]$ \eqref{idmirror} we get that the partition function of $T_{[2,1^2]}[SU(4)]$ coincides with that of $T^{[2,1^2]}[SU(4)]$ provided that $m_A\leftrightarrow i\frac{Q}{2}-m_A$ as expected from Mirror Symmetry \eqref{mAchange}
\be
\mathcal{Z}_{T_{[2,1^2]}[SU(4)]}(\vec X;\vec Y^{(2)},Y^{(1)};m_A)=\mathcal{Z}_{T^{[2,1^2]}[SU(4)]}(Y^{(1)},\vec Y^{(2)};\vec X;i\frac{Q}{2}-m_A)\, .\nn\\
\ee

\section{Partition function computations in \boldmath$4d$}
\subsection{Intriligator--Pouliot duality}
\label{IP}

The Intriligator--Pouliot duality was first proposed in \cite{Intriligator:1995ne} and it relates the two following $4d$ $\mathcal{N}=1$ theories:

\medskip
\noindent \textbf{Theory A}: $USp(2N_c)$ gauge theory with $2N_f$ fundamental chirals and no superpotential $\mathcal{W}=0$.

\medskip
\noindent \textbf{Theory B}: $USp(2N_f-2N_c-4)$ gauge theory with $2N_f$ fundamental chirals, $N_f(2N_f-1)$ singlets (collected in an antisymmetric matrix $M_{ab}$) and superpotential $\hat{\mathcal{W}}=M^{ab}q_aq_b$.

At the level of the $S^3\times S^1$ partition function, this translates into an integral identity proved in Theorem 3.1 of \cite{2003math......9252R}
\be
\oint\udl{\vec{z}_{N_c}} \prod_{i=1}^{N_c}\prod_{a=1}^{2N_f}\Gpq{v_a z_i^{\pm1}}=\prod_{a<b}^{2N_f}\Gpq{v_av_b}\oint\udl{\vec{z}_{N_f-N_c-2}} \prod_{i=1}^{N_f-N_c-2}\prod_{a=1}^{2N_f}\Gpq{(pq)^{1/2} v_a^{-1} z_i^{\pm1}}\, ,\nn\\
\label{IP}
\ee
which holds provided that
\be
\prod_{a=1}^{2N_f}v_a=(pq)^{N_f-N_c-1}
\label{balancingIP}
\ee
and where we defined the integration measure as
\be
\udl{\vec{z}_N}=\frac{\left[(p;p)(q;q)\right]^N}{2^N N!}\prod_{i=1}^N\frac{\udl{z_i}}{2\pi i\,z_i} \frac{1}{\prod_{n=1}^N\Gpq{z_n^{\pm2}}\prod_{n<m}^N\Gpq{z_n^{\pm1}z_m^{\pm1}}} \, ,
\ee
which includes the contribution of the vector multiplet.

Notice that for $N_c=N$ and $N_f=N+2$ the dual theory is a WZ model of $(N+2)(2N+3)$ chiral fields and the identity \eqref{IP} reduces to
\be
\oint\udl{\vec{z}_{N}} \prod_{i=1}^{N}\prod_{a=1}^{2N+4}\Gpq{v_az_i^{\pm1}}=\prod_{a<b}^{2N+4}\Gpq{v_av_b}\, ,\nn\\
\label{IPconf}
\ee
with the condition
\be
\prod_{a=1}^{2N+4}v_a=pq\, ,
\label{balancingIPconf}
\ee
which was first conjectured in \cite{10.1155/S1073792801000526}.
\\

\subsection{$4d$ flip-flip duality as repeated Intriligator--Pouliot duality}
\subsubsection{Derivation of $E[USp(6)] \leftrightarrow FFE[USp(6)]$}
\label{4dff}

Recall that the flip-flip duality of $T[SU(N)]$ in $3d$ can be realized as a repeated application of the Aharony duality and one-monopole duality. In $4d$, as claimed in the main text, the flip-flip duality of $E[USp(2 N)]$ can be also realized using the Intriligator--Pouliot duality only. In this appendix, we use the superconformal index to show how to obtain $FFE[USp(2 N)]$, the flip-flip dual of $E[USp(2 N)]$, by sequential Intriligator--Pouliot dualities. As an explicit example, we take $N = 3$, which requires the Intriligator--Pouliot duality three times in total to obtain the flip-flip dual.

The superconformal index of $E[USp(6)]$ is given by
\begin{align}
&\mathcal{I}_{E[USp(6)]}(\vec x;\vec y;t,c) \nn \\
&= \prod_{n = 1}^3 \frac{\Gpq{c\,y_3^{\pm1} x_n^{\pm1}}}{\Gpq{t^{-2} c^2} \Gpq{t^{-1} c^2}}\oint \udl{\vec z^{(1)}_{1}} \udl{\vec z^{(2)}_{2}} \Gpq{p q t^{-1}}^3 \prod_{i < j}^2 \Gpq{p q t^{-1} z^{(2)}_i{}^{\pm1} z^{(2)}_j{}^{\pm1}} \nn\\
&\quad \times \frac{\prod_{j = 1}^2 \Gpq{t^{1/2} z^{(1)}{}^{\pm1} z^{(2)}_j{}^{\pm1}}}{\Gpq{c y_2^{\pm1} z^{(1)}{}^{\pm1}}} \frac{\prod_{i=1}^2 \prod_{n = 1}^3 \Gpq{t^{1/2} z^{(2)}_i{}^{\pm1} x_n^{\pm1}}}{\prod_{i=1}^{2}\Gpq{t^{1/2} c y_3^{\pm1} z^{(2)}_i{}^{\pm1}}} \nn \\
&\quad \times \Gpq{t^{-1} c y_1^{\pm1} z^{(1)}{}^{\pm1}} \prod_{i = 1}^2 \Gpq{t^{-1/2} c y_2^{\pm1} z^{(2)}_i{}^{\pm1}} \,.
\end{align}
As a first step, we apply the Intriligator--Pouliot duality on the leftmost node, which corresponds to the following identity:
\begin{align}
&\oint \udl{\vec z^{(1)}_{1}} \Gamma_e(t^{-1} c y_1^{\pm1} z^{(1)}{}^{\pm1}) \Gamma_e(p q  c^{-1} y_2^{\pm1} z^{(1)}{}^{\pm1}) \prod_{j = 1}^{2} \Gamma_e(t^{1/2} z^{(1)}{}^{\pm1} z^{(2)}_j{}^{\pm1}) \nonumber \\
&= \Gamma_e(t^{-2} c^2) \Gamma_e(p q t^{-1} y_1^{\pm1} y_2^{\pm1}) \prod_{j = 1}^{2} \Gamma_e(t^{-1/2} c y_1^{\pm1} z^{(2)}_j{}^{\pm1}) \nonumber \\
&\quad \times \Gamma_e(p^{2} q^{2} c^{-2}) \prod_{j = 1}^{2} \Gamma_e(p q t^{1/2} c^{-1} y_2^{\pm1} z^{(2)}_j{}^{\pm1}) \Gamma_e(t)^2 \prod_{i < j}^2 \Gamma_e(t z^{(2)}_i{}^{\pm1} z^{(2)}_j{}^{\pm1}) \oint \udl{\vec z^{(1)}_{1}} \nonumber \\
&\quad \times \Gamma_e(p^{1/2} q^{1/2} t c^{-1} y_1^{\pm1} z^{(1)}{}^{\pm1}) \Gamma_e(p^{-1/2} q^{-1/2} c y_2^{\pm1} z^{(1)}{}^{\pm1}) \prod_{j = 1}^{2} \Gamma_e(p^{1/2} q^{1/2} t^{-1/2} z^{(1)}{}^{\pm1} z^{(2)}_j{}^{\pm1}) \,.
\end{align}
Next we apply the Intriligator--Pouliot duality on the middle gauge node. We thus collect the $z^{(2)}$ dependent factors and apply the following identity:
\begin{align}
&\oint \udl{\vec z^{(2)}_{2}} \prod_{i = 1}^2 \prod_{n = 1}^3 \Gamma_e(t^{1/2} z^{(2)}_i{}^{\pm1} x_n^{\pm1}) \nonumber \\
&\times \prod_{j = 1}^{2} \Gamma_e(p q t^{-1/2} c^{-1} y_3^{\pm1} z^{(2)}_j{}^{\pm1}) \prod_{j = 1}^{2} \Gamma_e(t^{-1/2} c y_1^{\pm1} z^{(2)}_j{}^{\pm1}) \prod_{j = 1}^{2} \Gamma_e(p^{1/2} q^{1/2} t^{-1/2} z^{(1)}{}^{\pm1} z^{(2)}_j{}^{\pm1}) \nonumber \\
&= \Gpq{t}^2 \prod_{m < n}^3 \Gpq{t x_m^{\pm1} x_n^{\pm1}} \prod_{n = 1}^3 \Gamma_e(p q c^{-1} x_n^{\pm1} y_3^{\pm1}) \prod_{n = 1}^3 \Gamma_e(c x_n^{\pm1} y_1^{\pm1}) \nn \\
&\quad \times \Gamma_e(p^2 q^2 t^{-1} c^{-2}) \Gamma_e(p q t^{-1} y_3^{\pm1} y_1^{\pm1}) \Gamma_e(p^{3/2} q^{3/2} t^{-1} c^{-1} y_3^{\pm1} z^{(1)}{}^{\pm1}) \nonumber \\
&\quad \times \Gamma_e(t^{-1} c^2) \Gamma_e(p^{1/2} q^{1/2} t^{-1} c y_1^{\pm1} z^{(1)}{}^{\pm1}) \oint \udl{\vec z'^{(2)}_2} \prod_{i = 1}^2 \prod_{n = 1}^3 \Gamma_e(p^{1/2} q^{1/2} t^{-1/2} z'^{(2)}_i{}^{\pm1} x_n^{\pm1}) \nonumber \\
&\quad \times \prod_{i = 1}^2 \Gamma_e(p^{-1/2} q^{-1/2} t^{1/2} c y_3^{\pm1} z'^{(2)}_i{}^{\pm1}) \prod_{i = 1}^2 \Gamma_e(p^{1/2} q^{1/2} t^{1/2} c^{-1} y_1^{\pm1} z'^{(2)}_i{}^{\pm1}) \prod_{i = 1}^2 \Gamma_e(t^{1/2} z^{(1)}{}^{\pm1} z'^{(2)}_i{}^{\pm1}) \,.
\end{align}
Lastly, we collect the $z^{(1)}$ dependent factors resulting from the previous two applications of the Intriligator--Pouliot duality, which become
\begin{align}
&\oint \udl{\vec z^{(1)}_{1}} \Gamma_e(p^{-1/2} q^{-1/2} c y_2^{\pm1} z^{(1)}{}^{\pm1}) \Gamma_e(p^{3/2} q^{3/2} t^{-1} c^{-1} y_3^{\pm1} z^{(1)}{}^{\pm1}) \prod_{i = 1}^2 \Gamma_e(t^{1/2} z^{(1)}{}^{\pm1} z'^{(2)}_i{}^{\pm1}) \nonumber \\
&= \Gamma_e(p^{-1} q^{-1} c^2) \Gamma_e(p q t^{-1} y_2^{\pm1} y_3^{\pm1}) \prod_{i = 1}^2 \Gamma_e(p^{-1/2} q^{-1/2} t^{1/2} c y_2^{\pm1} z'^{(2)}_i{}^{\pm1}) \Gamma_e(p^3 q^3 t^{-2} c^{-2}) \nonumber \\
&\quad \times \prod_{i = 1}^2 \Gamma_e(p^{3/2} q^{3/2} t^{-1/2} c^{-1} y_3^{\pm1} z'^{(2)}_i{}^{\pm1}) \Gamma_e(t)^2 \prod_{i < j}^2 \Gamma_e(t z'^{(2)}_i{}^{\pm1} z'^{(2)}_j{}^{\pm1}) \oint \udl{\vec z'^{(1)}_1} \nonumber \\
&\quad \times \Gamma_e(p q c^{-1} y_2^{\pm1} z'^{(1)}{}^{\pm1}) \Gamma_e(p^{-1} q^{-1} t c y_3^{\pm1} z'^{(1)}{}^{\pm1}) \prod_{i = 1}^2 \Gamma_e(p^{1/2} q^{1/2} t^{-1/2} z'^{(1)}{}^{\pm1} z'^{(2)}_i{}^{\pm1}) \,.
\end{align}
Combining all the remaining factors, we obtain the following expression for the entire superconformal index:
\begin{align}
&\mathcal{I}_{E[USp(6)]}(\vec x;\vec y;t,c) \nonumber \\
&= \prod_{m < n}^3 \Gpq{t x_m^\pm x_n^\pm}  \prod_{m < n}^3 \Gpq{p q t^{-1} y_m^\pm y_n^\pm} \nonumber \\
&\quad \times \frac{\prod_{n = 1}^3 \Gamma_e(c y_1^{\pm1} x_n^{\pm1})}{\Gamma_e(p^{-2} q^{-2} t^{2} c^{2}) \Gamma_e(p^{-1} q^{-1} t c^{2})} \oint \udl{\vec z'^{(1)}_1} \udl{\vec z'^{(2)}_2}\Gpq{t}^3 \prod_{i < j}^2 \Gpq{t z'^{(2)}_i{}^{\pm1} z'^{(2)}_j{}^{\pm1}}  \nonumber \\
&\quad \times \frac{\prod_{i = 1}^2 \Gamma_e(p^{1/2} q^{1/2} t^{-1/2} z'^{(1)}{}^{\pm1} z'^{(2)}_i{}^{\pm1})}{
\Gamma_e( c^2 y_2^{\pm1} z'^{(1)}{}^{\pm1})} \frac{\prod_{i = 1}^2 \prod_{n = 1}^3 \Gamma_e(p^{1/2} q^{1/2} t^{-1/2} z'^{(2)}_i{}^{\pm1} x_n^{\pm1})}{\prod_{i = 1}^2 \Gamma_e(p^{1/2} q^{1/2} t^{-1/2} c y_1^{\pm1} z'^{(2)}_i{}^{\pm1})} \nn \\
&\quad \times \Gamma_e(p^{-1} q^{-1} t c y_3^{\pm1} z'^{(1)}{}^{\pm1}) \prod_{i = 1}^2 \Gamma_e(p^{-1/2} q^{-1/2} t^{1/2} c y_2^{\pm1} z'^{(2)}_i{}^{\pm1}) \nonumber \\
&= \mathcal{I}_{FFE[USp(6)]}(\vec x;\vec y;p q/t,c) \,.
\end{align}
This proves the index equality of the flip-flip duality of $E[USp(6)]$ and the sequential applications of the Intriligator--Pouliot duality. Note that while the variables $y_n$ appear in the opposite way compared to the original definition, the index is invariant under such a shuffling of variables because it is a Weyl symmetry of the $USp(6)_y$ global symmetry. 
\\

\subsubsection{The case $\rho = [2,1]$}
\label{app:next-to-maximal}
In this appendix, we show how to obtain $E^{[2,1]}[USp(6)]$ from its flip-flip dual $\mathsf T^\vee$ by sequential applications of the Intriligator--Pouliot duality.
We start with the index of theory $\mathsf T^\vee$, which is given by \eqref{eq:T_B}. For $N = 3$, it is written as follows:
\begin{equation}
\makebox[\linewidth][c]{\scalebox{0.95}{$
\begin{split}
&\mathcal{I}_{\mathsf T^\vee}\left(\vec x;y^{(1)},y^{(2)};t,c\right) \\
&= \Gpq{p q t^{-1}}^2 \prod_{n < m}^3 \Gpq{p q t^{-1} x^{\pm1}_n x^{\pm1}_m} \Gpq{t^\frac{3}{2} y^{(1)}{}^{\pm1} y^{(2)}{}^{\pm1}} \prod_{i = 1}^2 \Gpq{t^i} \Gpq{t^{-1} c^2}\\
&\quad \times \frac{\Gpq{c\,x_3^{\pm1} y^{(2)}{}^{\pm1}} \prod_{n = 1}^2 \Gpq{c\,x_3^{\pm1} \left(t^{n-\frac{3}{2}} y^{(1)}\right)^{\pm1}}}{\Gpq{p^{-2} q^{-2} t^2 c^2} \Gpq{p^{-1} q^{-1} t c^2}}\oint \udl{\vec z^{(1)}_{1}} \udl{\vec z^{(2)}_{2}} \Gpq{t}^3 \prod_{i < j}^{2} \Gamma_e(t z^{(2)}_i{}^{\pm1} z^{(2)}_j{}^{\pm1})\\
&\quad \times \frac{\prod_{j = 1}^2 \Gpq{p^{1/2} q^{1/2} t^{-1/2} z^{(1)}{}^{\pm1} z^{(2)}_j{}^{\pm1}}}{\Gpq{c x_2^{\pm1} z^{(1)}{}^{\pm1}}} \frac{\prod_{i=1}^2 \Gpq{p^{1/2} q^{1/2} t^{-1} z^{(2)}_i{}^{\pm1} y^{(1)}{}^{\pm1}} \Gpq{p^{1/2} q^{1/2} t^{-1/2}z^{(2)}_i{}^{\pm1}y^{(1)}{}^{\pm1}}}{\prod_{i=1}^{2}\Gpq{p^{1/2} q^{1/2} t^{-1/2}c\,x_3^{\pm1}z^{(2)}_i{}^{\pm1}}} \\
&\quad \times \Gpq{p^{-1} q^{-1} t c x_1^{\pm1} z^{(1)}{}^{\pm1}} \prod_{i = 1}^2 \Gpq{p^{-1/2} q^{-1/2} t^{1/2} c x_2^{\pm1} z^{(2)}_i{}^{\pm1}} \,.
\end{split}$}}
\end{equation}
We first apply the Intriligator--Pouliot duality on the leftmost node, which corresponds to the following identity:
\begin{align}
&\oint \udl{\vec z^{(1)}_{1}} \Gamma_e(p^{-1} q^{-1} t c x_1^{\pm1} z^{(1)}{}^{\pm1}) \Gamma_e(p q  c^{-1} x_2^{\pm1} z^{(1)}{}^{\pm1}) \prod_{j = 1}^{2} \Gamma_e(p^{1/2} q^{1/2} t^{-1/2} z^{(1)}{}^{\pm1} z^{(2)}_j{}^{\pm1}) \nonumber \\
&= \Gamma_e(p^{-2} q^{-2} t^2 c^2) \Gamma_e(t x_1^{\pm1} x_2^{\pm1}) \prod_{j = 1}^{2} \Gamma_e(p^{-1/2} q^{-1/2} t^{1/2} c x_1^{\pm1} z^{(2)}_j{}^{\pm1}) \nonumber \\
&\quad \times \Gamma_e(p^{2} q^{2} c^{-2}) \prod_{j = 1}^{2} \Gamma_e(p^{3/2} q^{3/2} t^{-1/2} c^{-1} x_2^{\pm1} z^{(2)}_j{}^{\pm1}) \Gamma_e(p q t^{-1})^2 \prod_{i < j}^2 \Gamma_e(p q t^{-1} z^{(2)}_i{}^{\pm1} z^{(2)}_j{}^{\pm1}) \nonumber \\
&\quad \times \oint \udl{\vec z^{(1)}_{1}} \Gamma_e(p^{3/2} q^{3/2} t^{-1} c^{-1} x_1^{\pm1} z^{(1)}{}^{\pm1}) \Gamma_e(p^{-1/2} q^{-1/2} c x_2^{\pm1} z^{(1)}{}^{\pm1}) \prod_{j = 1}^{2} \Gamma_e(t^{1/2} z^{(1)}{}^{\pm1} z^{(2)}_j{}^{\pm1}) \,.
\end{align}
Next, we collect the $z^{(2)}$ dependent factors and apply the Intriligator--Pouliot duality again:
\begin{equation}
\makebox[\linewidth][c]{\scalebox{0.95}{$
\begin{split}
&\oint \udl{\vec z^{(2)}_{2}} \prod_{i = 1}^{2} \Gamma_e(p^{1/2} q^{1/2} t^{-1} z^{(2)}_i{}^{\pm1} y^{(1)\pm1}) \prod_{i = 1}^2\Gamma_e(p^{1/2} q^{1/2} t^{-1/2} z^{(2)}_i{}^{\pm1} y^{(2)\pm1}) \\
&\times \prod_{j = 1}^{2} \Gamma_e(p^{1/2} q^{1/2} t^{1/2} c^{-1} x_3^{\pm1} z^{(2)}_j{}^{\pm1}) \prod_{j = 1}^{2} \Gamma_e(p^{-1/2} q^{-1/2} t^{1/2} c x_1^{\pm1} z^{(2)}_j{}^{\pm1}) \prod_{j = 1}^{2} \Gamma_e(t^{1/2} z^{(1)}{}^{\pm1} z^{(2)}_j{}^{\pm1})  \\
&= \Gamma_e(p q t^{-2}) \Gamma_e(p q t^{-3/2} y^{(1)\pm1} y^{(2)\pm1}) \Gamma_e(p q t^{-1/2} c^{-1} y^{(1)\pm1} x_3^{\pm1}) \Gamma_e(t^{-1/2} c y^{(1)\pm1} x_1^{\pm1}) \Gamma_e(p^{1/2} q^{1/2} t^{-1/2} y^{(1)\pm1} z^{(1)}{}^{\pm1}) \\
&\quad \times \Gamma_e(p q c^{-1} y^{(2)\pm1} x_3^{\pm1}) \Gamma_e(c y^{(2)\pm1} x_1^{\pm1}) \Gamma_e(p q t c^{-2}) \Gamma_e(t x_3^{\pm1} x_1^{\pm1}) \Gamma_e(p^{1/2} q^{1/2} t c^{-1} x_3^{\pm1} z^{(1)}{}^{\pm1}) \\
&\quad \times \Gamma_e(p^{-1} q^{-1} t c^2) \Gamma_e(p^{-1/2} q^{-1/2} t c x_1^{\pm1} z^{(1)}{}^{\pm1}) \oint \udl{\vec z'^{(2)}_1} \Gamma_e(t z'^{(2)}{}^{\pm1} y^{(1)\pm1}) \Gamma_e(t^{1/2} z'^{(2)}{}^{\pm1} y^{(2)\pm1})  \\
&\quad \times \Gamma_e(t^{-1/2} c x_3^{\pm1} z'^{(2)}{}^{\pm1}) \Gamma_e(p q t^{-1/2} c^{-1} x_1^{\pm1} z'^{(2)}{}^{\pm1}) \Gamma_e(p^{1/2} q^{1/2} t^{-1/2} z^{(1)}{}^{\pm1} z'^{(2)}{}^{\pm1}) \,.
\end{split}$}}
\end{equation}
Lastly, we collect the $z^{(1)}$ dependent factors, which become
\begin{align}
&\oint \udl{\vec z^{(1)}_{1}} \Gamma_e(p^{-1/2} q^{-1/2} c x_2^{\pm1} z^{(1)}{}^{\pm1}) \Gamma_e(p^{1/2} q^{1/2} t c^{-1} x_3^{\pm1} z^{(1)}{}^{\pm1}) \nonumber \\
&\times \Gamma_e(p^{1/2} q^{1/2} t^{-1/2} z^{(1)}{}^{\pm1} y^{(1)\pm1}) \Gamma_e(p^{1/2} q^{1/2} t^{-1/2} z^{(1)}{}^{\pm1} z'^{(2)}{}^{\pm1}) \nonumber \\
&= \Gamma_e(p^{-1} q^{-1} c^2) \Gamma_e(t x_2^{\pm1} x_3^{\pm1}) \Gamma_e(t^{-1/2} c x_2^{\pm1} y^{(1)\pm1}) \Gamma_e(t^{-1/2} c x_2^{\pm1} z'^{(2)}{}^{\pm1}) \nonumber \\
&\quad \times \Gamma_e(p q t^2 c^{-2}) \Gamma_e(p q t^{1/2} c^{-1} x_3^{\pm1} y^{(1)\pm1}) \Gamma_e(p q t^{1/2} c^{-1} x_3^{\pm1} z'^{(2)}{}^{\pm1}) \Gamma_e(p q t^{-1})^2 \Gamma_e(p q t^{-1} y^{(1)\pm1} z'^{(2)}{}^{\pm1}) \nonumber \\
&\quad \times \oint \udl{\vec z'^{(1)}_1} \Gamma_e(p q c^{-1} x_2^{\pm1} z'^{(1)}{}^{\pm1}) \Gamma_e(t^{-1} c x_3^{\pm1} z'^{(1)}{}^{\pm1}) \Gamma_e(t^{1/2} z'^{(1)}{}^{\pm1} y^{(1)\pm1}) \Gamma_e(t^{1/2} z'^{(1)}{}^{\pm1} z'^{(2)}{}^{\pm1}) \,.
\end{align}
Combining all the remaining factors, we obtain the following expression for the entire superconformal index:
\begin{align}
&\Gamma_e(t^{-1/2} c x_1^{\pm1} y^{(1)\pm1}) \Gamma_e(t^{-1/2} c x_2^{\pm1} y^{(1)\pm1}) \Gamma_e(c x_1^{\pm1} y^{(2)\pm1})  \Gamma_e(p q t^2 c^{-2}) \nonumber \\
&\times \oint \udl{\vec z'^{(1)}_1} \udl{\vec z'^{(2)}_1} \Gpq{p q t^{-1}}^2 \Gamma_e(t^{1/2} z'^{(1)}{}^{\pm1} y^{(1)\pm1})  \nonumber \\
&\times \Gamma_e(p q c^{-1} x_2^{\pm1} z'^{(1)}{}^{\pm1}) \Gamma_e(t^{-1} c x_3^{\pm1} z'^{(1)}{}^{\pm1}) \Gamma_e(t^{1/2} z'^{(2)}{}^{\pm1} y^{(2)\pm1}) \nonumber \\
&\times \Gamma_e(p q t^{-1/2} c^{-1} x_1^{\pm1} z'^{(2)}{}^{\pm1}) \Gamma_e(t^{-1/2} c x_2^{\pm1} z'^{(2)}{}^{\pm1}) \Gamma_e(t^{1/2} z'^{(1)}{}^{\pm1} z'^{(2)}{}^{\pm1}) \nonumber \\
&= \mathcal{I}_{E[USp(6)]^{[2,1]}}\left(\vec x;y^{(1)},y^{(2)};p q/t,c\right) \,,
\end{align}
which completes the derivation.
\\

\subsection{Alternative derivation of $E_{[N-1,1]}[USp(2 N)] \leftrightarrow E^{[N-1,1]}[USp(2 N)]$}
\label{4dabelian}

In section \ref{sec:next-to-maximal}, we derived the mirror-like duality between $E_{[N-1,1]}[USp(2 N)]$ and $E^{[N-1,1]}[USp(2 N)]$ using the \eusp duality web. In this appendix  we provide an alternative derivation of this duality.

\begin{figure}[tbp]
\centering
\makebox[\linewidth][c]{\includegraphics[scale=0.34]{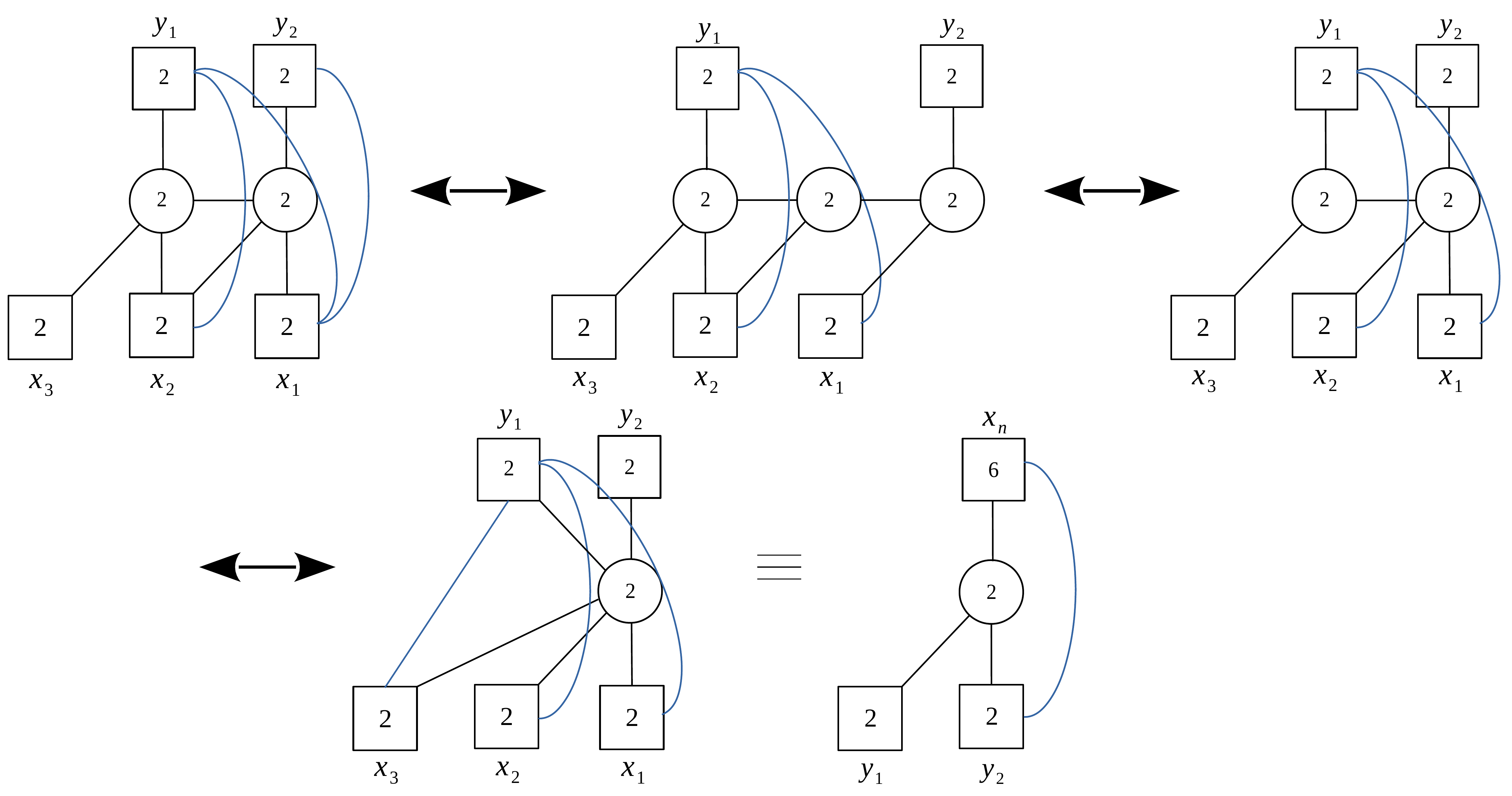}}
\caption{\label{fig:4d mirror} A direct derivation of the $4d$ mirror-like duality between $E_{[N-1,1]}[USp(2 N)]$ and $E^{[N-1,1]}[USp(2 N)]$ using the Intriligator--Pouliot duality. }
\end{figure}

First we note that the $3d$ counterpart of this duality is the abelian Mirror Symmetry which maps the $3d$ SQED with $N$ flavors to an abelian quiver of $N-1$ gauge nodes with one flavor attached to each end of the quiver. This abelian mirror can be obtained by sequential applications of the Aharony duality between the SQED with one flavor and the XYZ Wess-Zumino model \cite{Kapustin:1999ha}. Accordingly one can expect that the $4d$ mirror-like duality between $E_{[N-1,1]}[USp(2 N)]$ and $E^{[N-1,1]}[USp(2 N)]$ is also obtained by sequential applications of the Intriligator--Pouliot duality in the confining case, which indeed turns out to be true. For example, this procedure for $N = 3$ is shown in Figure \ref{fig:4d mirror}.
In this appendix, we also exhibit the derivation of the duality in terms of the $4d$ superconformal index.

Let us start with $E^{[2,1]}[USp(6)]$, whose superconformal index is given by
\begin{align}
\label{eq:abelian quiver}
&\mathcal{I}_{E^{[2,1]}[USp(6)]}\left(y^{(1)},y^{(2)};\vec x;p q/t,c\right) \nn \\
&=\Gamma_e(t^{-1/2} c x_1^{\pm1} y^{(1)\pm1}) \Gamma_e(t^{-1/2} c x_2^{\pm1} y^{(1)\pm1}) \Gamma_e(c x_1^{\pm1} y^{(2)\pm1})  \Gamma_e(p q t^2 c^{-2}) \nonumber \\
&\quad \times \oint \udl{\vec z^{(1)}_1} \udl{\vec z^{(2)}_1} \Gpq{p q t^{-1}}^{2} \Gamma_e(t^{1/2} z^{(1)}{}^{\pm1} y^{(1)\pm1})  \nonumber \\
&\quad \times \Gamma_e(p q c^{-1} x_2^{\pm1} z^{(1)}{}^{\pm1}) \Gamma_e(t^{-1} c x_3^{\pm1} z^{(1)}{}^{\pm1}) \Gamma_e(t^{1/2} z^{(2)}{}^{\pm1} y^{(2)\pm1}) \nonumber \\
&\quad \times \Gamma_e(p q t^{-1/2} c^{-1} x_1^{\pm1} z^{(2)}{}^{\pm1}) \Gamma_e(t^{-1/2} c x_2^{\pm1} z^{(2)}{}^{\pm1}) \Gamma_e(t^{1/2} z^{(1)}{}^{\pm1} z^{(2)}{}^{\pm1}) \,.
\end{align}
We can apply the Intriligator--Pouliot duality relating a WZ model with 15 chirals to the $USp(2)$ theory with six chirals
to trade some of the chirals in  \eqref{eq:abelian quiver} for a new $USp(2)$ gauge node: 
\begin{align}
&\Gamma_e(c x_1^{\pm1} y^{(2)}{}^{\pm1}) \Gamma_e(p q t^{-1/2} c^{-1} x_1^{\pm1} z^{(2)}{}^{\pm1}) \Gamma_e(t^{1/2} z^{(2)}{}^{\pm1} y^{(2)}{}^{\pm1}) \nonumber \\
&= \Gamma_e(p^2 q^2 t^{-1} c^{-2}) \Gamma_e(t) \Gamma_e(c^2) \oint \udl{\vec z'^{(1)}_1} \Gamma_e(p^{-1/2} q^{-1/2} t^{1/2} c z'^{(1)}{}^{\pm1} y^{(2)}{}^{\pm1}) \nonumber \\
& \quad \times \Gamma_e(p^{1/2} q^{1/2} t^{-1/2} z'^{(1)}{}^{\pm1} x_1^{\pm1}) \Gamma_e(p^{1/2} q^{1/2} c^{-1} z'^{(1)}{}^{\pm1} z^{(2)}{}^{\pm1}) \,,
\end{align}
in this way we obtain the second quiver in Figure \ref{fig:4d mirror}.

We then observe that collecting the factors depending on $z^{(2)}$, we can apply the Intriligator--Pouliot duality to confine the second node in the  second quiver in Figure \ref{fig:4d mirror}:
\begin{align}
&\oint \udl{\vec z^{(2)}_1} \Gamma_e(t^{1/2} z^{(1)}{}^{\pm1} z^{(2)}{}^{\pm1}) \Gamma_e(t^{-1/2} c x_2^{\pm1} z^{(2)}{}^{\pm1}) \Gamma_e(p^{1/2} q^{1/2} c^{-1} z'^{(1)}{}^{\pm1} z^{(2)}{}^{\pm1}) \nn \\
&= \Gamma_e(c z^{(1)}{}^{\pm1} x_2^{\pm1}) \Gamma_e(p^{1/2} q^{1/2} t^{-1/2} x_2^{\pm1} z'^{(1)}{}^{\pm1}) \Gamma_e(p^{1/2} q^{1/2} t^{1/2} c^{-1} z^{(1)}{}^{\pm1} z'^{(1)}{}^{\pm1}) \nn \\
&\quad \times \Gamma_e(t) \Gamma_e(t^{-1} c^2) \Gamma_e(p q c^{-2})\,,
\end{align}
we then arrive at the third quiver in Figure \ref{fig:4d mirror}.

Then we collect the factors depending on $z^{(1)}$ and apply again Intriligator--Pouliot duality to confine this node:
\begin{align}
&\oint \udl{\vec z^{(1)}_1} \Gamma_e(t^{1/2} z^{(1)}{}^{\pm1} y^{(1)}{}^{\pm1}) \Gamma_e(t^{-1} c z^{(1)}{}^{\pm1} x_3^{\pm1}) \Gamma_e(p^{1/2} q^{1/2} t^{1/2} c^{-1} z^{(1)}{}^{\pm1} z'^{(1)}{}^{\pm1})  \nonumber \\
&= \Gamma_e(t^{-1/2} c x_3^{\pm1} y^{(1)}{}^{\pm1}) \Gamma_e(p^{1/2} q^{1/2} t^{-1/2} x_3^{\pm1} z'^{(1)}{}^{\pm1}) \Gamma_e(p^{1/2} q^{1/2} t c^{-1} y^{(1)}{}^{\pm1} z'^{(1)}{}^{\pm1}) \nn \\
&\quad \times \Gamma_e(t) \Gamma_e(t^{-2} c^2) \Gamma_e(p q t c^{-2}) \,.
\end{align}
 Collecting the remaining factors, we obtain the partition function of the last quiver in  Figure \ref{fig:4d mirror}:
\begin{align}
&\mathcal I_{E^{[2,1]}[USp(6)]}\left(y^{(1)},y^{(2)},\vec x,t,c\right) \nn \\
&= \frac{\prod_{n=1}^3 \Gpq{t^{-1/2} c y^{(1)}{}^{\pm1} x_n^{\pm1}}}{\Gpq{p^{-1} q^{-1} t c^2}} \oint\udl{\vec z'^{(1)}_{1}} \Gpq{t} \Gpq{p^{1/2} q^{1/2} t c^{-1} y^{(1)}{}^{\pm1} z'^{(1)}{}^{\pm1}} \nn\\
&\quad \times \Gpq{p^{-1/2} q^{-1/2} t^{1/2} c y^{(2)}{}^{\pm1} z'^{(1)}{}^{\pm1}}\prod_{n=1}^3 \Gpq{p^{1/2} q^{1/2} t^{-1/2} x_n^{\pm1}z'^{(1)}{}^{\pm1}} \nn \\
&= \mathcal I_{E_{[2,1]}[USp(6)]}\left( \vec x,y^{(2)},y^{(1)},t,c\right) \,,
\end{align}
which coincides with the superconformal index of $E_{[2,1]}[USp(6)]$ given in \eqref{eq:T_C}. Applying this procedurefor generic $N$ we prove the identity between  the indices of $\mathcal I_{E_{[N-1,1]}[USp(2 N)]}$ and $\mathcal I_{E^{[N-1,1]}[USp(2 N)]}$.
\\

\bibliographystyle{ytphys}

\end{document}